\documentclass[12pt]{article}
\usepackage{epsf,epsfig,here,cite}
\newcommand{\rhobar}{\bar {\rho}}
\newcommand{\etabar}{\bar{\eta}}
\newcommand{\epsilonk}{\varepsilon_K}

\newcommand{\dmd}{\Delta m_d}
\newcommand{\dms}{\Delta m_s}
\newcommand{\snb}{\sin 2\beta}
\newcommand{\BK}{B_K}
\newcommand{\fbdsqbd}{F_{B_d} \sqrt{\hat B_{B_d}}}
\newcommand{\fbssqbs}{F_{B_s} \sqrt{\hat B_{B_s}}}
\newcommand{\vcb}{\left | {V_{cb}} \right |}
\newcommand{\vub}{\left | {V_{ub}} \right |}
\def\utfit{{\bf{U}}\kern-.24em{\bf{T}}\kern-.21em{\it{fit}}\@}

\begin{document}
\pagestyle{empty}
\pagenumbering{arabic}
\begin{center}
  \begin{LARGE}
    \textbf{The {\utfit} Collaboration Report}\\
    \textbf{on the Status of the Unitarity Triangle}\\
    \textbf{beyond the Standard Model}\\\vspace{0.2cm}
    \textbf{I. Model-independent Analysis and}\\
    \textbf{Minimal Flavour Violation}\\
  \end{LARGE}  
\end{center}
\vspace{0.4cm}
\begin{figure}[htb!]
  \begin{center}
    \includegraphics[width=2.5cm]{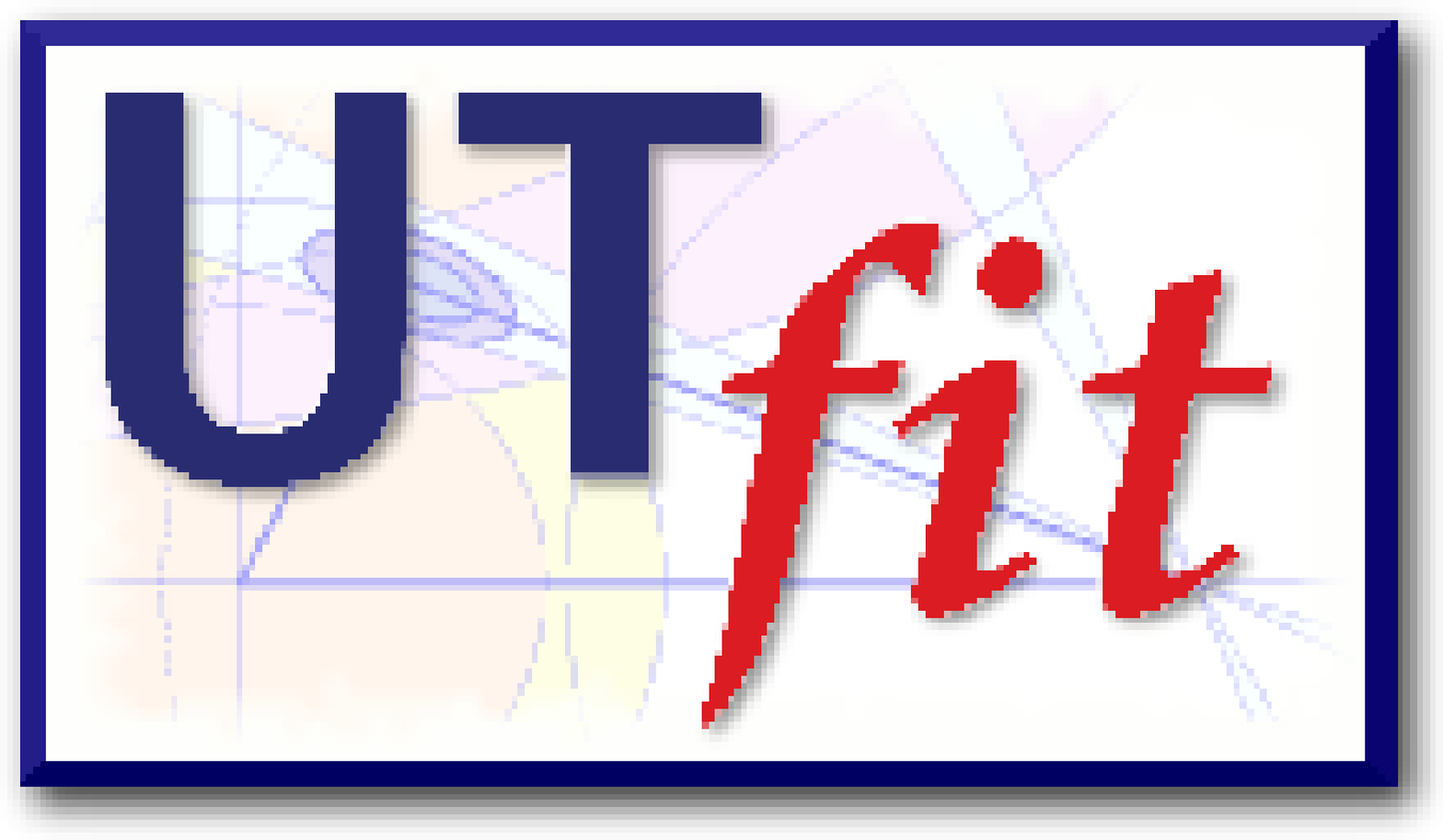}
  \end{center}
\end{figure}
\vspace{-1.0cm}
\begin{center}
  \Large{\textbf{UT}}\large{\textit{fit}}\large{~Collaboration :} \\
\end{center}
\begin{center}
  \begin{large}
  \textbf{M.~Bona$^{(a)}$, M.~Ciuchini$^{(b)}$, E.~Franco$^{(c)}$,
    V.~Lubicz$^{(b)}$, } \\
  \textbf{G. Martinelli$^{(c)}$, F. Parodi$^{(d)}$, M.
    Pierini$^{(e)}$, P.
    Roudeau$^{(f)}$, }\\
  \textbf{C. Schiavi$^{(d)}$, L.~Silvestrini$^{(c)}$, A.
    Stocchi$^{(f)}$, and V.~Vagnoni$^{(g)}$}
  \end{large}
\end{center}
\begin{center}
  \noindent 
  \begin{scriptsize}
    \noindent
    \textbf{$^{(a)}$   INFN,  Sez. di Torino,}\\
    \hspace*{0.5cm}{Via P. Giuria 1, I-10125  Torino, Italy}\\
    \textbf{$^{(b)}$ Dip. di Fisica, Universit{\`a} di Roma Tre
      and INFN,  Sez. di Roma III,}\\
    \hspace*{0.5cm}{Via della Vasca Navale 84, I-00146 Roma, Italy}\\
    \noindent
    \textbf{$^{(c)}$ Dip. di Fisica, Universit\`a di Roma ``La Sapienza'' and INFN, Sez. di Roma,}\\
    \hspace*{0.5cm}{Piazzale A. Moro 2, I-00185 Roma, Italy}\\
    \noindent
    \textbf{$^{(d)}$ Dip. di Fisica, Universit\`a di Genova and INFN,}\\
    \hspace*{0.5cm}{Via Dodecaneso 33, I-16146 Genova, Italy}\\
    \noindent
    \textbf{$^{(e)}$ Department of Physics, University of Wisconsin,}\\
    \hspace*{0.5cm}{Madison, WI 53706, USA}\\
    \textbf{$^{(f)}$ Laboratoire de l'Acc\'el\'erateur Lin\'eaire,}\\
    \hspace*{0.5cm}{IN2P3-CNRS et Universit\'e de Paris-Sud, BP 34,
      F-91898 Orsay Cedex}\\
    \noindent
    \textbf{$^{(g)}$INFN, Sez. di Bologna,}\\
    \hspace*{0.5cm}{Via Irnerio 46, I-40126 Bologna, Italy}\\
  \end{scriptsize}
\end{center}

\vspace*{0.5cm}

\begin{abstract}
  Starting from a (new physics independent) tree level determination
  of $\rhobar$ and $\etabar$, we perform the Unitarity Triangle
  analysis in general extensions of the Standard Model with arbitrary
  new physics contributions to loop-mediated processes. Using a simple
  parameterization, we determine the allowed ranges of non-standard
  contributions to $|\Delta F|=2$ processes.  Remarkably, the recent
  measurements from $B$ factories allow us to determine with good
  precision the shape of the Unitarity Triangle even in the presence
  of new physics, and to derive stringent constraints on non-standard
  contributions to $|\Delta F|=2$ processes. Since the present
  experimental constraints favour models with Minimal Flavour
  Violation, we present the determination of the Universal Unitarity
  Triangle that can be defined in this class of extensions of the
  Standard Model.  Finally, we perform a combined fit of the Unitarity
  Triangle and of new physics contributions in Minimal Flavour
  Violation, reaching a sensitivity to a new physics scale of about 5
  TeV.  We also extrapolate all these analyses into a ``year 2010''
  scenario for experimental and theoretical inputs in the flavour
  sector.  All the results presented in this paper are also available
  at the URL \texttt{http://www.utfit.org}, where they are
  continuously updated.
\end{abstract}

\newpage \pagestyle{plain}

\section{Introduction}

With the increasing precision of the experimental results,
the Unitarity Triangle (UT) analysis shows the impressive success of
the CKM picture in describing CP violation in the Standard Model (SM).
UT parameters have been consistently determined using both
CP-conserving ($\vert V_{ub}/V_{cb}\vert$, $\dmd$ and ${\dmd}/{\dms}$) and
CP-violating ($\epsilonk$ and $\sin 2 \beta$) processes
\cite{utfit2005}.  Additional measurements of several combinations of
angles of the UT, especially $\gamma$ from $B \to DK$ and $\alpha$
from charmless $B$ decays, confirm this picture \cite{utfit2005}.

This success becomes a puzzle once, as a possible solution to the
gauge hierarchy problem, the SM is considered as an effective theory
valid up to energies not much higher than the electroweak scale.
Indeed, even in the favourable case in which the theory above the
cutoff is weakly coupled, such as the Minimal Supersymmetric Standard
Model, large contributions to Flavour Changing Neutral Current (FCNC)
and CP-violating processes are expected to arise \cite{pellicani},
clashing with the large amount of accurate experimental data now
available on these transitions.  This is due to the presence of
additional sources of flavour and CP violation beyond the CKM matrix
(see for example Ref.~\cite{Hall} for the supersymmetry case).

In general, the flavour puzzle admits two classes of possible
solutions. The first one contains models in which a flavour symmetry
is invoked to explain the hierarchy of quark masses and mixing angles.
These models are based on different theoretical approaches
(supersymmetry, grand unification, extra dimensions,~\dots) leading to
different levels of agreement with the data and to different
low-energy signals.  However, low-energy processes generally receive
sizable additional contributions which jeopardize the validity of the
SM UT analysis (see Ref.~\cite{piai} for a supersymmetric example). A
generalized UT fit allowing for the presence of arbitrary New Physics
(NP) contributions is therefore very useful for model building, since
it provides at the same time the allowed ranges for the SM CKM
parameters and for the NP contributions to $|\Delta F|=2$ processes.
This will be the subject of the first part of the present work
(Secs.~\ref{sec:generalUT}-\ref{sec:ds2}).

The second class of solutions to the flavour puzzle contains models
with Minimal Flavour Violation (MFV). The basic idea of MFV is that
the only source of flavour violation is in the SM Yukawa couplings, so
that all FCNC and CP-violating phenomena can be expressed in terms of
the CKM matrix and the top quark Yukawa coupling
\cite{uut,gino,MFVgen}. This leads to strong correlations between
different observables, and allows for a detailed study of low-energy
phenomena. While there are several implementations of MFV in different
contexts (two-Higgs doublet models, supersymmetry~\cite{MFVsusy},
extra dimensions~\cite{MFVextra},~\dots), it is possible to perform
two very general analyses under the MFV hypothesis. The first is the
determination of the so-called Universal Unitarity Triangle (UUT)
\cite{uut}, which is a UT fit performed using only quantities that are
independent of NP contributions within MFV models. The second is a
simultaneous fit of the UT and of NP contributions in the $|\Delta F|=2$
sector.  These two analyses will be presented in the second part of
this work (Sec.~\ref{sec:uut}), and they serve as the starting point
for the study of rare decays and CP violation in MFV models~\cite{BurasMFV}.

Finally, in Sec.~\ref{sec:ckm2010} we present the possible future
improvements in the above analyses by considering a ``year 2010''
scenario for experimental data and theoretical inputs in the flavour
sector.

While to our knowledge the determination of the UUT is presented in
this work for the first time, several attempts have been previously
made in the study of the UT in the presence of NP. Considering only
model-independent analyses, in Ref.~\cite{UTfitNPold} the case of NP
contributions to $|\Delta B|=2$ or $|\Delta S|=2$ transitions was
analyzed: this corresponds to the discussion in Section~\ref{sec:ds2}
of the present work. A first version of the present analysis, with
some experimental constraints missing, was presented in
Ref.~\cite{MaurizioSUSY}. Constraints on NP in the $|\Delta B|=2$ sector
using $B$ physics only were considered in
Refs.~\cite{ASLNP,ligeti,CKMfitter}. The determination of the UT from
tree-level processes only was presented in Ref.~\cite{utfit2005}. A
general analysis was recently performed in Ref.~\cite{branco}, but not
all available constraints were used. With respect to these previous
studies, we improve several theoretical aspects and perform a
simultaneous determination of UT and NP parameters using all the
available constraints in all sectors.

A compilation of the experimental and theoretical inputs to our
analyses is presented in Table~\ref{tab:inputs}. 

\begin{table}[ht]
\begin{footnotesize}
\begin{center}
\begin{tabular}{@{}llll}
\hline\hline
\\
         Parameter                          &  Value                            
     & Gaussian ($\sigma$)      &   Uniform             \\
                                            &                                   
     &                          & (half-width)          \\ \hline\hline
         $\lambda$                          &  0.2258                      
     &  0.0014                  &    -                  \\ \hline
$\left |V_{cb} \right |$(excl.)             & $ 41.3 \times 10^{-3}$            
     & $1.0 \times 10^{-3}$     &   $1.8 \times 10^{-3}$                   \\
$\left |V_{cb} \right |$(incl.)             & $ 41.6 \times 10^{-3}$            
     & $0.7 \times 10^{-3}$     & -  \\ 
$\left |V_{ub} \right |$(excl.)             & $ 38.0  \times 10^{-4}$           
     & $2.7 \times 10^{-4}$     & $4.7 \times 10^{-4}$  \\  
$\left |V_{ub} \right |$(incl.)        & $ 43.9  \times 10^{-4}$           
     & $2.0 \times 10^{-4}$     &  $2.7 \times 10^{-4}$             \\ \hline
$\Delta m_d$ [ps$^{-1}$]         & $0.501$            
     & $0.005$   &        -              \\
$\Delta m_s$ [ps$^{-1}$]                                & $>$ 14.5 at 95\% C.L.   
     & \multicolumn{2}{c}{sensitivity 18.3}   \\ \hline
$\fbssqbs$ [MeV]                             & $276$                       
     & $38$                &          -            \\
$\xi=\frac{\fbssqbs}{\fbdsqbd}$             & 1.24                              
     & 0.04                     &  0.06            \\\hline
$\hat B_K$                                  & 0.79                              
     & 0.04                     &     0.09              \\
$\epsilonk$                                 & $2.280 \times 10^{-3}$            
     & $0.013 \times 10^{-3}$   &          -            \\
$f_K$ [MeV]                                       & 160                         
     & \multicolumn{2}{c}{fixed}                        \\
$\Delta m_K$   [ps$^{-1}$]                             & 0.5301 $\times
10^{-2}$ & \multicolumn{2}{c}{fixed}
\\ \hline 
$\snb$                                      &  0.687              
     &  0.032                   &          -            \\ \hline
$\overline m_t$ [GeV]                                      & $163.8$
     & $3.2$         &          -            \\
$\overline m_b$ [GeV]                                       & 4.21 
     & 0.08           &          -            \\
$\overline m_c$  [GeV]                                     & 1.3
     & 0.1            &          -            \\
$\alpha_s(M_Z)$                                  & 0.119                             
     & 0.003                    &          -            \\
$G_F $  [GeV$^{-2}$]                                    & 1.16639 $\times 10^{-5}$
     & \multicolumn{2}{c}{fixed}                        \\
$ m_{W}$  [GeV]                                   & 80.425 
     & \multicolumn{2}{c}{fixed}                        \\
$ m_{B^0_d}$   [GeV]                              & 5.279 
     & \multicolumn{2}{c}{fixed}                        \\
$ m_{B^0_s}$   [GeV]                              & 5.375
     & \multicolumn{2}{c}{fixed}                        \\
$ m_K^0$    [GeV]                                & 0.4977
     & \multicolumn{2}{c}{fixed}                        \\ \hline\hline
\end{tabular} 
\end{center}
\end{footnotesize}
\caption{\textit{Values of the relevant quantities used in the UT fit.
    The Gaussian and the flat contributions to the
    uncertainty are given in the third and fourth columns 
    respectively (for details on the statistical treatment see
    Ref.~\protect\cite{hep-ph/0012308}). Several branching ratios and
    CP asymmetries have been used. Their values and errors can be
    found in Ref.~\cite{hfag} and have been updated to Summer 2005.
}}
\label{tab:inputs} 
\end{table}

\section{Unitarity clock, unitarity hands:\\
A model-independent determination of the UT}
\label{sec:generalUT}

\begin{figure}[t]
\begin{center}
\includegraphics[width=0.8\textwidth]{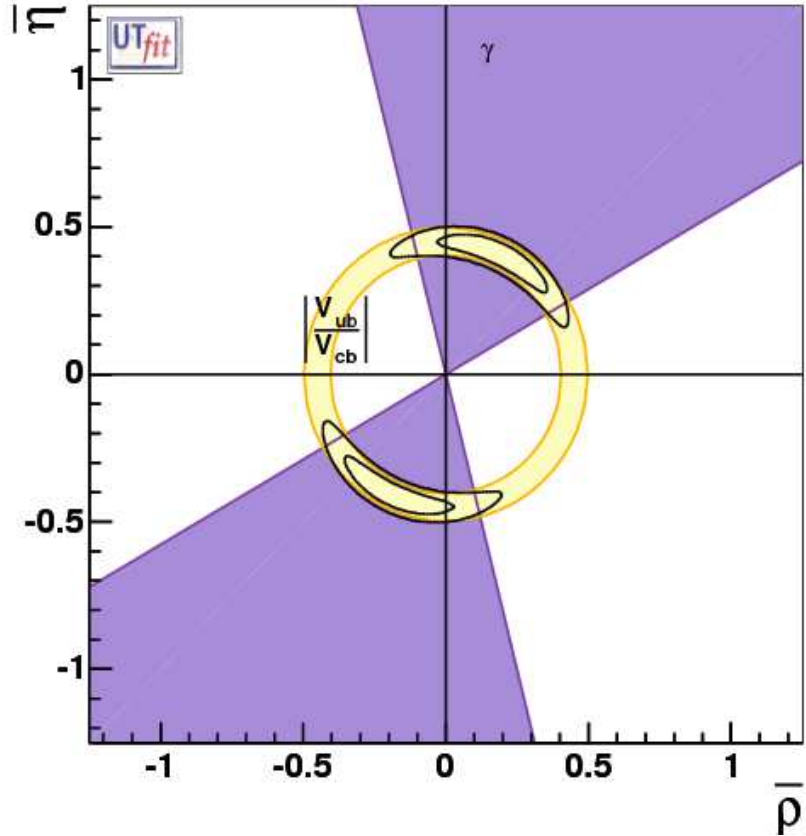} 
\caption{%
  \textit{The selected region on $\rhobar$-$\etabar$ plane obtained from
  the determination of $\vert V_{ub}/V_{cb}\vert$ and $\gamma$ (using $DK$ final
  states).  Selected regions corresponding to $68\%$ and $95\%$
  probability are shown, together with $95\%$ probability regions for
  $\gamma$ and $\vert V_{ub}/V_{cb}\vert$.}}
\label{fig:vubgamma}
\end{center}
\end{figure}

Let us first of all discuss the shape of the UT in the presence of
arbitrary NP contributions. All the available experimental data
exclude the possibility of sizable contributions to tree-level SM
processes, so that extensions of the SM in which NP enters low-energy
processes at the tree level are strongly disfavoured. We can therefore
safely assume in this work that NP enters observables in the flavour
sector only at the loop level. It is then possible to determine two
regions in the $\bar \rho-\bar \eta$ plane independently of NP
contributions, using only tree-level $B$ decays.  The CKM elements
$V_{ub}$ and $V_{cb}$ are determined using semileptonic inclusive and
exclusive $B$ decays. The angle $\gamma$ is obtained by measuring the
phase of $V_{ub}$ appearing in the interference between $b \to c$ and
$b \to u$ transitions to $DK$ final states.\footnote{We neglect
  possible NP contributions to $D^0$--$\bar D^0$ mixing, since their
  contribution is expected to be well below the present experimental
  accuracy \cite{ddbar1}. In the future, it might become necessary to
  take them into account following Ref.~\cite{ddbar2}.}  As shown in
Fig.~\ref{fig:vubgamma}, it is now lunchtime ($\sim$13:35) on
Andrzej's unitarity clock
\cite{hep-ph/0101336}.\footnote{Fig.~\ref{fig:vubgamma} first appeared
  in Ref.~\cite{utfit2005}. Similar results were recently obtained in
  Ref.~\cite{branco}.}

The results of this analysis, reported in Tab.~\ref{tab:vubgamma}, can
be used as a reference for model-building and phenomenology in any
extension of the SM with loop-mediated contributions to FCNC
processes. The present precision is expected to improve considerably
in the near future, as discussed in Sec.~\ref{sec:ckm2010}.

\begin{table}[ht]
\begin{center}
\begin{tabular}{ccc}
\hline
\multicolumn{3}{c}{UT fit - using only $\vert V_{ub}/V_{cb}\vert$ and $\gamma$} \\
\hline
& SM Solution  &   2$^\mathrm{nd}$ Solution \\
\hline
$\rhobar$  & 0.18 $\pm$ 0.12 & -0.18 $\pm$ 0.12 \\
$\etabar$ & 0.41 $\pm$ 0.05 & -0.41 $\pm$ 0.05 \\ 
\hline
$\sin 2 \beta$ & 0.782 $\pm$ 0.065 & -0.641 $\pm$ 0.087 \\ 
$\gamma\,[^\circ]$ & 65 $\pm$ 18 & -115 $\pm$ 18 \\
$\alpha\,[^\circ]$ & 87$\pm$ 15 & -46 $\pm$ 15 \\
$2\beta+\gamma\,[^\circ]$ & 122 $\pm$ 13 & -152 $\pm$ 13   \\
\hline
\end{tabular}
\caption{\textit{Results for several UT parameters, obtained using the
    constraints from $\vert V_{ub}/V_{cb}\vert$ and $\gamma$ (using
    $DK$ final states).}}  \label{tab:vubgamma} \end{center}
\end{table}
Beyond the Standard Model, one can include the information
from other constraints, taking into account the effect of NP in a
general way. In particular, one has to consider two effects:
\begin{itemize}
\item The contribution of new operators in the
  $|\Delta F| = 2$ Hamiltonian, which affects mixing processes and, as
  a consequence, the determination of $\Delta m_{d,s}$, $\epsilonk$
  and of the angles $\beta$ and $\alpha$.
\item The effect of NP in
  the $|\Delta F| = 1$ Hamiltonian, for all those processes occurring
  through penguin transitions. In our case, this concerns the
  determination of $\alpha$ from charmless $B$ decays and
  the CP asymmetry in semileptonic $B$ decays $A_\mathrm{SL}$.
\end{itemize}

\section{Model-independent constraints on New Physics in
  $\mathbf{|\Delta F|}$=2 transitions} \label{sec:generalC}
Our goal
in this Section is to use the available experimental information on
loop-mediated processes to constrain the NP contributions to $|\Delta
F|$=2 transitions. In general, NP models introduce a large number of
new parameters: flavour changing couplings, short distance
coefficients and matrix elements of new local operators. The specific
list and the actual values of these parameters can only be determined
within a given model. Nevertheless, each of the mixing processes
listed in Tab.~\ref{tab:schema} is described by a single amplitude and
can be parameterized, without loss of generality, in terms of two
parameters, which quantify the difference of the complex amplitude
with respect to the SM one~\cite{cfactors}. Thus, for instance, in the
case of $B^0_q-\bar{B}^0_q$ mixing we define
\begin{equation} C_{B_q}
  \, e^{2 i \phi_{B_q}} = \frac{\langle
    B^0_q|H_\mathrm{eff}^\mathrm{full}|\bar{B}^0_q\rangle} {\langle
    B^0_q|H_\mathrm{eff}^\mathrm{SM}|\bar{B}^0_q\rangle}\,, \qquad
  (q=d,s) \label{eq:paranp}
\end{equation}
where $H_\mathrm{eff}^\mathrm{SM}$ includes only the SM box diagrams,
while $H_\mathrm{eff}^\mathrm{full}$ includes also the NP
contributions.\footnote{$C_{B_q}$ and $\phi_{B_q}$ parameterize NP
  effects in the dispersive part of the effective Hamiltonian only.}
In the absence of NP effects, $C_{B_q}=1$ and $\phi_{B_q}=0$ by
definition. The experimental quantities determined from the
$B^0_q-\bar{B}^0_q$ mixings and listed in Tab.~\ref{tab:schema} are
related to their SM counterparts and the NP parameters by the
following relations:
\begin{equation} \Delta m_d^\mathrm{exp} = C_{B_d}
  \Delta m_d^\mathrm{SM} \,,\; \sin 2 \beta^\mathrm{exp} = \sin (2
  \beta^\mathrm{SM} + 2\phi_{B_d})\,,\; \alpha^\mathrm{exp} =
  \alpha^\mathrm{SM} - \phi_{B_d}\,, \label{eq:NPangles}
\end{equation}
in a self-explanatory notation. 
\begin{table}[t]
  \begin{center} \begin{tabular}{cccc} \hline Tree-Level & B$^0_d$
      mixing & K$^0$ mixing & B$^0_s$ mixing \\ \hline $\vert
      V_{ub}/V_{cb}\vert$ & $\Delta {m_d}$ & $\epsilon_K$
      & $\Delta m_s$          \\
      $\gamma$ ($DK$) & $A_{CP}(B \to J/\psi K)$ &
      & $A_{CP}(B^0_s \to J/\psi \phi)$  \\
      &    $A_{CP}(B\to \pi \pi,\rho \pi, \rho \rho)$   \\
      &    $A_\mathrm{SL}$   \\
      \hline
\end{tabular}
\caption{\textit{Different processes and
        corresponding measurements contributing to the determination
        of $\rhobar$, $\etabar$, $C_{B_d}$, $\phi_{B_d}$,
        $C_{B_s}$,$\phi_{B_s}$ and $C_{\epsilonk}$.  $\Delta m_K$ is
        not considered due to the fact that the long distance effects
        are not well under control.}}  \label{tab:schema} \end{center}
\end{table}
As far as the $K^0-\bar{K}^0$ mixing is concerned, we find
it convenient to introduce a single parameter which relates the
imaginary part of the amplitude to the SM one: \begin{equation}
  C_{\epsilon_K} = \frac{\mathrm{Im}[\langle
    K^0|H_{\mathrm{eff}}^{\mathrm{full}}|\bar{K}^0\rangle]}
  {\mathrm{Im}[\langle
    K^0|H_{\mathrm{eff}}^{\mathrm{SM}}|\bar{K}^0\rangle]}\,.
  \label{eq:ceps} \end{equation} This definition implies in fact a
simple relation for the measured value of $\epsilonk$,
\begin{eqnarray} & \epsilonk^{\mathrm{exp}} = C_{\epsilon_K
  }\,\epsilonk^\mathrm{SM} & \qquad (K^0-\bar{K}^0~\mathrm{mixing})\,.
  \label{eq:cfactk} \end{eqnarray} $\Delta m_K$ is not considered
because the long distance effects are not well under control.
Therefore, all NP effects which enter the present analysis are
parameterized in terms of three real quantities, $C_{B_d}$,
$\phi_{B_d}$, and $C_{\epsilon_K}$.  NP in the $B_s$ sector is not
considered in this case, due to the lack of experimental information,
since both $\dms$ and $A_\mathrm{CP}(B_s \to J/\psi \phi)$ are not
measured yet.
\subsection{New Physics effects in the extraction of
  $\mathbf{\alpha}$ from $\mathbf{|\Delta F|}$=1 processes}
\label{subsec:dFeq1}
In principle, the extraction of $\alpha$ from $B\to \pi \pi,~\rho
\pi,~\rho \rho$ decays is affected by NP effects in $|\Delta F|$=1
transitions. Actually, in the presence of NP in the strong $b \to d$
penguins, the decay amplitudes for $B$ mesons decaying into $\pi \pi$,
$\rho \pi$ and $\rho \rho$ are a simple generalization of the SM ones
(given for example in Eqs.~(17) and (18) of Ref.~\cite{utfit2005}). We
assume that NP modifies significantly only the ``penguin'' amplitude
$P$ without changing its isospin quantum numbers (i.e.~barring large
isospin-breaking NP effects). Then, instead of a complex penguin
amplitude with vanishing weak phase, we have two independent arbitrary
complex penguin amplitudes for $B$ and $\bar B$ decays. For example,
the amplitudes of $B\to \pi \pi (\rho \rho)$ can be written as
\begin{eqnarray}
  A^{+-} &=& -Te^{-i\alpha} + P e^{i \phi_{P}}e^{i \delta_{P}}
  \nonumber \\  
  \bar A^{+-} &=& -Te^{i\alpha} + \bar P e^{-i \phi_{P}}e^{i \delta_{P}}
  \nonumber \\  
  A^{+0} &=&
  -\frac{1}{\sqrt{2}}\left[ e^{-i\alpha}\left( T + T_c e^{i 
        \delta_{T_c}} \right) \right]  \nonumber \\
  \bar A^{-0} &=&
  -\frac{1}{\sqrt{2}}\left[ e^{i\alpha}\left( T + T_c e^{i 
        \delta_{T_c}} \right) \right]  \nonumber \\
  A^{00} &=&
  -\frac{1}{\sqrt{2}}\left[T_c e^{-i\alpha} e^{i \delta_{T_c}} +
    P e^{i \phi_{P}}e^{i
      \delta_{P}}\right] \,, \nonumber \\
  \bar A^{00} &=&
  -\frac{1}{\sqrt{2}}\left[T_c e^{i\alpha} e^{i \delta_{T_c}} +
    P e^{-i \phi_{P}}e^{i
      \delta_{P}}\right] \,, 
  \label{eq:ppampli}
\end{eqnarray}
where $T$, $T_c$, $P$ and $\bar P$ are real
parameters, $\delta_{P}$ and $\delta_{T_{c}}$ are strong phases,
$\alpha$ is the angle of the UT, and $\phi_{P}$ is an additional weak
phase. 

\begin{figure}[htb!]
\begin{center}
\includegraphics[width=0.80\textwidth]{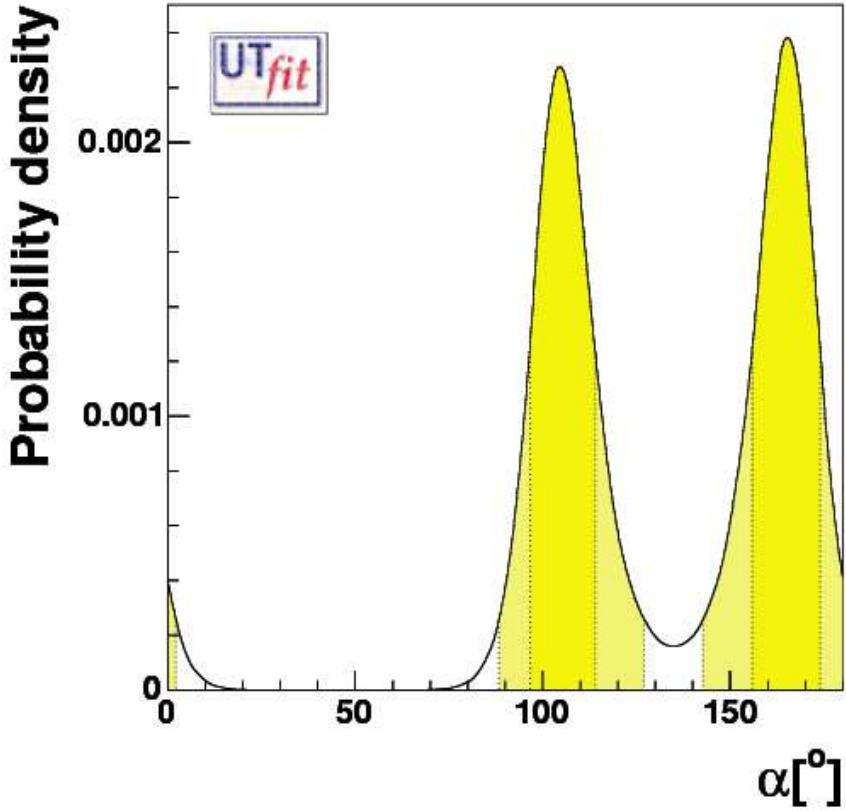}
\caption{%
\textit{P.d.f. of $\alpha$ from the combination of isospin
        analyses of $\pi \pi$, $\rho \pi$ and $\rho \rho$ decay modes,
        including NP effects in the $|\Delta F| = 1$ Hamiltonian.}}
\label{fig:df1alpha}
\end{center}
\end{figure}

The procedure to extract $\alpha$ is exactly the same as in the
SM~\cite{utfit2005}, since we assume that NP does not affect the
$\Delta I=3/2$ amplitudes. However, in the NP fit we lose the
knowledge of the weak phase of the penguins. In spite of this
additional free parameter, the experimental information available
nowadays is sufficient to constrain $\alpha$ even in the presence of
NP, as shown in Fig.~\ref{fig:df1alpha}. We take $T$, $T_c$, $P$ and
$\bar P$ to be flatly distributed in a range larger than the one
determined from the fit, and all the phases to be flatly distributed
in the $[0,2\pi)$ range. Here and in the following figures, dark
(light) areas correspond to the $68\%$ ($95\%$) probability region.
One should notice that these analyses bound $\alpha$ through the
quantity $\pi - \beta - \gamma$, where $\gamma$ comes from the decay
amplitudes and $\beta$ from $B_d - \bar B_d$ mixing. Therefore, in the
presence of NP effects in the $|\Delta F| = 2$ Hamiltonian, this bound
should be regarded as a constraint on $\alpha^\mathrm{SM}-\phi_{B_d}$
(see Eq.~\ref{eq:NPangles}).

The reader might notice a contradiction between the discussion above
and the results of ref.~\cite{alphagen,CKMfitter}, in which it is
stated that the NP parameters introduced above can be eliminated by a
redefinition of the $T$, $T_c$ and $P$ parameters. Explicitly, one has
\begin{eqnarray}
  A^{+-} &=& -T_X e^{-i\alpha} + P_X \nonumber \\
  A^{+0} &=& -\frac{1}{\sqrt{2}}\left[ e^{-i\alpha}\left( T_X + T_{cX}
    \right) 
  \right]  \nonumber \\
  A^{00} &=& -\frac{1}{\sqrt{2}}\left[T_{cX} e^{-i\alpha} + P_X \right] \,,
  \label{eq:ppamplicazzoni}
\end{eqnarray}
with
\begin{eqnarray}
  \label{eq:TXPX}
  T_X&=&T+\frac{P e^{i \phi_P}-\bar P e^{- i \phi_P}}{2 i \sin \alpha}
  e^{i \delta_P} \\ 
  T_{cX}&=&T_c e^{i \delta_{T_c}}-\frac{P e^{i \phi_P}-\bar P e^{- i \phi_P}}{2 i \sin \alpha}
  e^{i \delta_P} \nonumber \\ 
  P_X 
  &=&\frac{P e^{i(\alpha + \phi_P)}- \bar P e^{-i(\alpha + \phi_P)}}{2
    i \sin \alpha} e^{i \delta_P}\,.\nonumber
\end{eqnarray}
However, the transformations (\ref{eq:TXPX}) are singular for $\alpha
\to 0$. This implies that there is no limited \emph{a-priori} range
for the $X$ parameters. For this reason, the parameterization in
eq.~(\ref{eq:ppampli}) is of no use for our purpose.

\subsection{$\mathbf{A_\mathrm{SL}}$: general considerations and the
  inclusion of $\mathbf{|\Delta F|}$=1 New Physics effects}\label{subsec:ASL}
One can also add the constraint coming from the CP
asymmetry in semileptonic $B$ decays $A_\mathrm{SL}$, defined as
\begin{equation} \label{eq:ASLdef}
A_\mathrm{SL}\equiv
  \frac{\Gamma(\bar B^0 \to \ell^+ X)- \Gamma(B^0 \to \ell^-
    X)}{\Gamma(\bar B^0 \to \ell^+ X)+ \Gamma(B^0 \to \ell^- X)}\,.
\end{equation}
It has been noted in Ref.~\cite{ASLNP} that, even
though the present experimental bound is not precise enough to bound
$\rhobar$ and $\etabar$ in the Standard Model, $A_\mathrm{SL}$ is a
crucial ingredient of the UT analysis once the formulae are
generalized according to Eq.~(\ref{eq:NPangles}), since this is the
only constraint that depends on both $C_{B_d}$ and $\phi_{B_d}$:
\begin{equation} \label{eq:ASLCphi}
A_\mathrm{SL}=- \mathrm{Re}
  \left(\frac{\Gamma_{12}}{M_{12}}\right)^\mathrm{SM} \frac{\sin 2
    \phi_{B_d}}{C_{B_d}} + \mathrm{Im}
  \left(\frac{\Gamma_{12}}{M_{12}}\right)^\mathrm{SM} \frac{\cos 2
    \phi_{B_d}}{C_{B_d}}\,,
\end{equation}
where $\Gamma_{12}$ and
$M_{12}$ are the absorptive and dispersive parts of the
$B^0_d-\bar{B}^0_d$ mixing amplitude.  At the leading order,
$A_\mathrm{SL}$ is independent of penguin operators, and therefore it
is also independent of NP in $\vert \Delta F \vert =1$ processes.
However, at the NLO, the penguin contribution should be taken into
account. In the SM, the effect of penguin operators is GIM suppressed
since their CKM factor is aligned with $M_{12}$: both are proportional
to $(V_{tb}^*V_{td})^2$.  This is not true anymore in the presence of
NP, so that the effects of penguins are amplified beyond the SM and
the approximation made in Ref.~\cite{ASLNP} of neglecting this
contribution is questionable.  For our analysis of $A_\mathrm{SL}$, we
therefore start from the full NLO calculation of
Ref.~\cite{Ciuchini:2003ww}, allowing for an additional NP
contribution to the penguin term in the $\vert \Delta F \vert =1$
amplitude. This introduces two additional parameters ($C_\mathrm{Pen}$
and $\phi_\mathrm{Pen}$), encoding NP contributions to the penguin
part in analogy to what $C_{B_d}$ and $\phi_{B_d}$ do for the box
contribution. Since the penguin amplitude is $\mathcal{O}(\alpha_{s})$
with respect to the leading contribution, these parameters introduce a
smearing in the theoretical determination of $A_\mathrm{SL}$.  The
generalized expression of $A_\mathrm{SL}$ is given by
\begin{eqnarray}
  A_\mathrm{SL} &=& - \frac{2 \kappa}{C_{B_d}} \left\{\sin\left(2
      \phi_{B_d}\right) \left(n_1+\frac{n_6 B_2+n_{11}}{B_1}\right)
    -\frac{\sin\left(\beta+2 \phi_{B_d}\right)}{R_t}
    \left(n_2+\frac{n_7
        B_2+n_{12}}{B_1}\right)\right.  \nonumber \\
  &&+\frac{\sin\left(2 (\beta+\phi_{B_d})\right)}{R_t^2}
  \left(n_3+\frac{n_8 B_2 + n_{13}}{B_1}\right)
  +\sin\left(\phi_\mathrm{Pen}+2 \phi_{B_d}\right) C_\mathrm{Pen}
  \left(n_4+n_9 \frac{ B_2}{B_1}\right)
  \nonumber \\
  &&\left. -\sin\left(\beta+\phi_\mathrm{Pen}+2 \phi_{B_d}\right)
    \frac{C_\mathrm{Pen}}{R_t}
    \left(n_5+n_{10}\frac{B_2}{B_1}\right)\right\} \label{eq:ASLfull}
\end{eqnarray}
where $B_1$ corresponds to the usual $B_d$ parameter
for $B^0 - \bar B^0$ mixing, $B_2 = 0.84 \pm 0.07$ (flat)
\cite{hep-lat/0110091}, $R_t=\sqrt{(1-\rhobar)^2+\etabar^2}$ is the
length of one of the UT sides, $\kappa$ is defined in
Ref.~\cite{Ciuchini:2003ww} and the magic numbers $n_i$ are given in
Tab.~\ref{tab:magic}.  The Standard Model expression can be recovered
in the limit $C_X\to 1$ and $\phi_X \to 0$ (where $X=B_d$, Pen).
Eq.~(\ref{eq:ASLfull}) contains NLO QCD and $1/m_b$ corrections; the
latter have been estimated using matrix elements computed in the
vacuum insertion approximation, since lattice results are not
available.
\begin{table} \begin{center} \begin{tabular}{|cc|cc|cc|}
      \hline $n_{1 }$ & $ 0.1797 \pm 0.0017$ & $n_{2 }$ & $ 0.1391\pm
      0.0193$&
      $n_{3  }$ & $  -0.0012\pm  0.0014$\\
      $n_{4 }$ & $ -0.0074\pm 0.0020$& $n_{5 }$ & $ 0.0020\pm 0.0007$&
      $n_{6  }$ & $   1.0116\pm     0.0826$\\
      $n_{7 }$ & $ 0.0455\pm 0.0144$& $n_{8 }$ & $ -0.0004\pm 0.0046$&
      $n_{9  }$ & $  -0.0714\pm     0.0170$\\
      $n_{10 }$ & $ -0.0041\pm 0.0016$& $n_{11 }$ & $ -0.3331\pm
      0.2178$&
      $n_{12 }$ & $   0.0028\pm   0.0101$\\
      $n_{13 }$ & $  -0.0036\pm  0.0033$& & & & \\
      \hline
\end{tabular}
\caption{\textit{Magic numbers for the
        calculation of $A_{\mathrm{SL}}$.  The quoted errors
        correspond to Gaussian distributions.}}  \label{tab:magic}
  \end{center}
\end{table}

To display the main phenomenological
consequences of $A_{SL}$, let us consider a simplified formula
obtained by setting all magic numbers to their central values, and
dropping all those smaller than $10^{-2}$.  In this way we get
\begin{eqnarray} A_\mathrm{SL} &\sim& \frac{2 \kappa}{C_{B_d}}
  \left\{\sin\left(2 \phi_{B_d}\right) \left(0.18+\frac{1.01
        B_2-0.33}{B_1}\right) -\frac{\sin\left(\beta+2
        \phi_{B_d}\right)}{R_t} \left(0.14+0.05
      \frac{B_2}{B_1}\right)\right.  \nonumber \\
  && \left.+\sin\left(\phi_\mathrm{Pen}+2 \phi_{B_d}\right)
    C_\mathrm{Pen} (-0.07) \frac{ B_2}{B_1}\right\}\,.
  \label{eq:ASLeff}
\end{eqnarray}
The SM penguin contribution
vanishes at this level of accuracy.  It is evident from the simplified
expression in Eq.~(\ref{eq:ASLeff}) that the phase $\phi_{B_d}$ can
induce an order-of-magnitude enhancement of $A_\mathrm{SL}$ relative
to the SM, while the penguin phase $\phi_\mathrm{Pen}$ can induce
corrections comparable to the SM contribution. To be conservative, for
our analysis we varied $C_\mathrm{Pen}$ in the range $[0,2]$ with
$\phi_\mathrm{Pen} \in [0,2 \pi]$. This produces only a minor smearing
of the dominant effects due to NP in $|\Delta B|=2$ transitions.

\subsection{Results of the analysis and constraints on
  $\mathbf{|\Delta F|=2}$ NP contributions} \label{subsec:NPresults}
\begin{figure}[t]
\begin{center}
\includegraphics[width=0.32\textwidth]{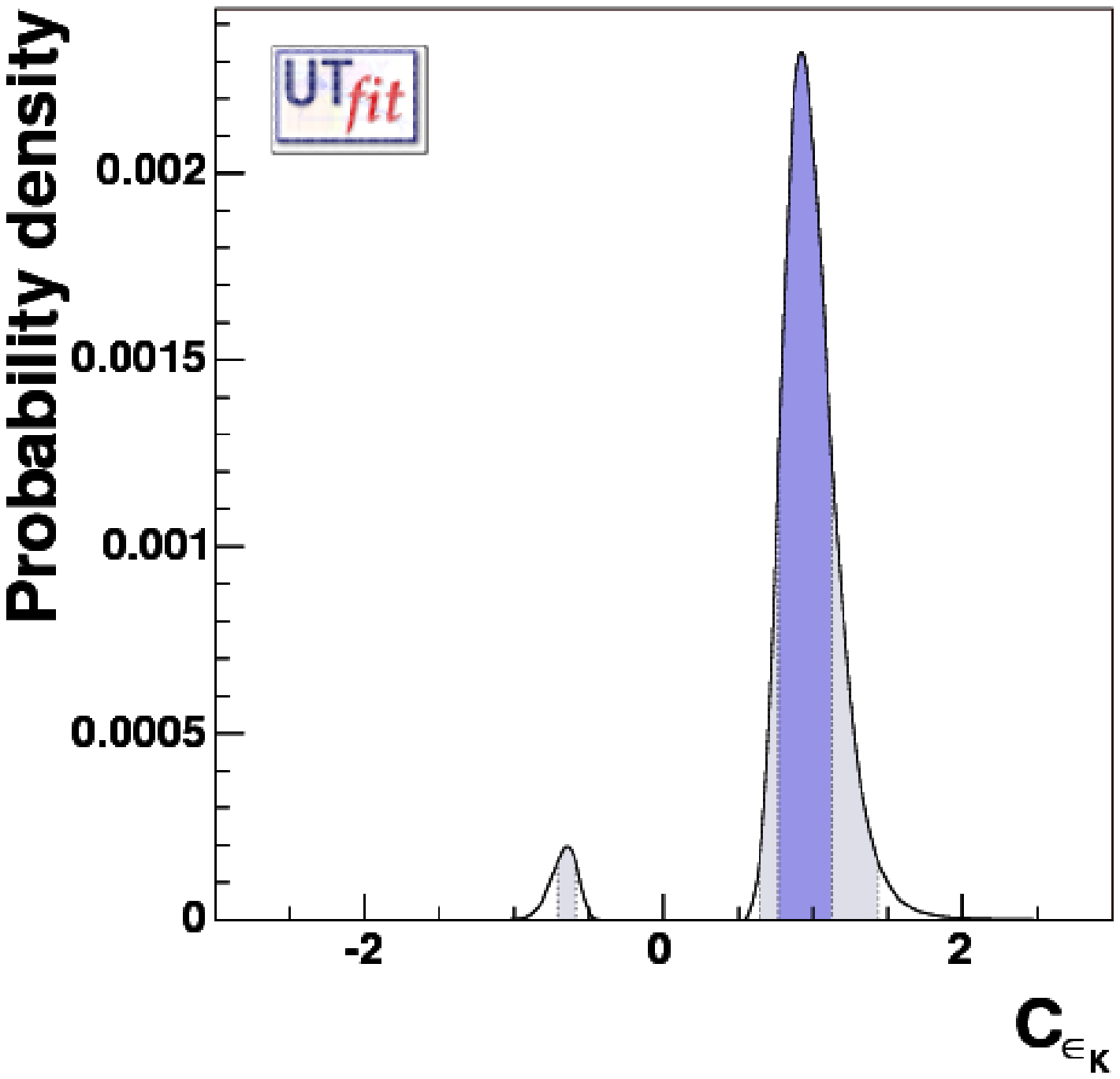}
\includegraphics[width=0.32\textwidth]{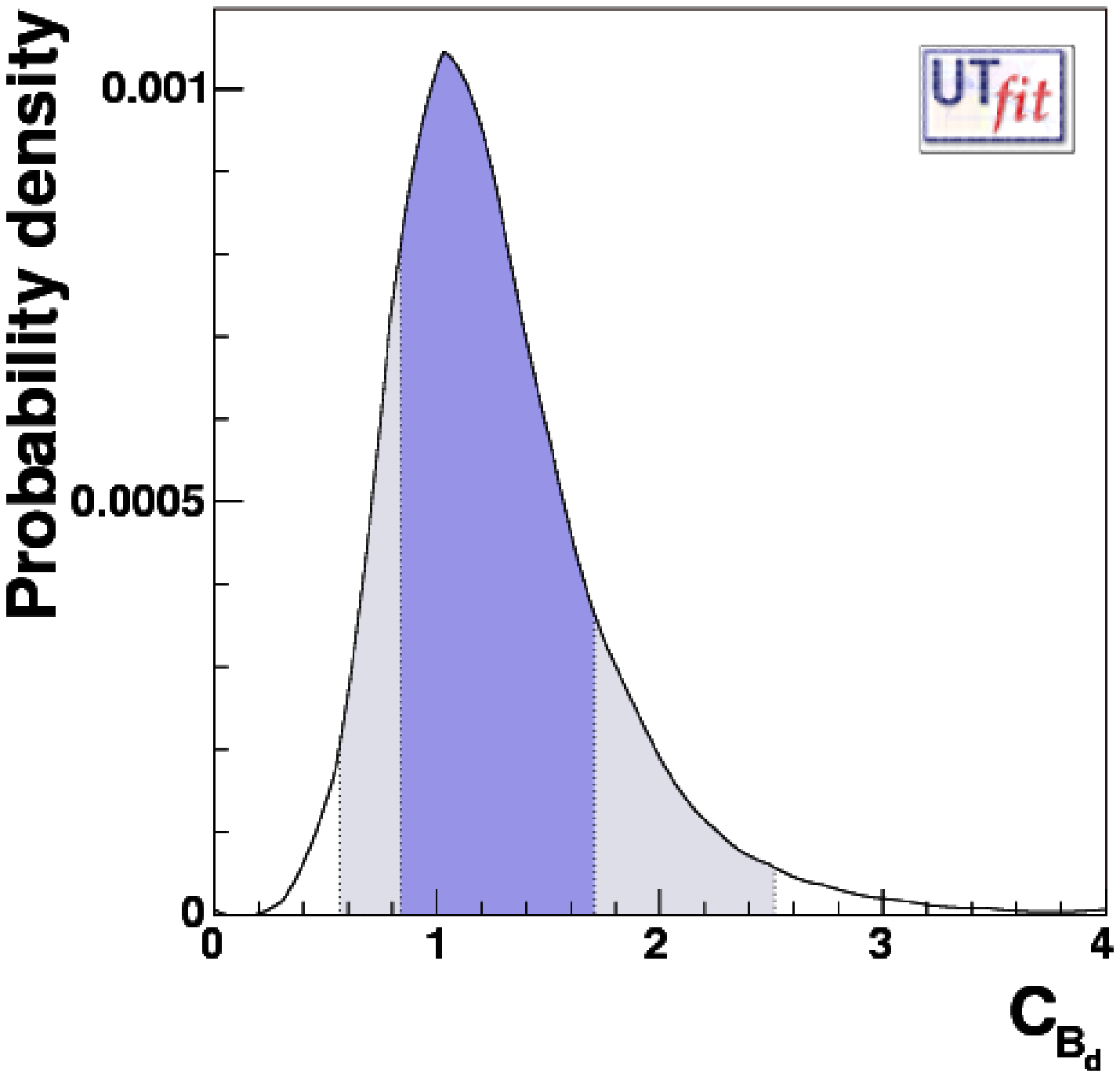}
\includegraphics[width=0.32\textwidth]{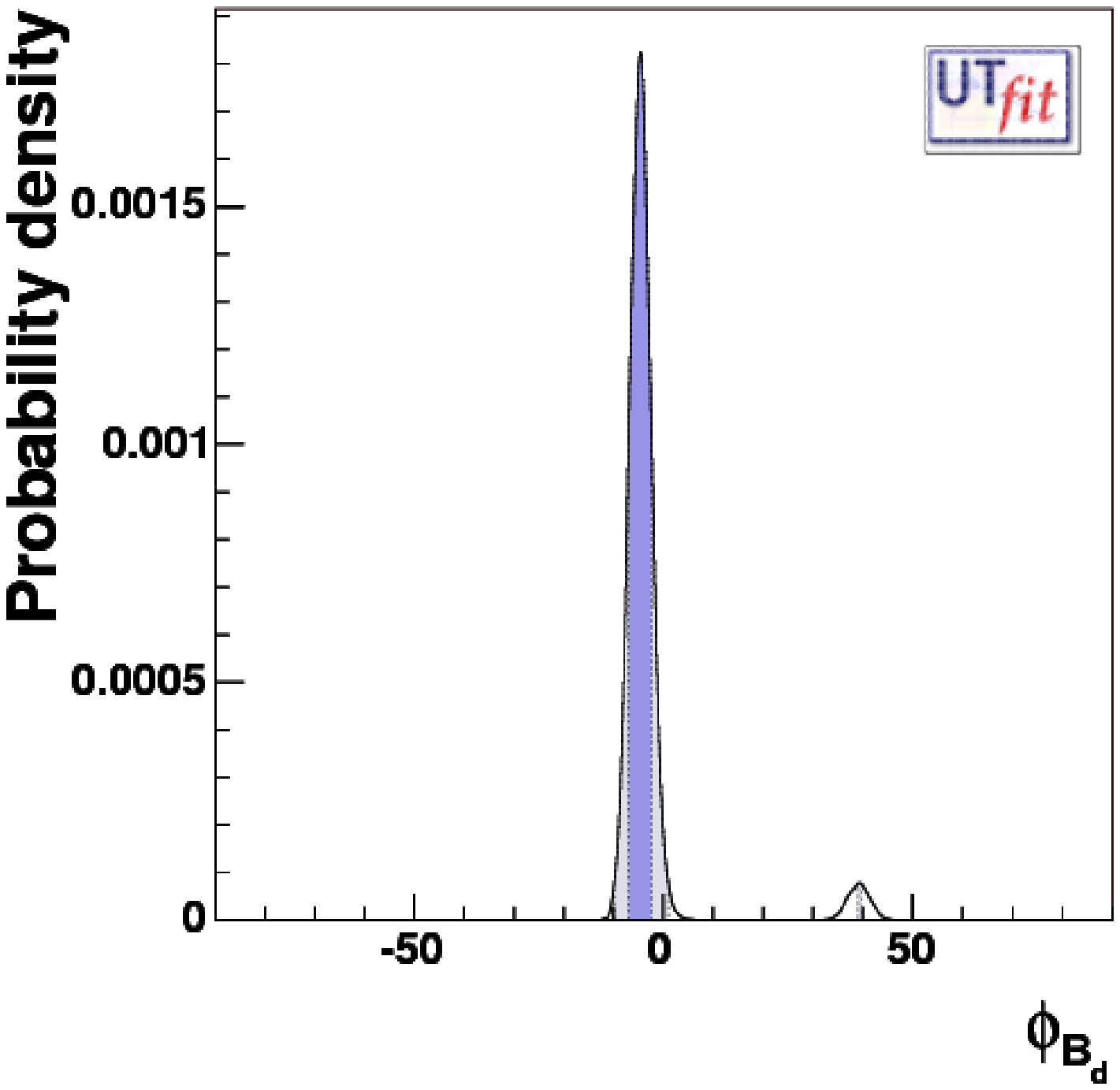}\\
\includegraphics[width=0.45\textwidth]{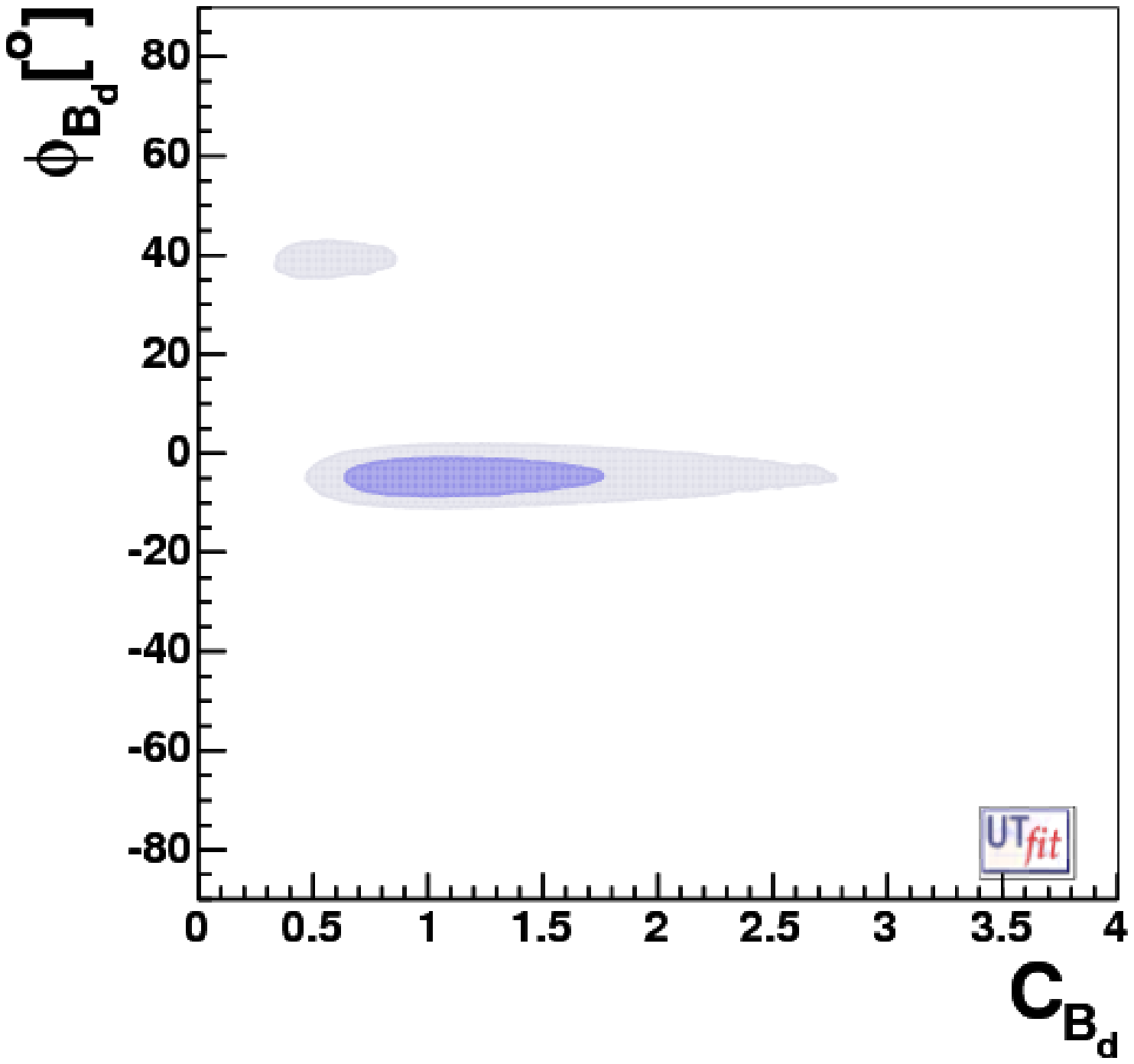}
\includegraphics[width=0.45\textwidth]{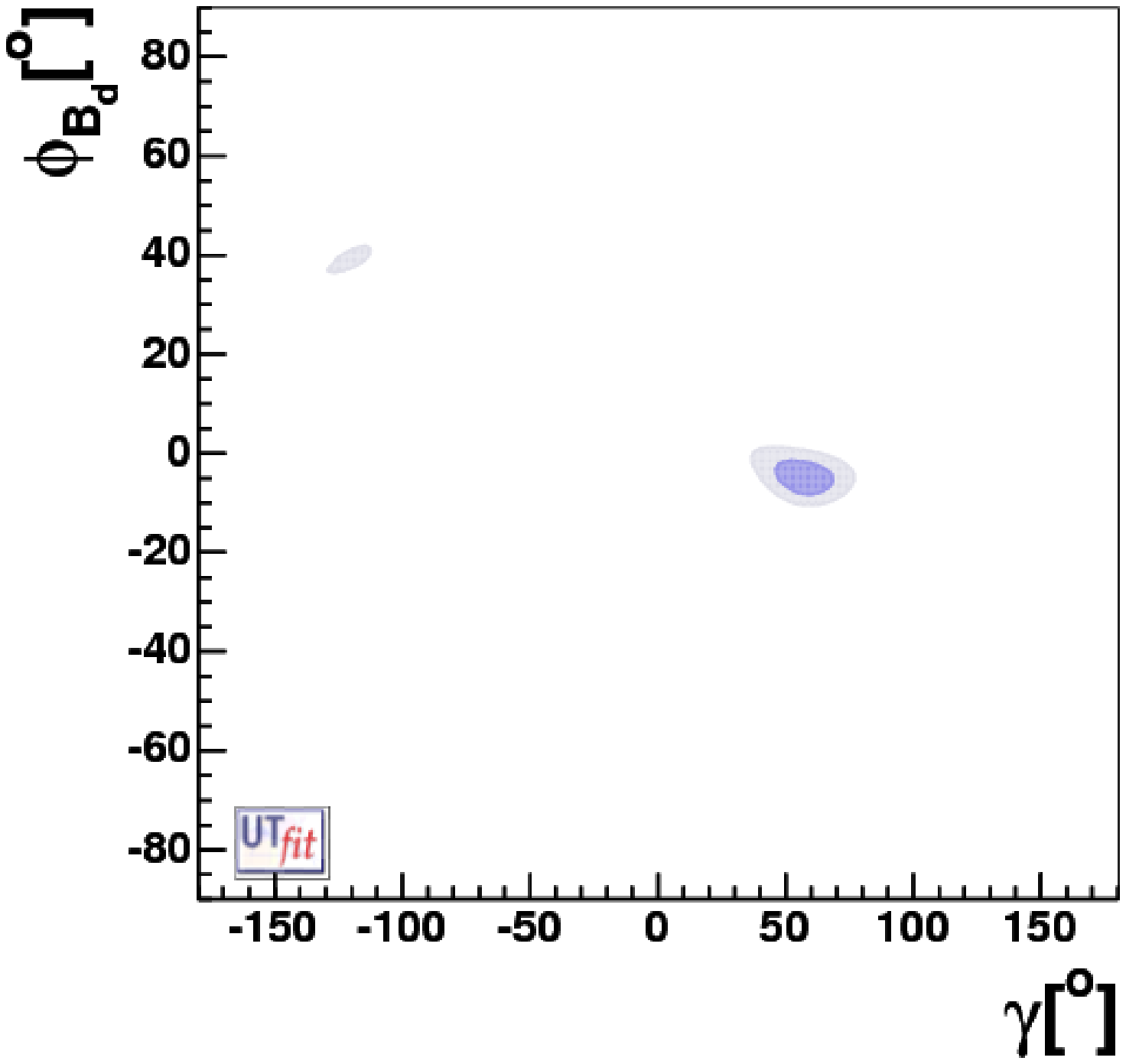}
\caption{%
  \textit{Output P.d.f.'s for $C_{\epsilon_K}$ (top-left),$C_{B_d}$
  (top-center), $\phi_{B_d}$ (top-right), and 2D distributions of
  $\phi_{B_d}\,vs.\,C_{B_d}$ (bottom-left) and
  $\phi_{B_d}\,vs.\,\gamma$.  Dark (light) areas correspond to the
  $68\%$ ($95\%$) probability region.}}
\label{fig:NP}
\end{center}
\end{figure}

\begin{figure}[t]
\begin{center}
\includegraphics[width=0.8\textwidth]{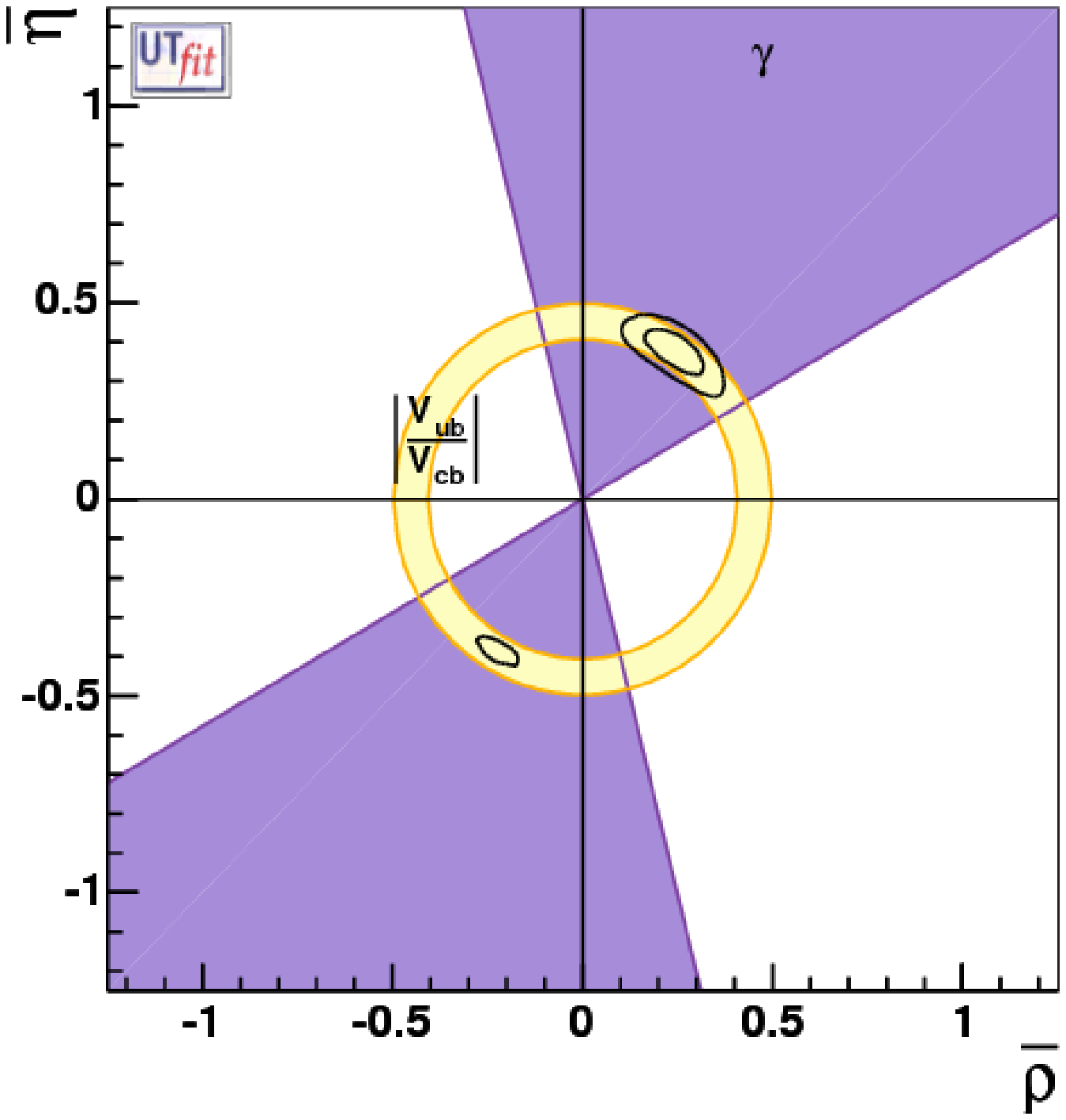} 
\caption{%
  \textit{The selected region on $\rhobar$-$\etabar$ plane obtained
    from the NP generalized analysis.  Selected regions corresponding
    to $68\%$ and $95\%$ probability are shown, together with $95\%$
    probability regions for $\gamma$ (from $DK$ final states) and
    $\vert V_{ub}/V_{cb}\vert$.}}
\label{fig:rhoetanp}
\end{center}
\end{figure}

To obtain the constraints on NP we extract $C_{B_d}$,
$C_{\epsilon_K}$, $\rhobar$ and $\etabar$ with a flat distribution in
a range much larger than the experimentally allowed region.  The phase
$\phi_{B_d}$ is taken to be flatly distributed in the range $[0,\pi]$.
The generated events are weighted using the experimental information
on $\vert V_{ub}/V_{cb}\vert$, $B \to D K$ decays ($\gamma$),
$\epsilon_K$, $B \to \rho \rho$, $\rho \pi$, and $\pi \pi$ decays
($\alpha$), $B \to J/\Psi K^{(*)}$ and $B \to D^0 h^0$
decays~\cite{hep-ph/0503174,D0h0} ($\beta$), and
$A_\mathrm{SL}$~\cite{hfag}, using the technique described in
ref.~\cite{hep-ph/0012308}. The output p.d.f.'s for $C_{B_d}$,
$C_{\epsilon_K}$, $C_{B_d}$ vs. $\phi_{B_d}$, and $\gamma$ vs.
$\phi_{B_d}$ are shown in Fig.~\ref{fig:NP}, and the corresponding
regions in the $\rhobar$--$\etabar$ plane are presented in
Fig.~\ref{fig:rhoetanp}.

It is important to remark that the constraints coming from the
experimental observables allow for an increase in the precision on
$\rhobar$ and $\etabar$ with respect to the pure tree-level
determination. This is clear comparing Fig.~\ref{fig:vubgamma} to
Fig.~\ref{fig:rhoetanp}.

\begin{figure}[ht!]
\begin{center}
 \includegraphics[width=0.32\textwidth]{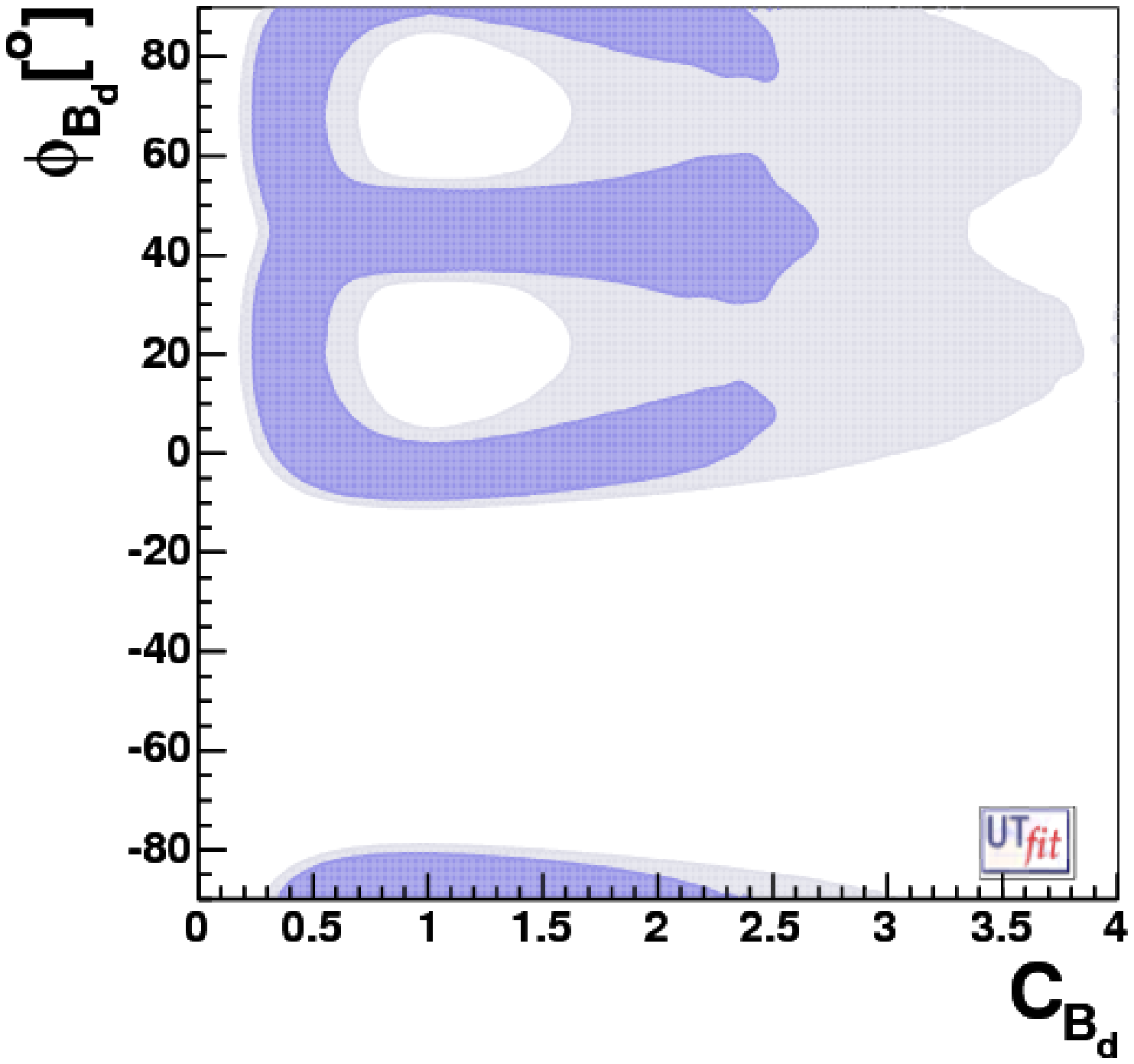}
 \includegraphics[width=0.32\textwidth]{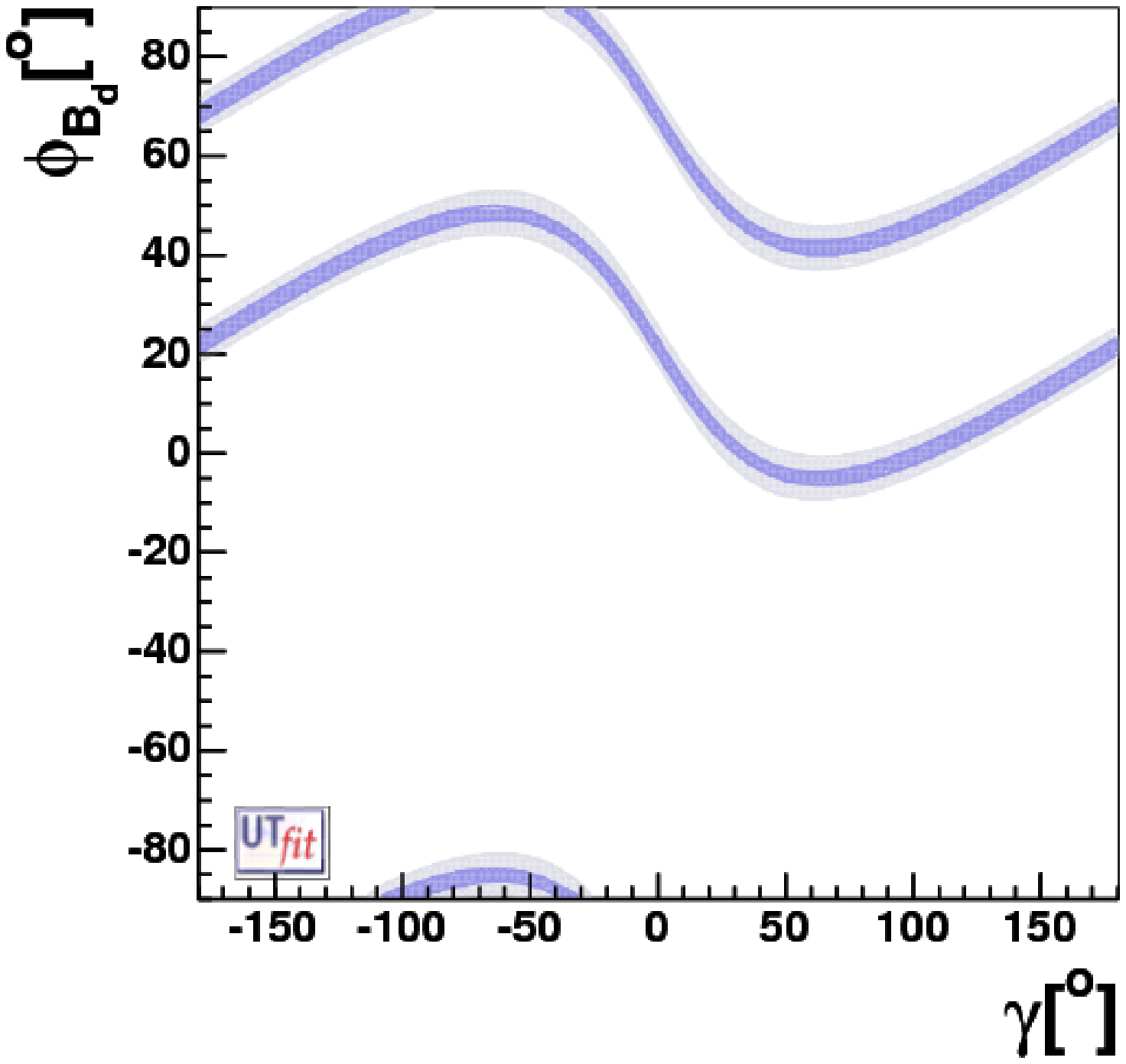}\\
 \includegraphics[width=0.32\textwidth]{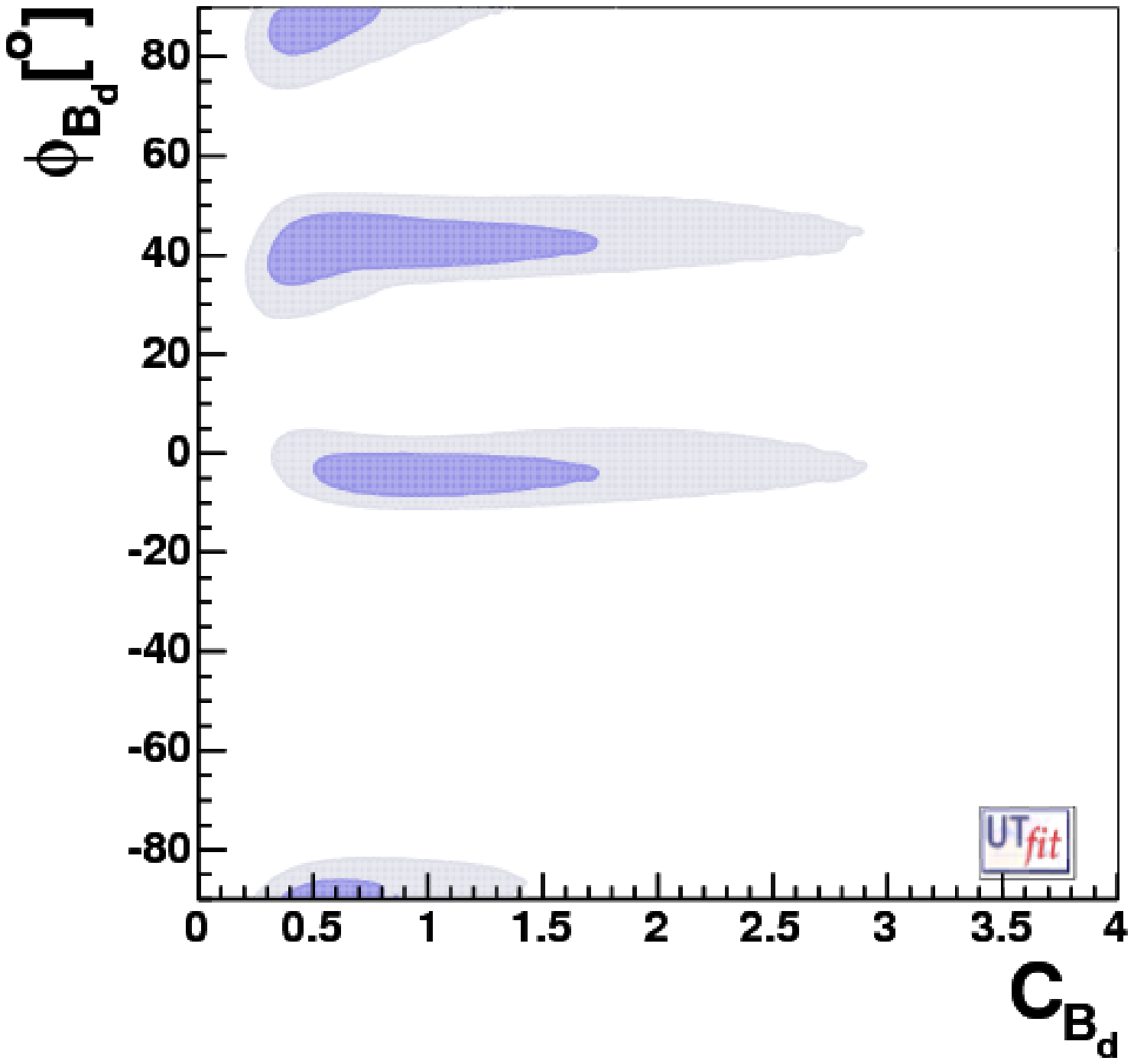}
 \includegraphics[width=0.32\textwidth]{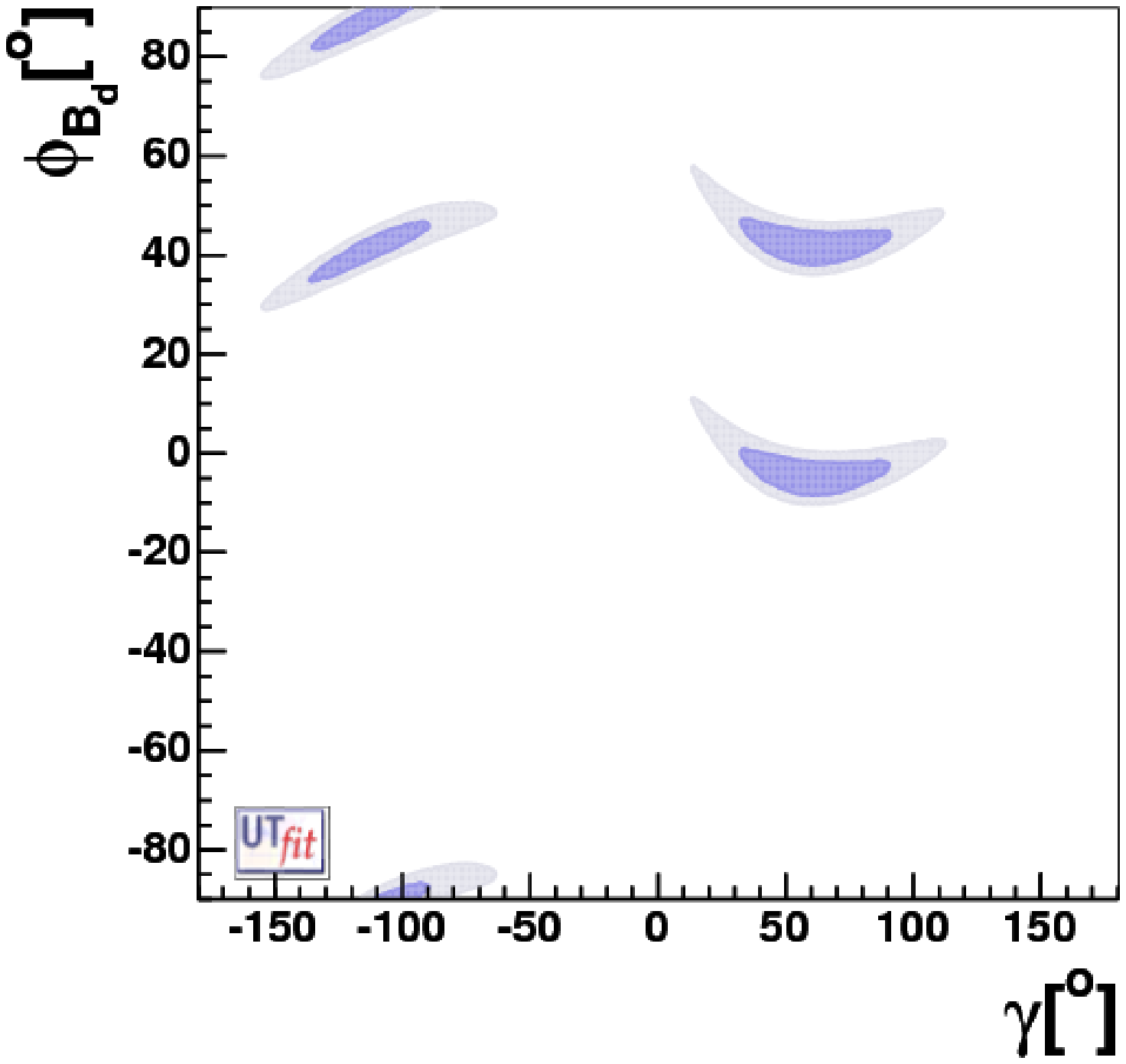}\\
 \includegraphics[width=0.32\textwidth]{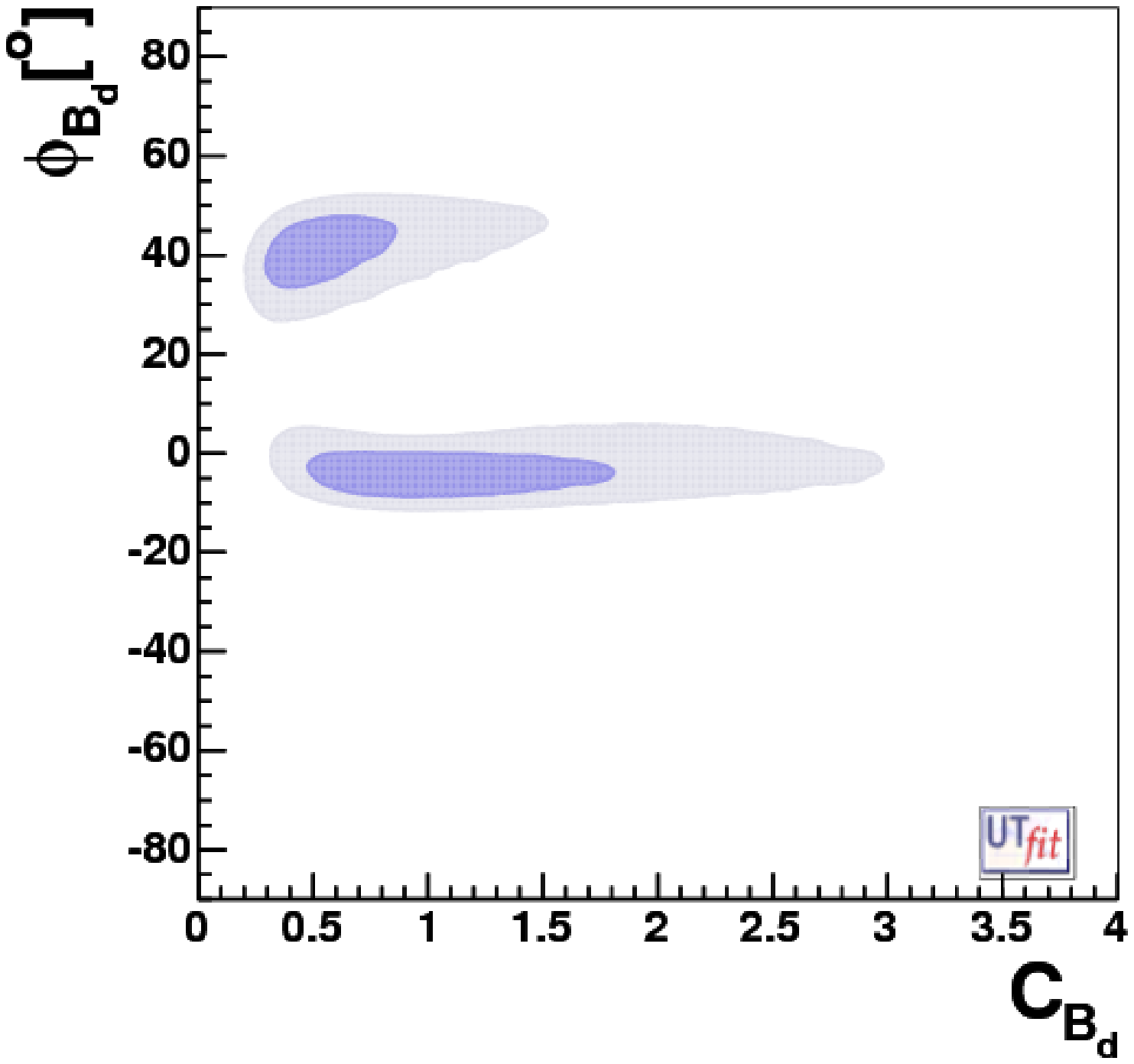}
 \includegraphics[width=0.32\textwidth]{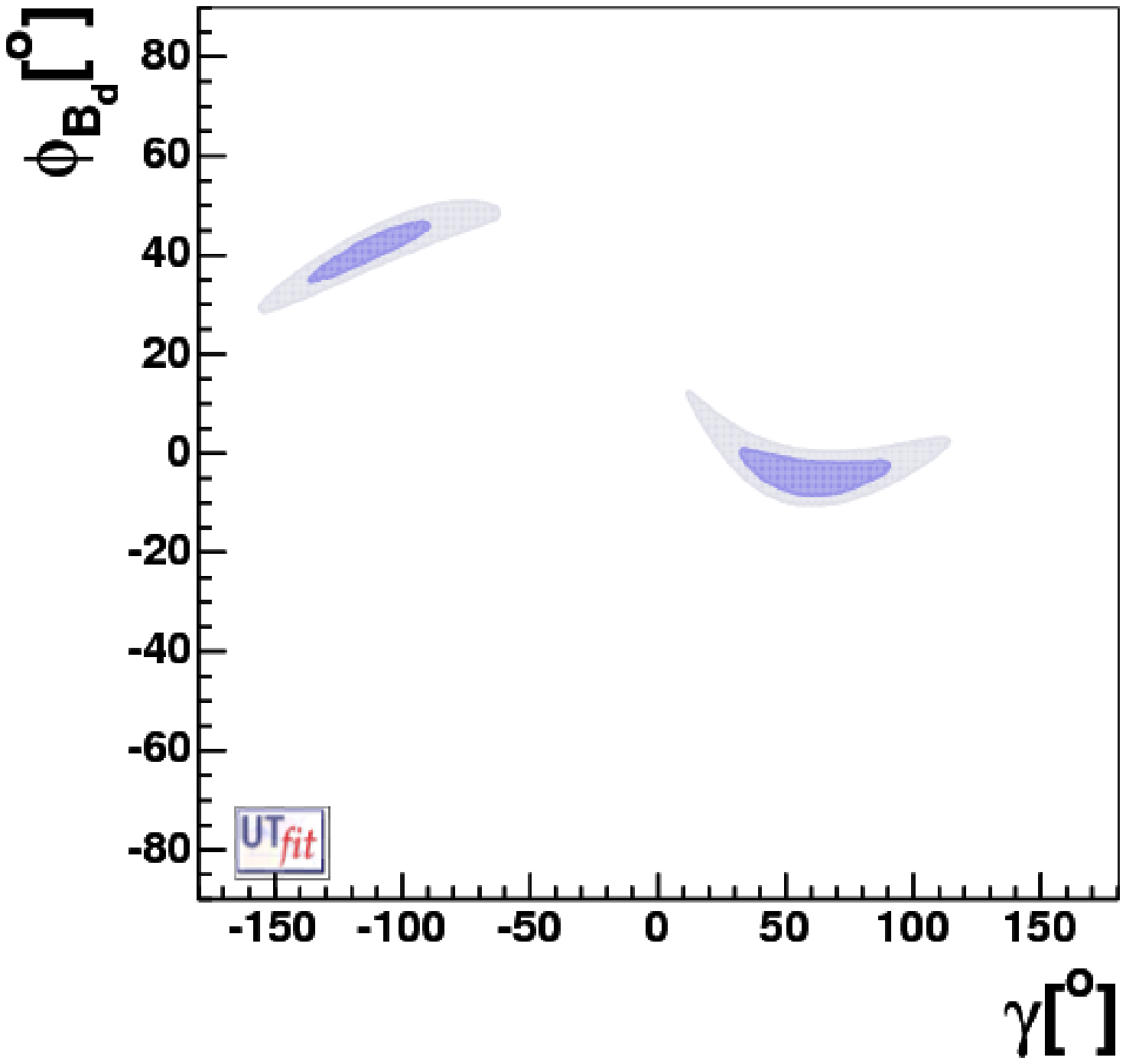}\\
 \includegraphics[width=0.32\textwidth]{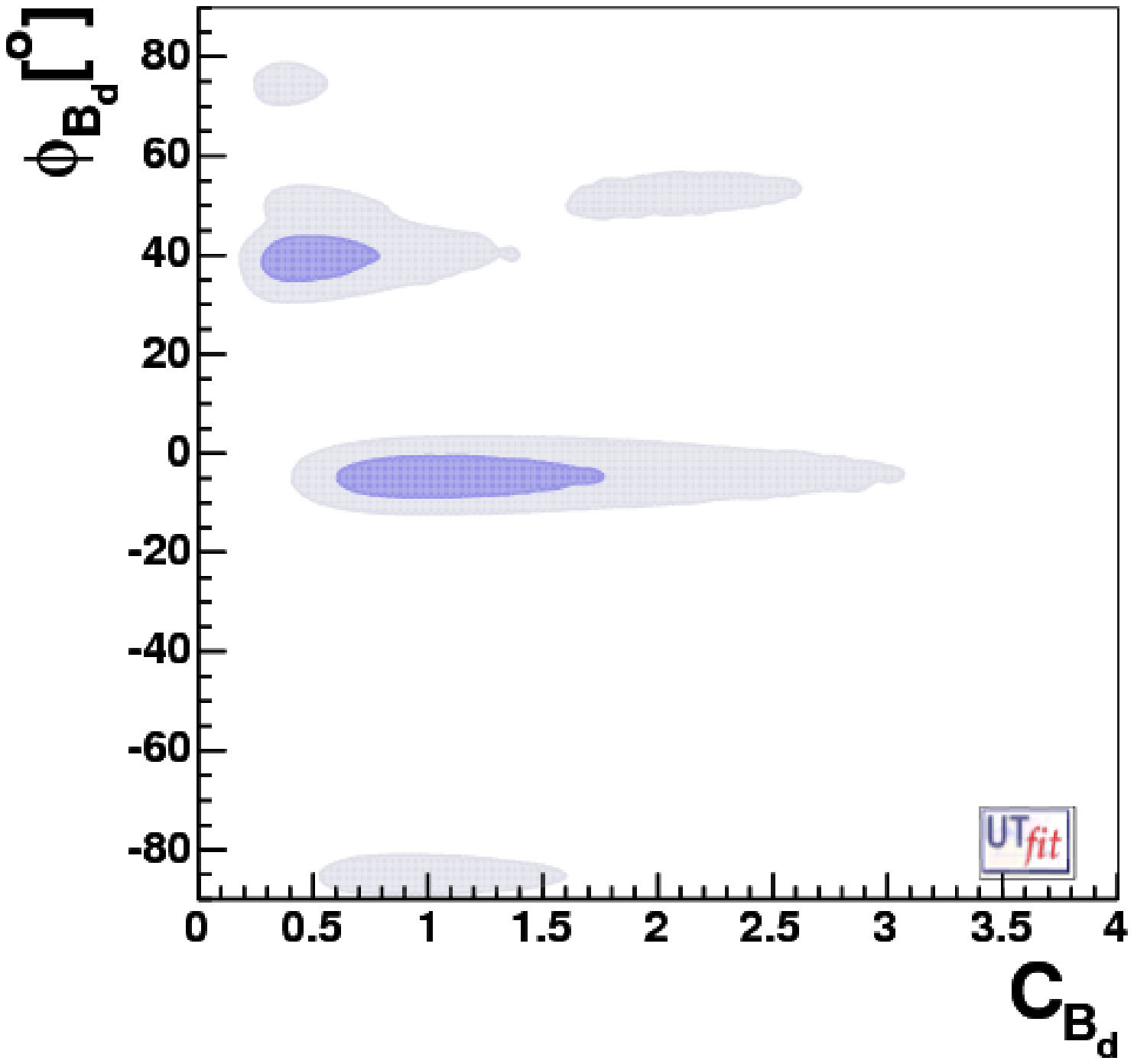}
 \includegraphics[width=0.32\textwidth]{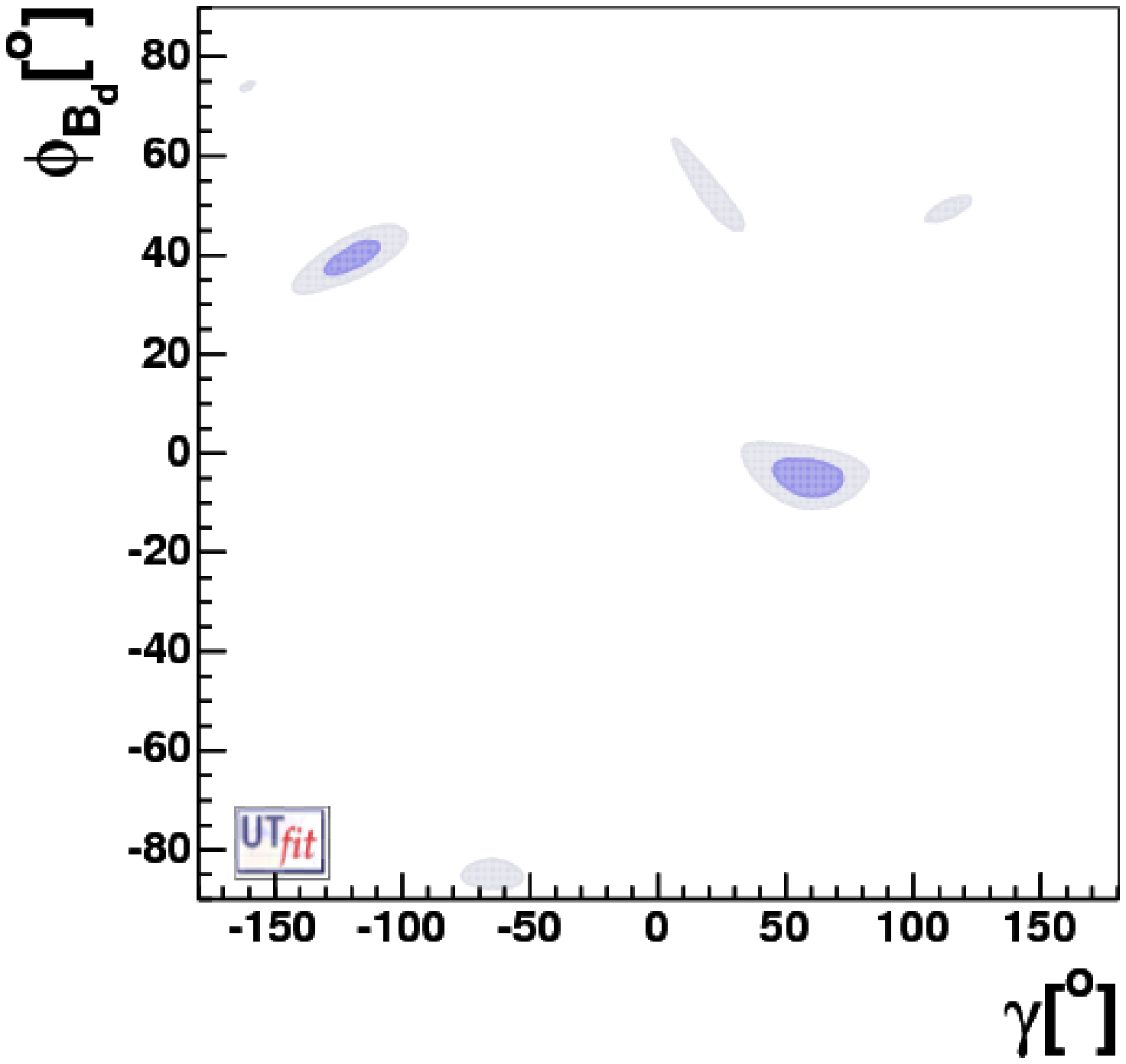}
\caption{%
  \textit{2D distributions of $\phi_{B_d}$ vs. $C_{B_d}$ (left) and
  $\phi_{B_d}$vs. $\gamma$ (right) using the following constraints:
  i) $|V_{ub}/V_{cb}|$, $\Delta m_d$, $\varepsilon_K$ and $\sin 2
  \beta$ (first row); ii) the constraints in i) plus $\gamma$ (second
  row); iii) the constraints in ii) plus $\cos 2 \beta$ from $B_d \to
  J/\psi K^*$ and $\beta$ from $B \to D h^0$ (third row);
  iv) the constraints in ii) plus $\alpha$ (fourth row).}}
\label{fig:individual}
\end{center}
\end{figure}

To illustrate the impact of each experimental constraint on the analysis,
in Fig.~\ref{fig:individual} we show the selected regions
in the $\phi_{B_d}$ vs. $C_{B_d}$ and $\phi_{B_d}$ vs. $\gamma$ planes
using different combinations of constraints. The first row represents
the pre-2004 situation, when only $|V_{ub}/V_{cb}|$, $\Delta m_d$,
$\varepsilon_K$ and $\sin 2 \beta$ were available, selecting a
continuous band for $\phi_{B_d}$ as a function of $\gamma$ and a broad
region for $C_{B_d}$. Adding the determination of $\gamma$ (second
row), only four regions in the $\phi_{B_d}$ vs. $\gamma$ plane
survive, two of which overlap in the $\phi_{B_d}$ vs. $C_{B_d}$ plane.
Two of these solutions have values of $\cos 2 (\beta + \phi_{B_d})$
and $\alpha - \phi_{B_d}$ different from the SM predictions, and are
therefore disfavoured by $(\cos 2 \beta)^\mathrm{exp}$ and
by the measurement of $(2 \beta)^\mathrm{exp}$ from $B \to D h^0$
decays, and by $\alpha^\mathrm{exp}$ (third and fourth row
respectively). On the other hand, the third solution has a very large
value for $A_\mathrm{SL}$ and is therefore disfavoured by
$A_\mathrm{SL}^\mathrm{exp}$, leading to the final results already
presented in Fig.~\ref{fig:NP}.

\begin{table}[ht]
\begin{center}
\begin{tabular}{ccc}
\hline
\multicolumn{3}{c}{Generalized {\utfit} analysis in the presence of NP}
\\
\hline
& Standard Solution $(\gamma>0)$  &   Non-Standard Solution $(\gamma<0)$\\
\hline
\multicolumn{3}{c}{UT parameters} \\
\hline
$\rhobar$ & 0.246 $\pm$ 0.053 ([0.115,\,0.370] @95$\%$) & [-0.230,\,-0.212] @95$\%$ \\
$\etabar$ & 0.379 $\pm$ 0.039 ([0.277,\,0.463] @95$\%$) & [-0.398,\,-0.381] @95$\%$ \\
\hline
$\sin 2 \beta$ & 0.799 $\pm$ 0.037 ([0.694,\,0.880] @95$\%$) & [-0.588,\,-0.574] @95$\%$  \\ 
$\gamma$ $[^\circ]$ & $57.1 \pm 7.8$ ([37.9,\,75.4] @95$\%$) & [-121.5,\,-118.4] @95$\%$  \\
$\alpha$ $[^\circ]$ & $96.0 \pm 8.4$ ([78.3,\,116.5] @95$\%$) & [-44.5,\, -40.0] @95$\%$  \\
$2\beta+\gamma$ $[^\circ]$ & $110.9 \pm 9.2$ ([88.8,\, 128.6] @95$\%$)& [-158.5,\,-153.0] @95$\%$ \\
Im $\lambda_t$ [$\times 10^{-5}$]& \multicolumn{2}{c}{$14.9\pm1.5$ ([11.7,\,17.6] @95$\%$)}\\
$\Delta m_s$ [ps$^{-1}$] & \multicolumn{2}{c}{$18.0 \pm 5.3$([8.9,\,29.6] @95$\%$)} \\
\hline 
\multicolumn{3}{c}{NP related parameters} \\
\hline
$C_{B_d}$ & \multicolumn{2}{c}{$1.27 \pm 0.44$ ([0.56,\,2.51] @95$\%$)}\\
$\phi_{B_d} [^\circ]$ & $-4.7 \pm 2.3$ ([-9.9,\,1.0] @95$\%$) & [39.0,\,39.8] @95$\%$ \\
\hline
$C_{\epsilon_K}$ & $0.95 \pm 0.18$ ([0.64,\,1.44] @95$\%$) & [-0.71,\,-0.59] @95$\%$ \\
\hline
\end{tabular}
\caption{\textit{Results of the NP generalized analysis on UT
    parameters. The values for $C_{B_d}$, $\phi_{B_d}$ and
    $C_{\epsilon_K}$ are reported.  The second solution is excluded at
    68$\%$ probability level so we quote the $95\%$ ranges only.}}
\label{tab:np} 
\end{center}
\end{table}

\begin{figure}[hp]
\begin{center}
{\includegraphics[width=0.45\textwidth]{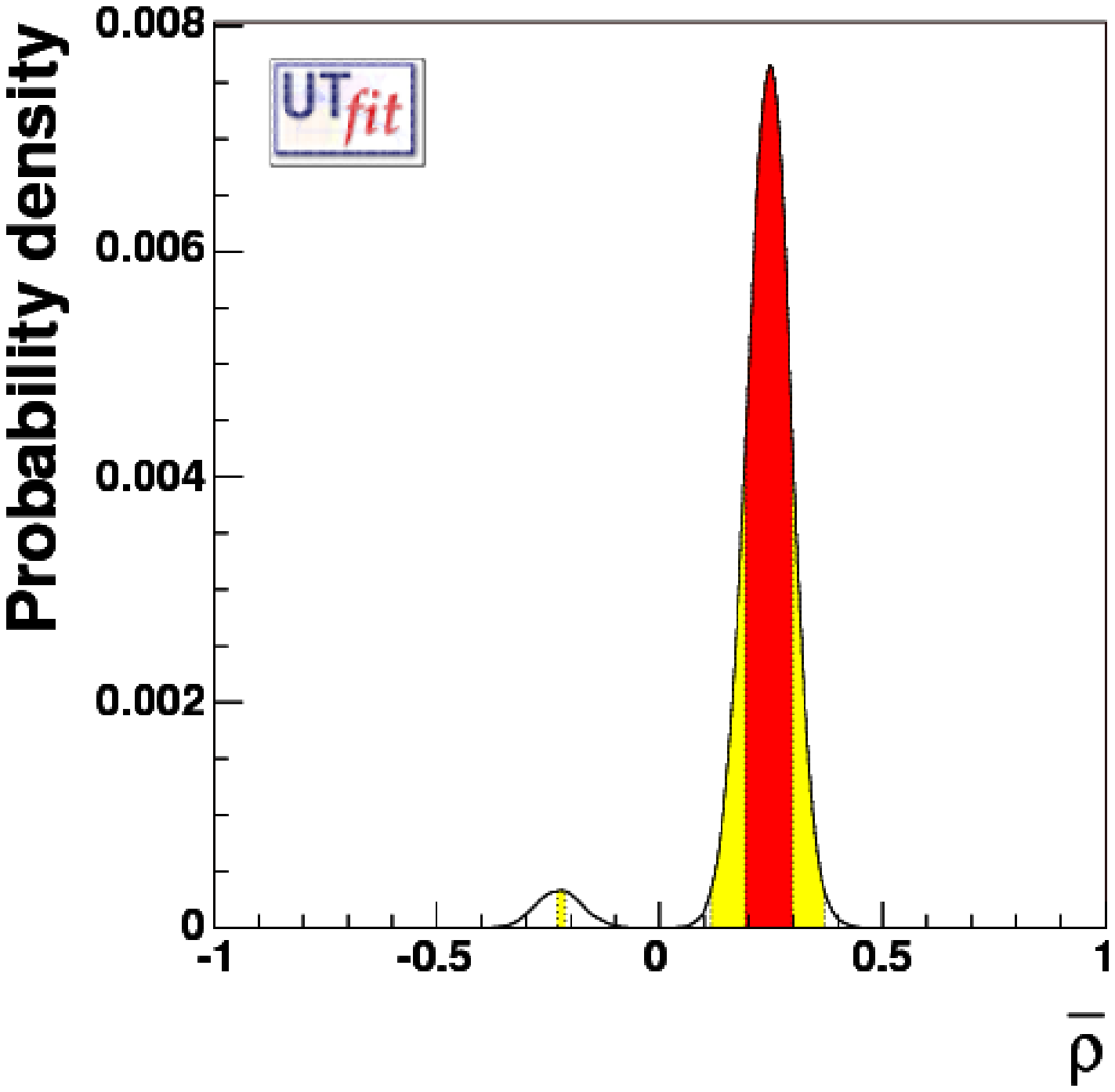}}
{\includegraphics[width=0.45\textwidth]{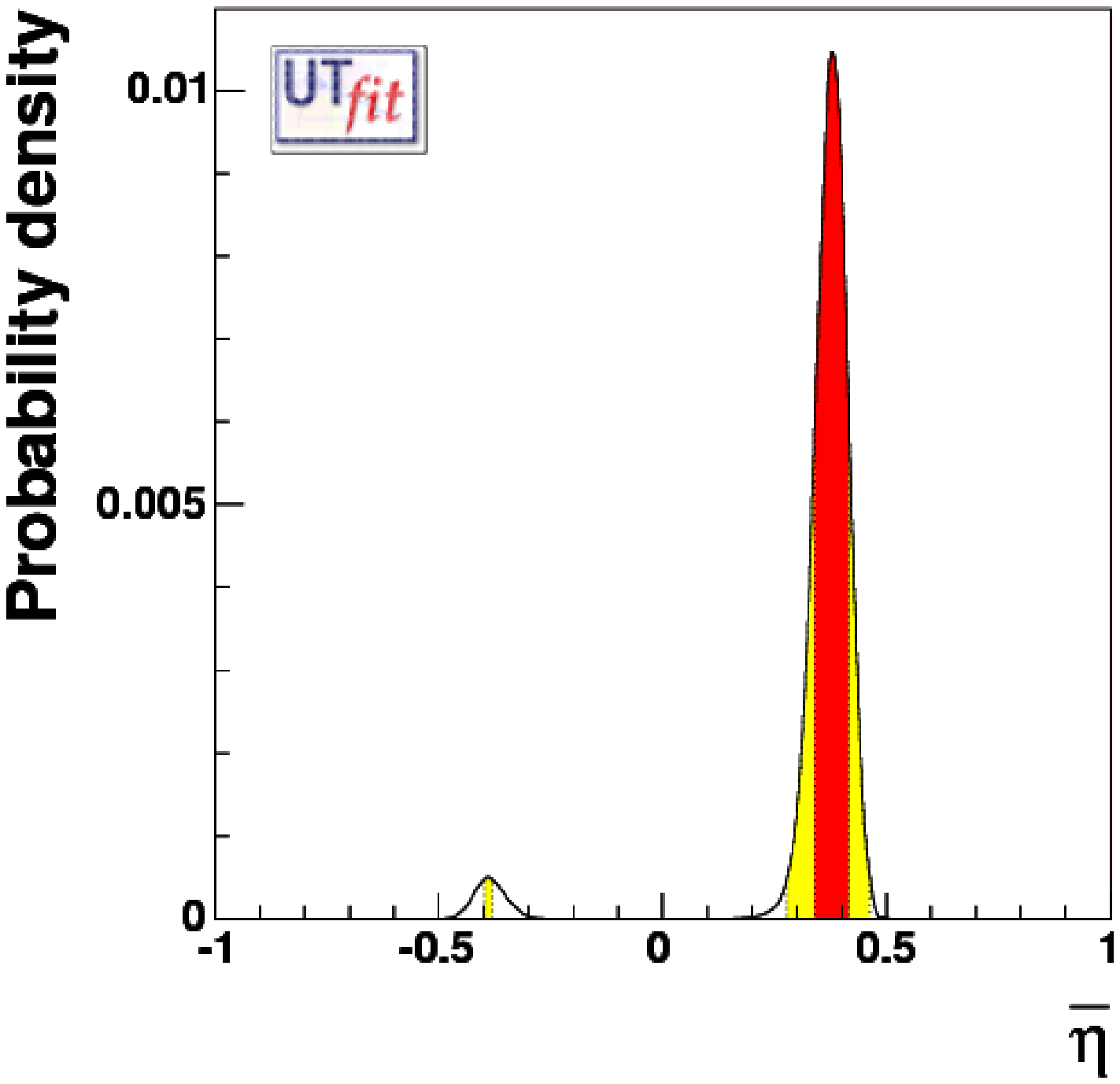}}
{\includegraphics[width=0.45\textwidth]{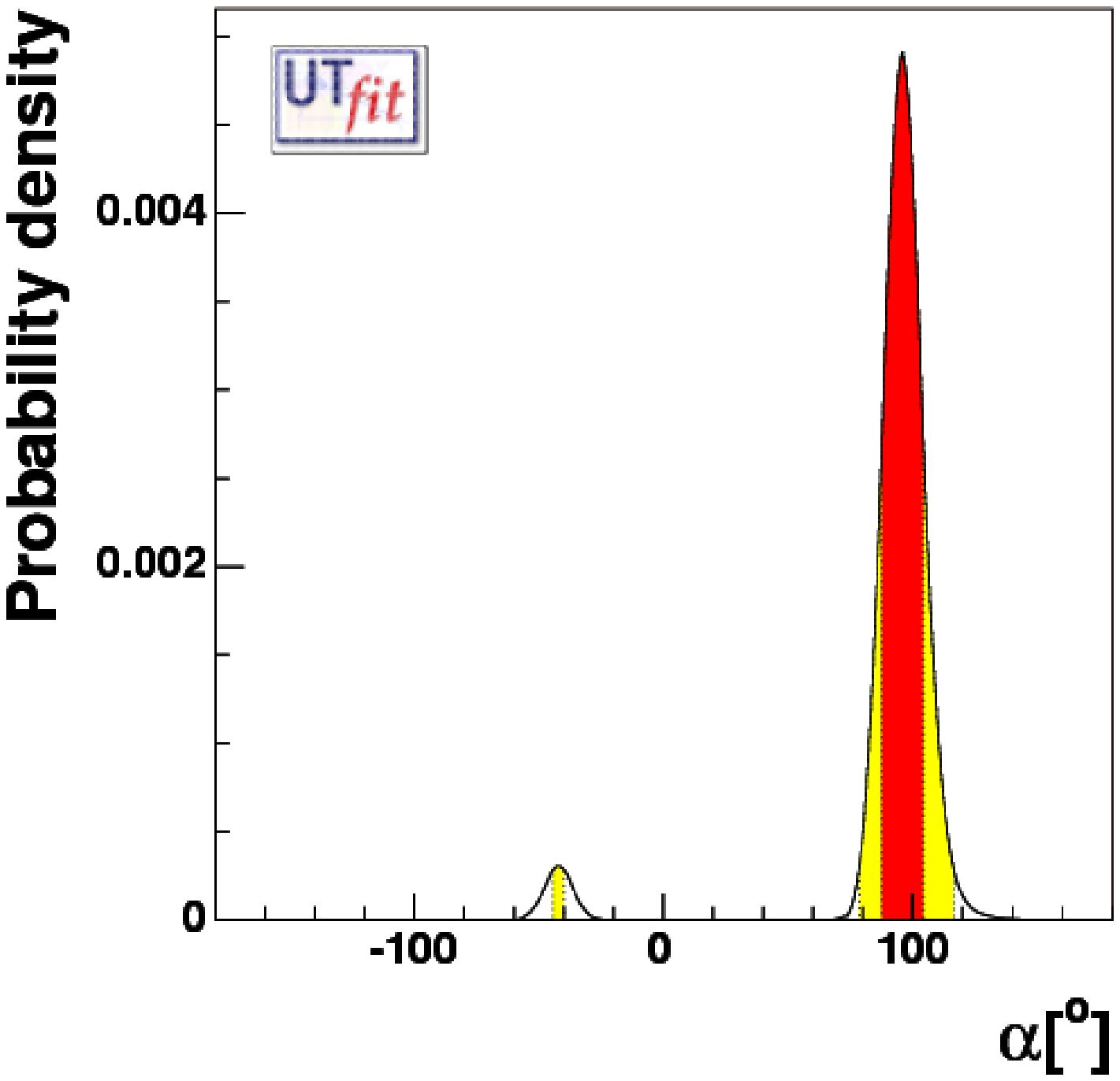}}
{\includegraphics[width=0.45\textwidth]{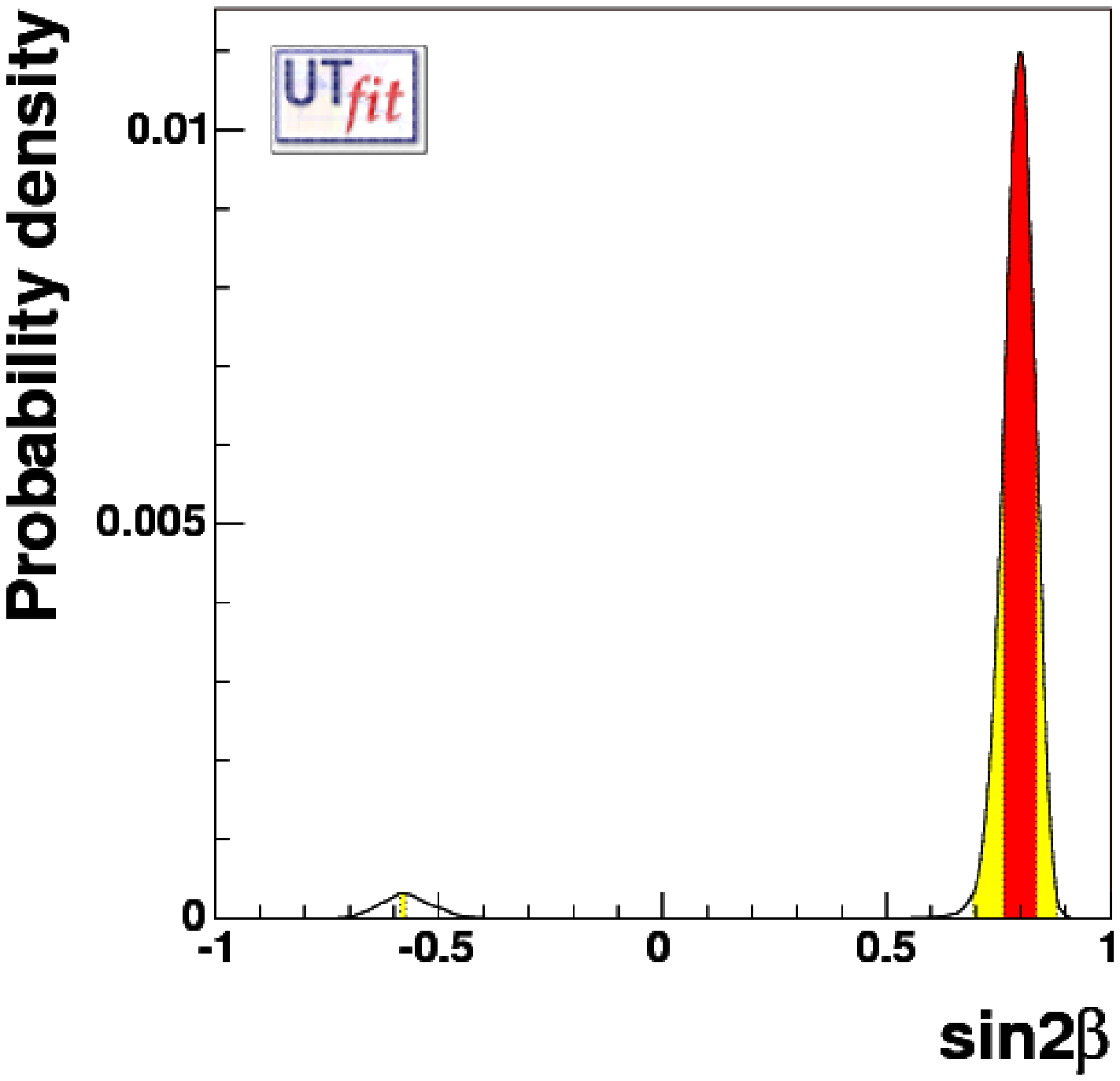}}
{\includegraphics[width=0.45\textwidth]{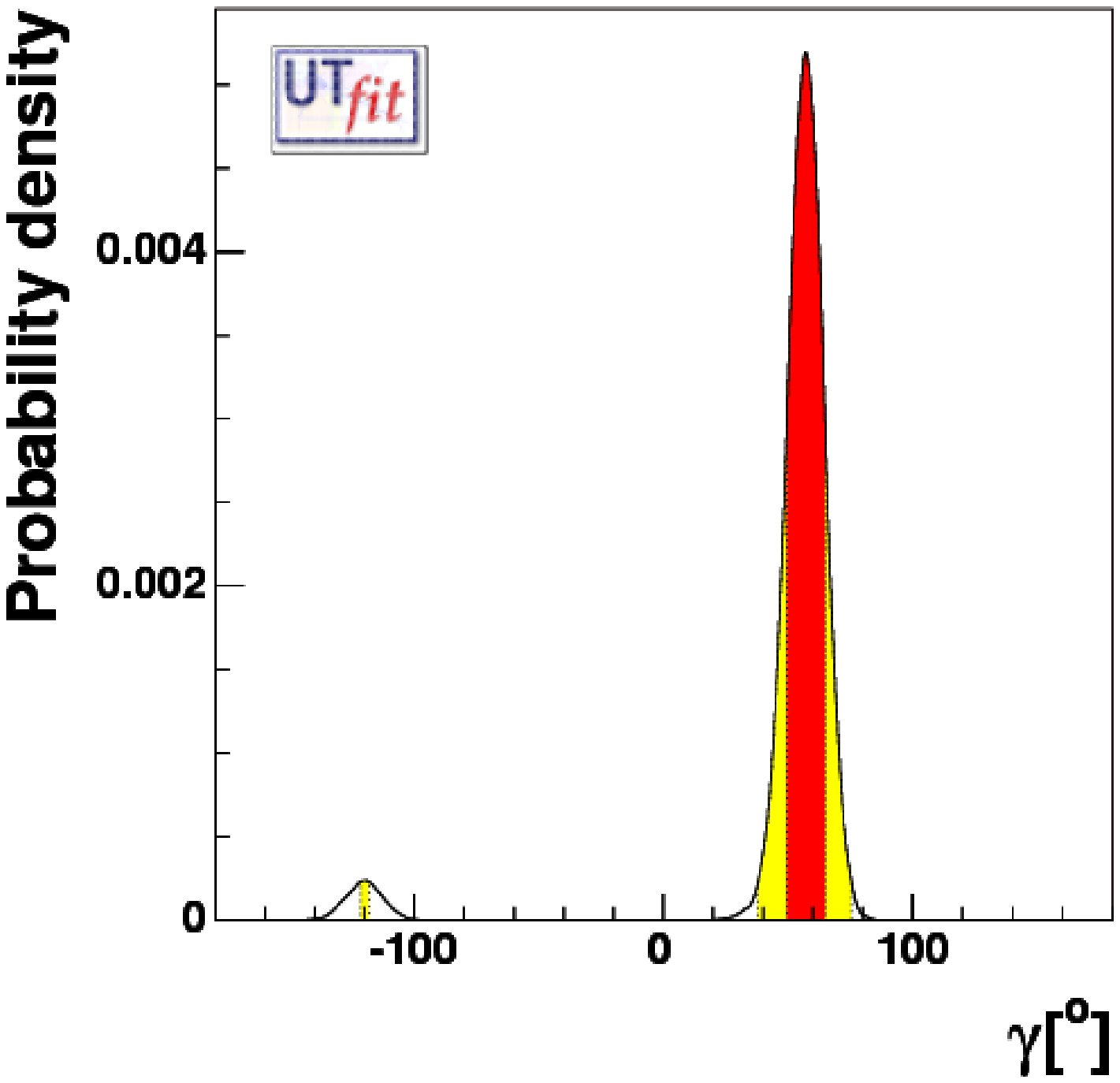}}
{\includegraphics[width=0.45\textwidth]{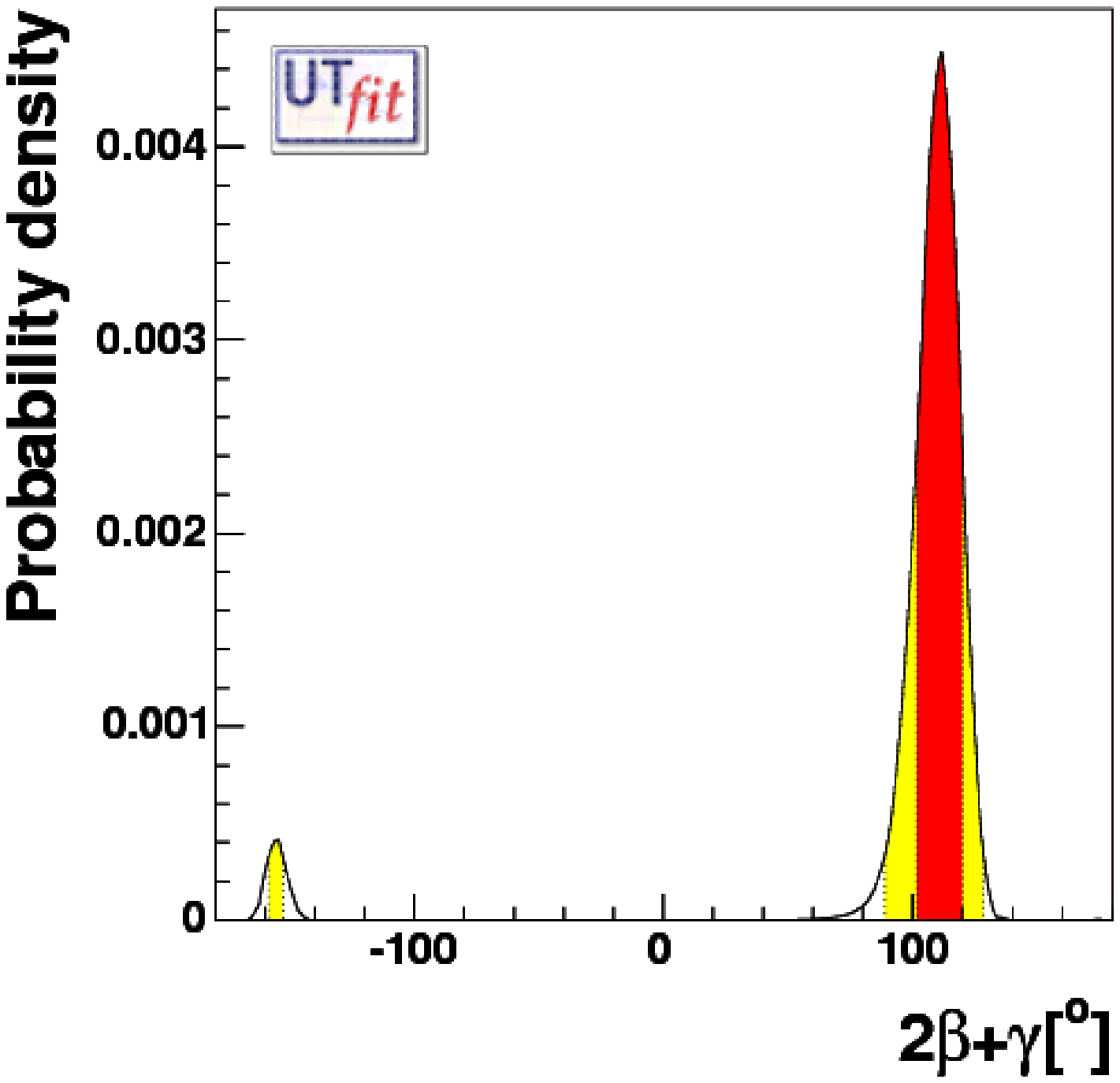}}
\caption{\textit{From top to bottom and from left to right, the
    p.d.f.'s for $\rhobar$, $\etabar$, $\alpha$, $\snb$, $\gamma$ and
    $2\beta+\gamma$. The red (darker) and the yellow (lighter) zones
    correspond respectively to 68\% and 95\% of the area. These
    results are obtained in the presence of NP in all the processes
    entering the UT analysis.}}
\label{fig:NP-1dim}
\end{center}
\end{figure}

\begin{figure}[thb!]
\begin{center}
\includegraphics[width=0.45\textwidth]{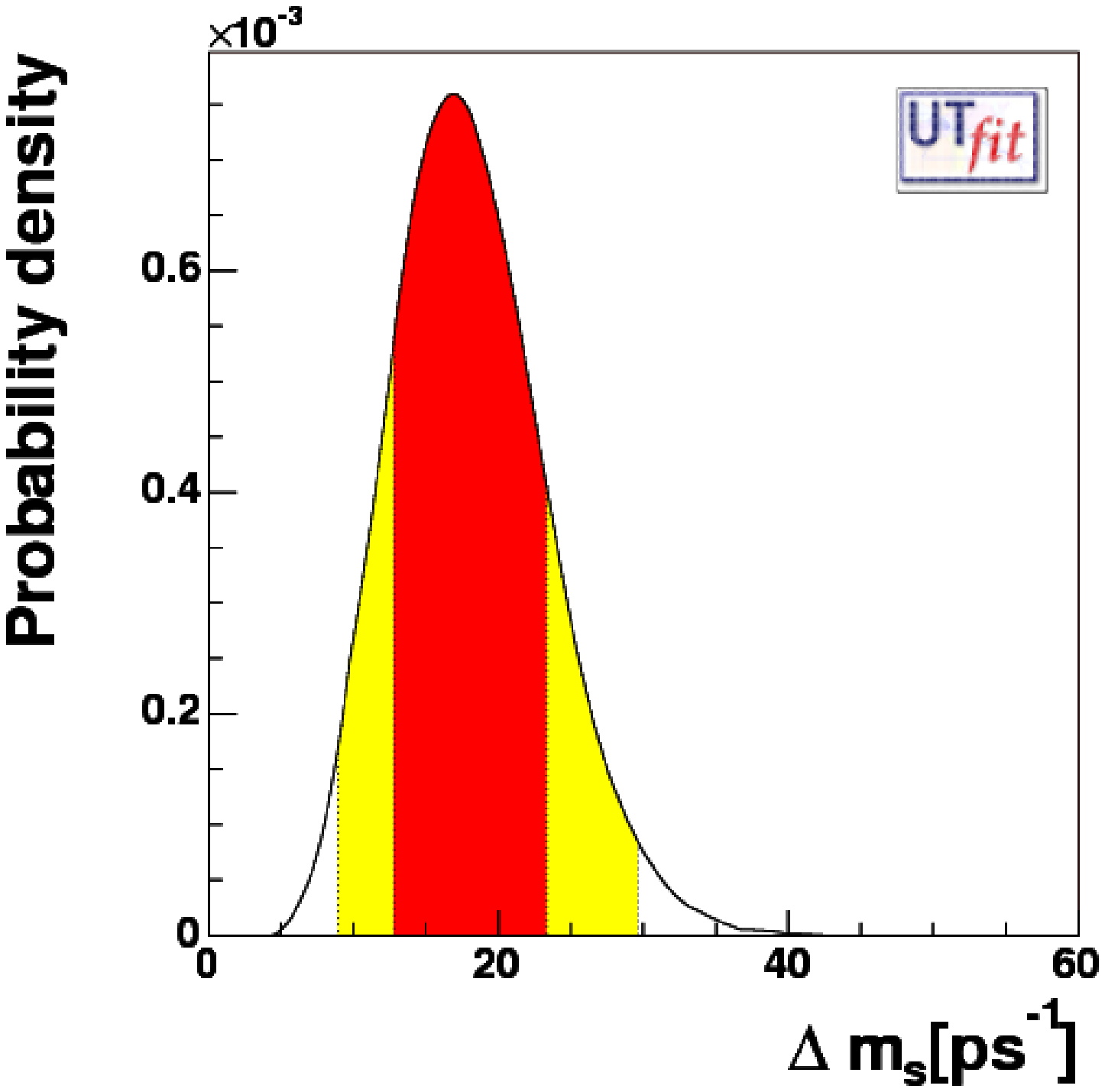} 
\includegraphics[width=0.45\textwidth]{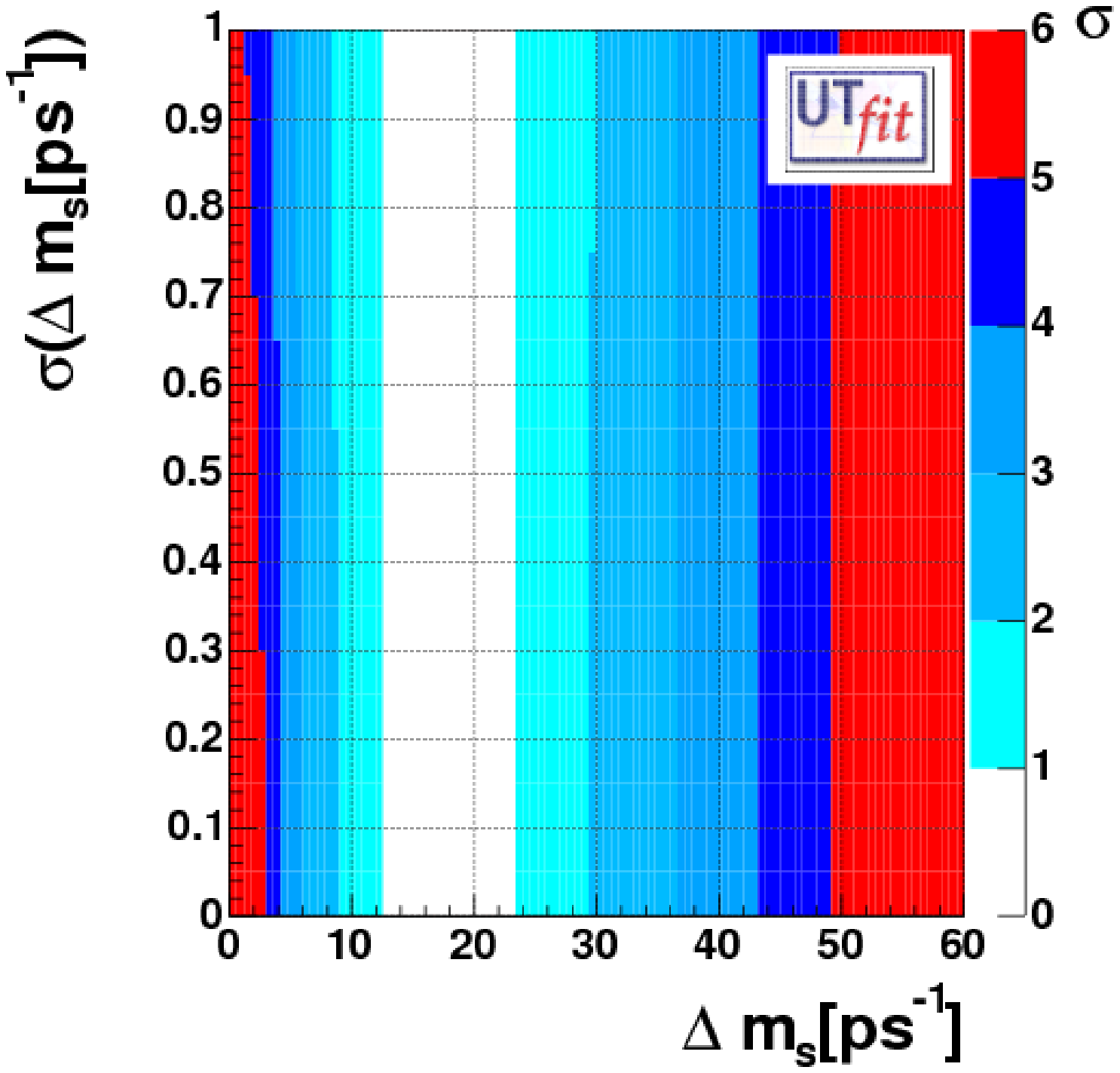} 
\caption[]{\textit{P.d.f. (left) and compatibility plot (right) 
for the SM contribution to $\dms$ in the presence of NP in all
the quantities entering the UT analysis, setting $C_{B_s}=1$.}}
\label{fig:dmsNP}
\end{center}
\end{figure}

In Tab.~\ref{tab:np} we give the numerical results for the NP
parameters and some of the relevant UT quantities, for which we
show the output distributions in Fig.~\ref{fig:NP-1dim}. A comment is
needed for the case of $\dms$: the output distribution reported in
Fig.~\ref{fig:dmsNP} represents the SM contribution only
(\textit{i.e.} it corresponds to $C_{B_s}=1$).
Therefore this numerical result should not be taken as a prediction for
$\dms$ in a general NP scenario in which $C_{B_s} \neq 1$. The
conclusion that we can draw from the output distribution of $\dms$ is
most easily read from the compatibility plot\footnote{The method used
  to calculate the level of agreement in the compatibility plot is
  explained in \cite{utfit2005}.}  shown in Fig.~\ref{fig:dmsNP}: a
value of $\dms > 30\, (36)$ ps$^{-1}$ would imply the presence of NP
in $B_s-\bar B_s$ mixing at the $2\, (3)\, \sigma$ level.  On the
other hand, from the similar result in the contest of the Standard
Model~\cite{utfit2005} one can still conclude that $\dms > 29\, (34)$
ps$^{-1}$ would imply the presence of NP at the $2\, (3)\, \sigma$
level (but not necessarily in the $B_s$ sector).

\begin{figure}[thb!]
\begin{center}
\includegraphics[width=0.8\textwidth]{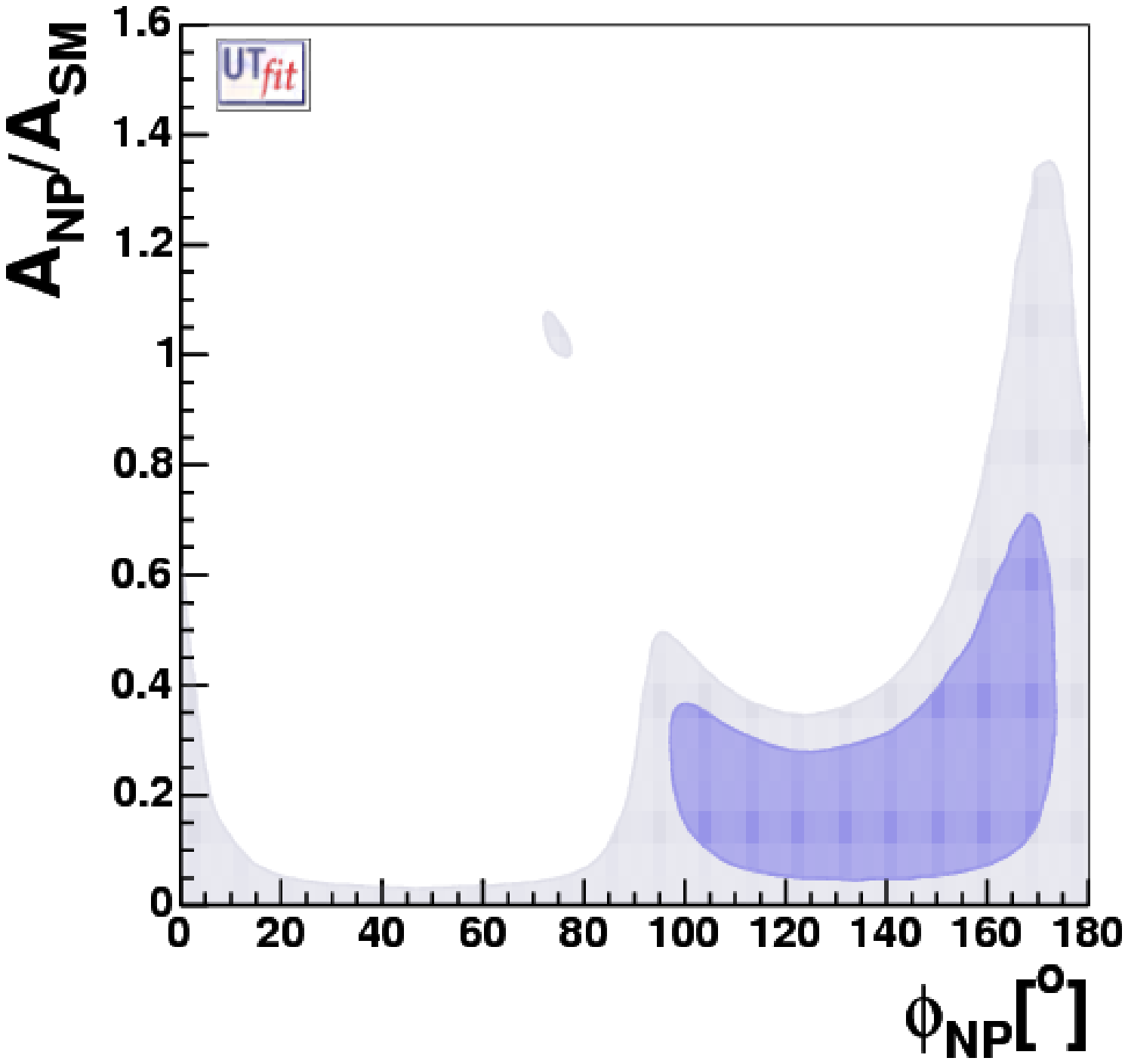} 
\caption[]{\textit{P.d.f. in the $(A_\mathrm{NP}/A_\mathrm{SM})$ vs.
    $\phi_\mathrm{NP}$ plane for NP in the $|\Delta B|=2$ sector (see
    Eq.~(\ref{eq:asmanp})).}}
\label{fig:achilleplot}
\end{center}
\end{figure}

Before concluding this section, let us analyze more in detail the
results in Fig.~\ref{fig:NP}. Writing
\begin{equation}
  \label{eq:asmanp}
  C_{B_d}e^{2 i \phi_{B_d}} = \frac{A_\mathrm{SM} e^{2 i \beta} +
    A_\mathrm{NP} e^{2 i (\beta + \phi_\mathrm{NP})}}{A_\mathrm{SM}
    e^{2 i \beta}}\,, 
\end{equation}
and given the p.d.f. for $C_{B_d}$ and $\phi_{B_d}$, we can derive the
p.d.f. in the $(A_\mathrm{NP}/A_\mathrm{SM})$ vs. $\phi_\mathrm{NP}$
plane. The result is reported in Fig.~\ref{fig:achilleplot}. We see
that the NP contribution can be substantial if its phase is close to
the SM phase, while for arbitrary phases its magnitude has to be much
smaller than the SM one. Notice that, with the latest data, the SM
($\phi_{B_d}=0$) is disfavoured at $68\%$ probability due to a slight
disagreement between $\sin 2\beta$ and $|V_{ub}/V_{cb}|$.  This
requires $A_\mathrm{NP}\neq 0$ and $\phi_\mathrm{NP}\neq 0$. For the
same reason, $\phi_\mathrm{NP}>90^\circ$ at $68\%$ probability and the
plot is not symmetric around $\phi_\mathrm{NP}=90^\circ$. 

A similar parameterization has been used in ref.~\cite{pirgiolo}.
Comparing our Fig.~\ref{fig:achilleplot} with Fig.~5 of
ref.~\cite{pirgiolo}, one notices small differences.  However, since
they are using the statistical method of ref.~\cite{CKMfitter}, they
are plotting areas corresponding to ``at least'' confidence levels, so
that their areas are expected to be larger than ours.

Assuming that the small but non-vanishing value for $\phi_{B_d}$ we
obtained is just due to a statistical fluctuation, the result of our
analysis points either towards models in which new sources of flavour
and CP violation are only present in $b \to s$ transitions, a
well-motivated possibility in flavour models and in grand-unified
models, or towards models with no new source of flavour and CP
violation beyond the ones present in the SM (Minimal Flavour
Violation). This second possibility will be studied in detail in
Section~\ref{sec:uut}.

\section{Constraints on NP from $\mathbf{\vert \Delta S \vert=2}$ or 
  $\mathbf{\vert \Delta  B \vert=2}$ transitions only}
\label{sec:ds2}

A complementary information to the one presented in the previous
section is obtained by allowing NP contributions to be present only in
$|\Delta S|=2$ or $|\Delta B|=2$ transitions. This can be useful to test
models beyond the SM in which NP contributions are expected to affect
dominantly only one of these two sectors, and is also the starting
point to update previous analyses of NP in $|\Delta S|=2$ or $|\Delta
B|=2$ processes in supersymmetry \cite{SUSYDS2,SUSYDB2} or in any other given
model.

\begin{figure}[ht!]
\begin{center}
{\includegraphics[width=.45\textwidth]{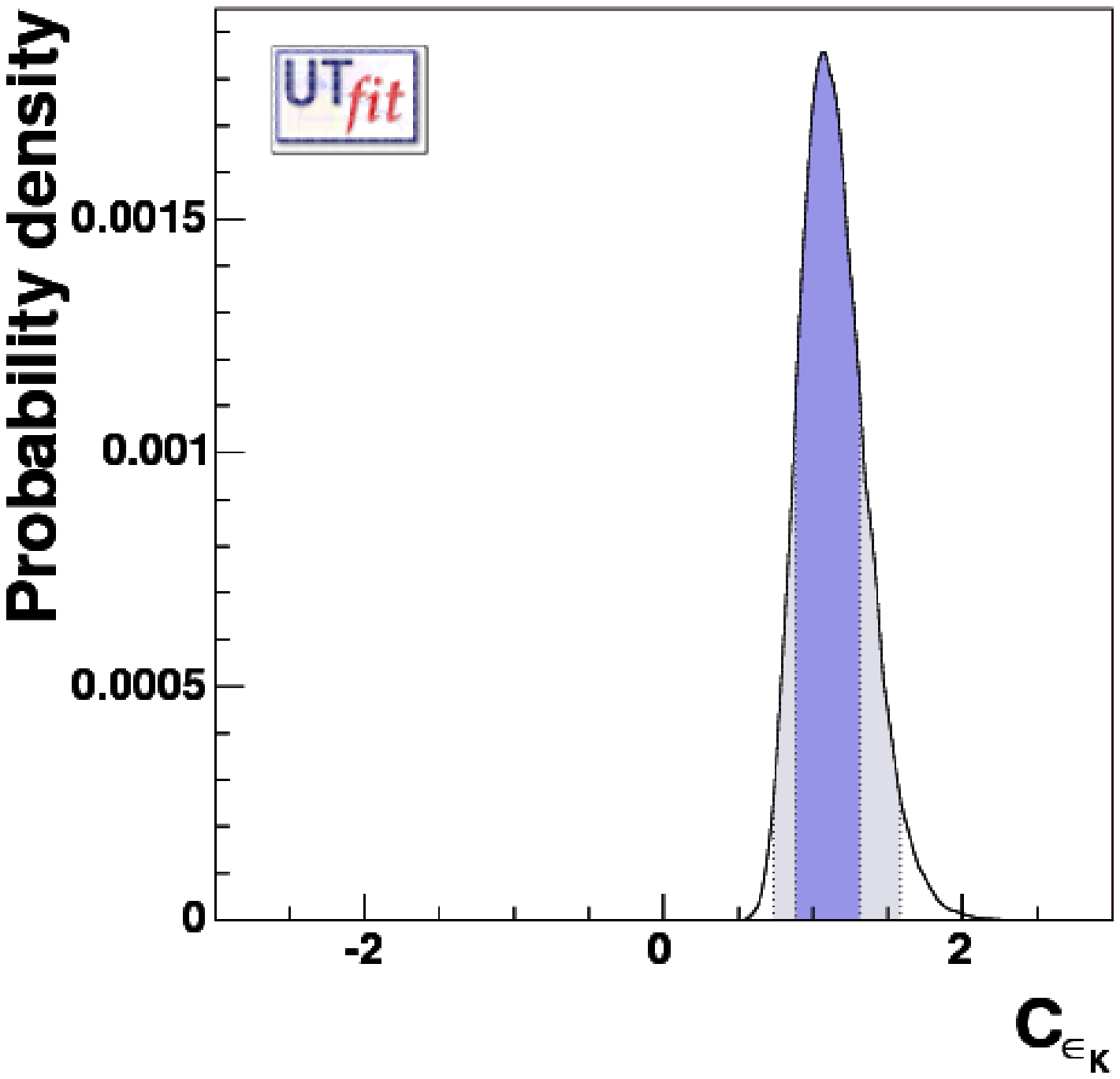}} 
{\includegraphics[width=.45\textwidth]{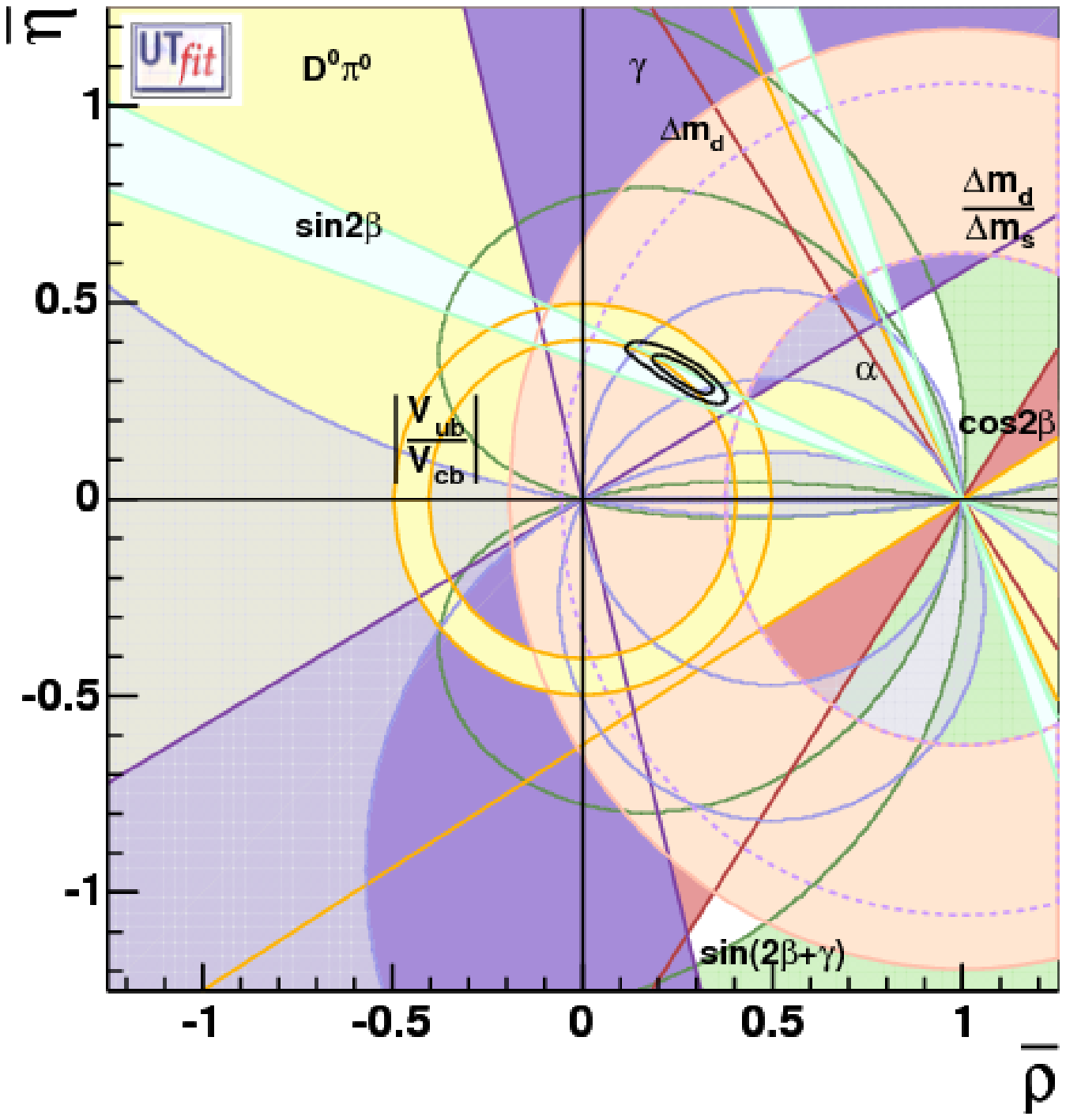}} 
\caption{\textit{The
    p.d.f. for $C_{\varepsilon_K}$ (left) and the
    $\rhobar-\etabar$ plane (right). In the left plot, the 
    darker and the lighter zones
    correspond respectively to 68\% and 95\% of the area. These
    results are obtained in the presence of NP in $K^0-\bar{K}^0$
    mixing only.}}
\label{fig:epskonly}
\end{center}
\end{figure}

Allowing NP to affect only $C_{\varepsilon_K}$, we obtain the results
for the UT parameters, for $C_{\varepsilon_K}$ and for $\rhobar$ and
$\etabar$ reported in Fig.~\ref{fig:epskonly} and in
Tab.~\ref{tab:npseparate}. The determination of the UT is essentially
equivalent to the SM one, since only $\varepsilon_K$ is missing in
this case.

\begin{figure}[hp]
\begin{center}
{\includegraphics[width=.45\textwidth]{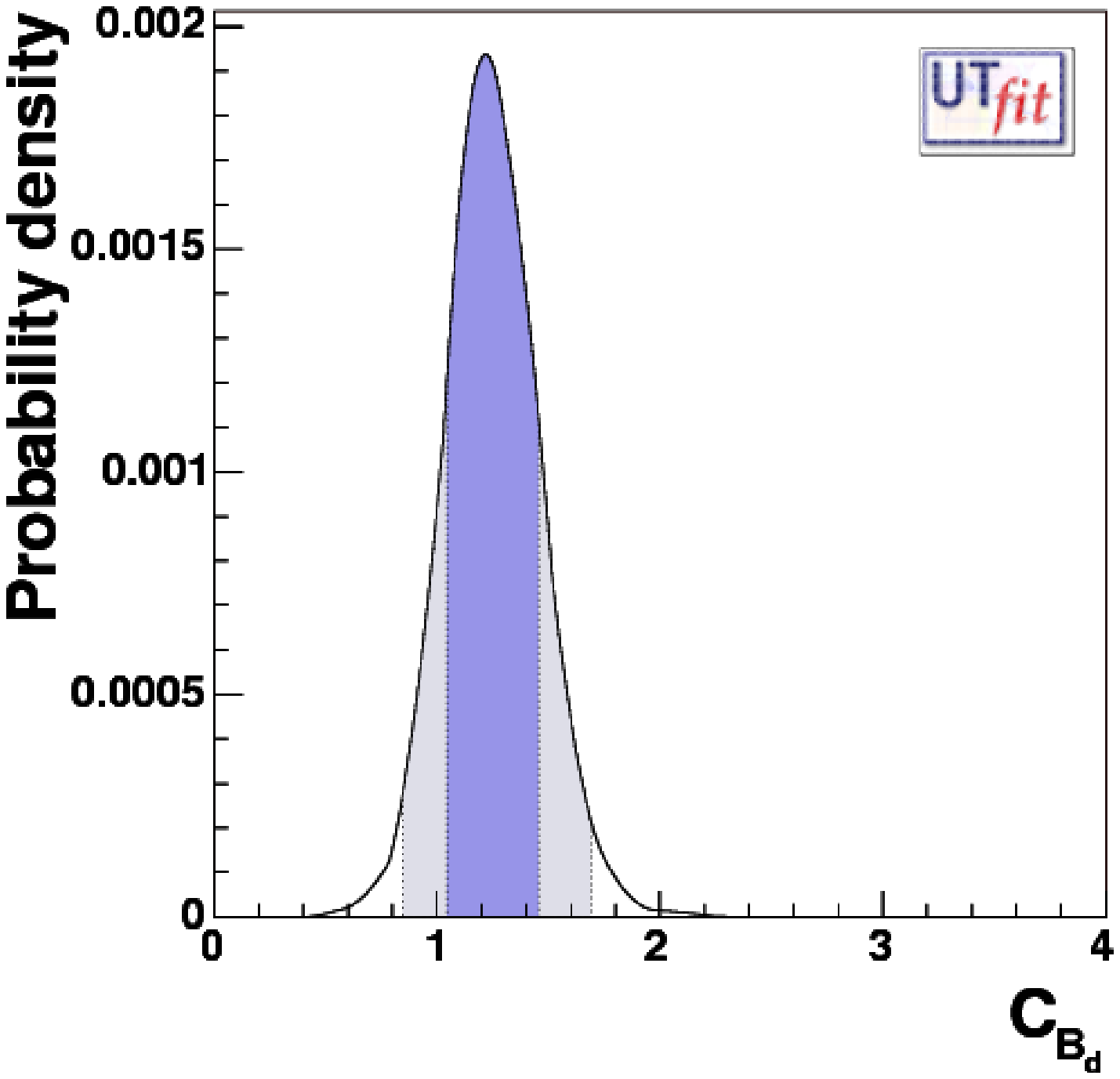}}
{\includegraphics[width=.45\textwidth]{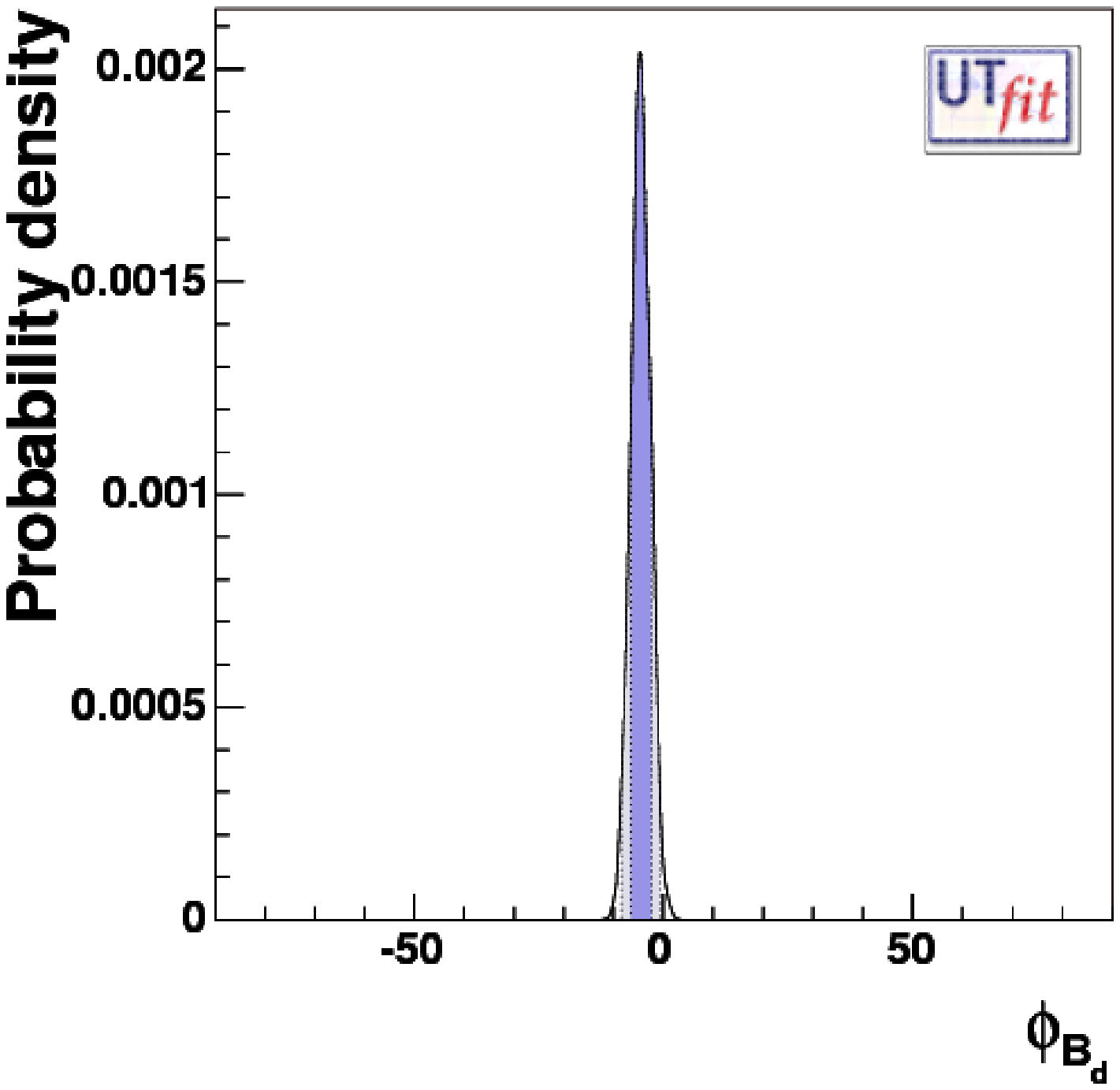}} 
{\includegraphics[width=.45\textwidth]{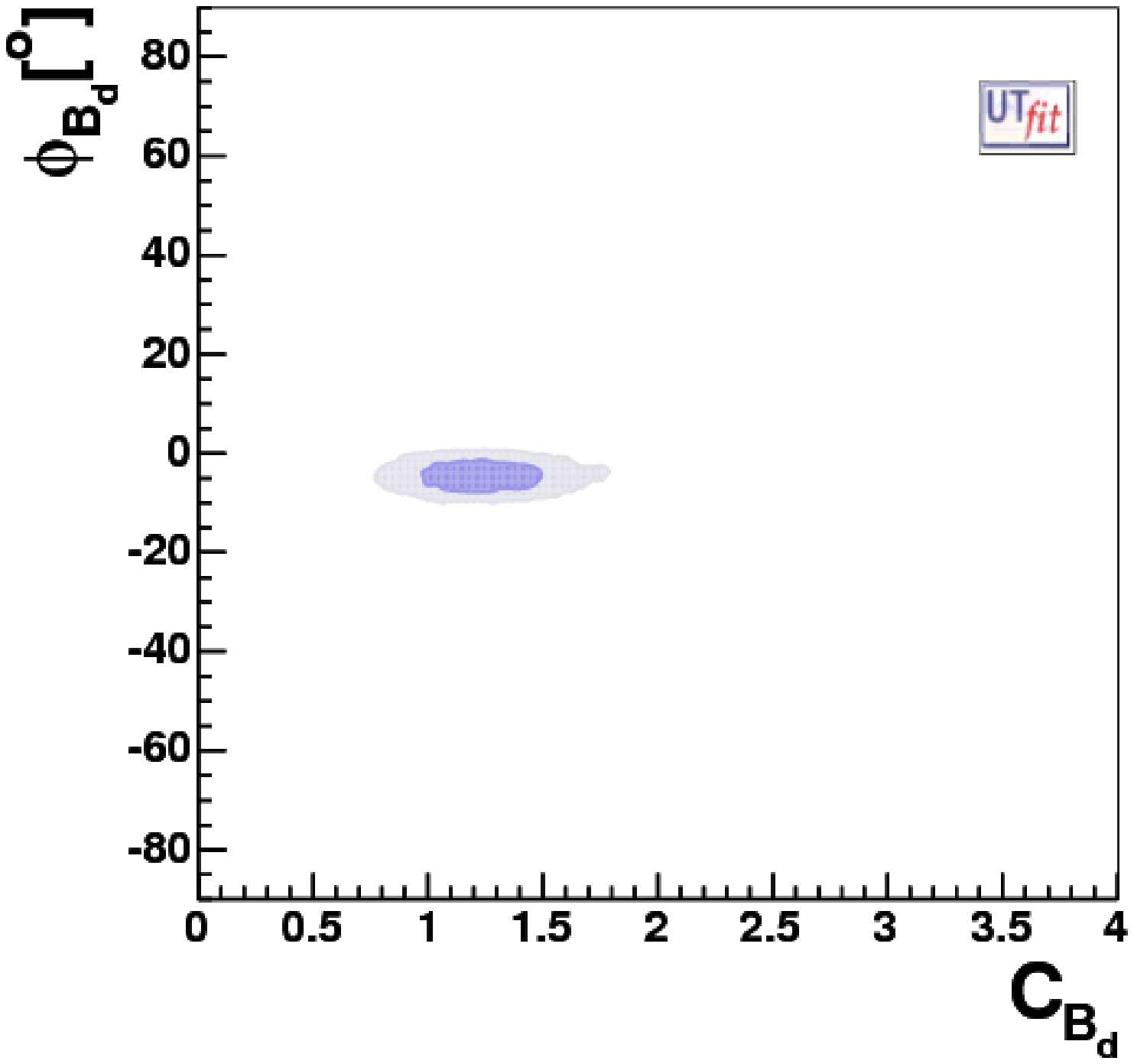}}
{\includegraphics[width=.45\textwidth]{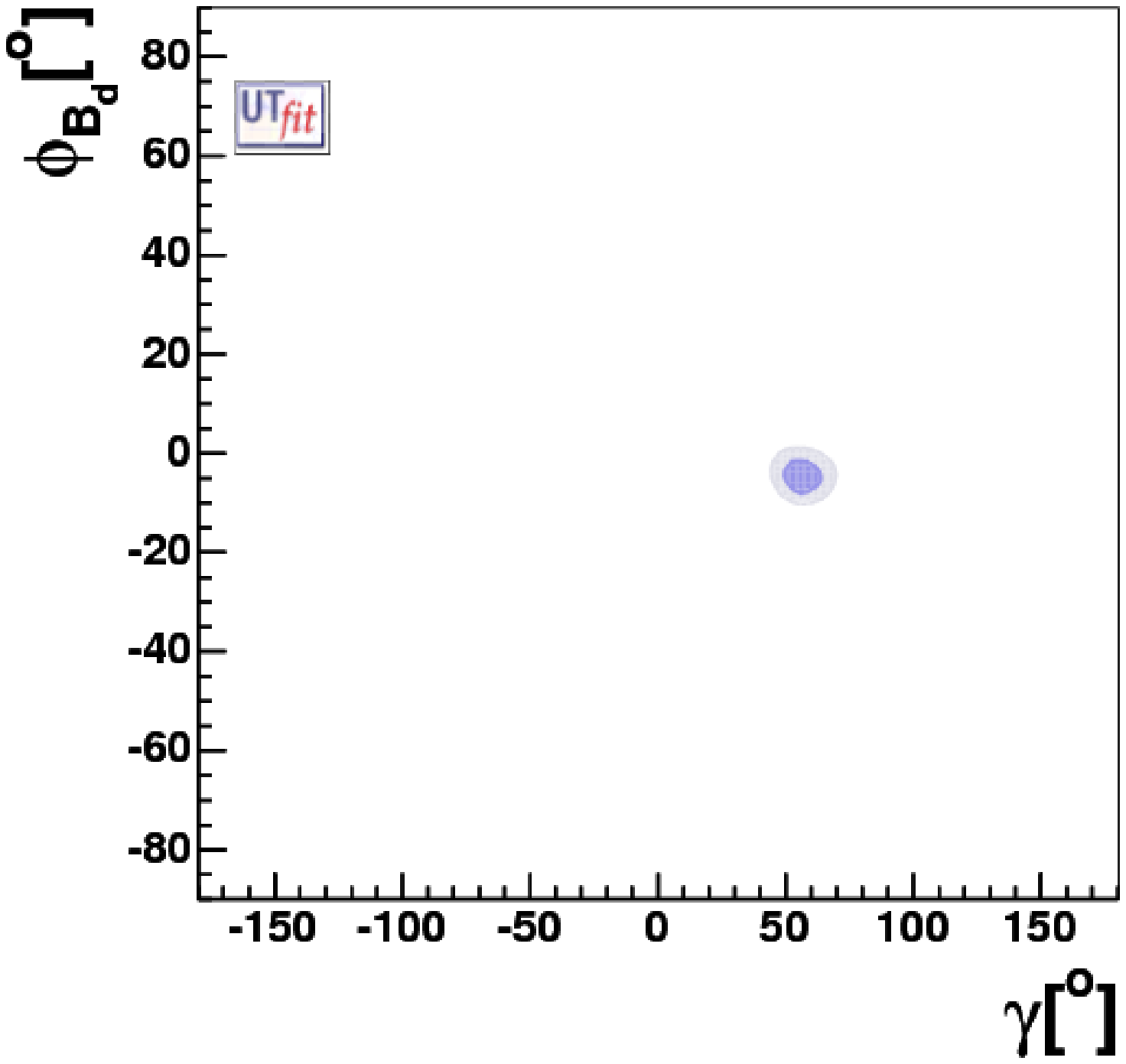}} 
{\includegraphics[width=.45\textwidth]{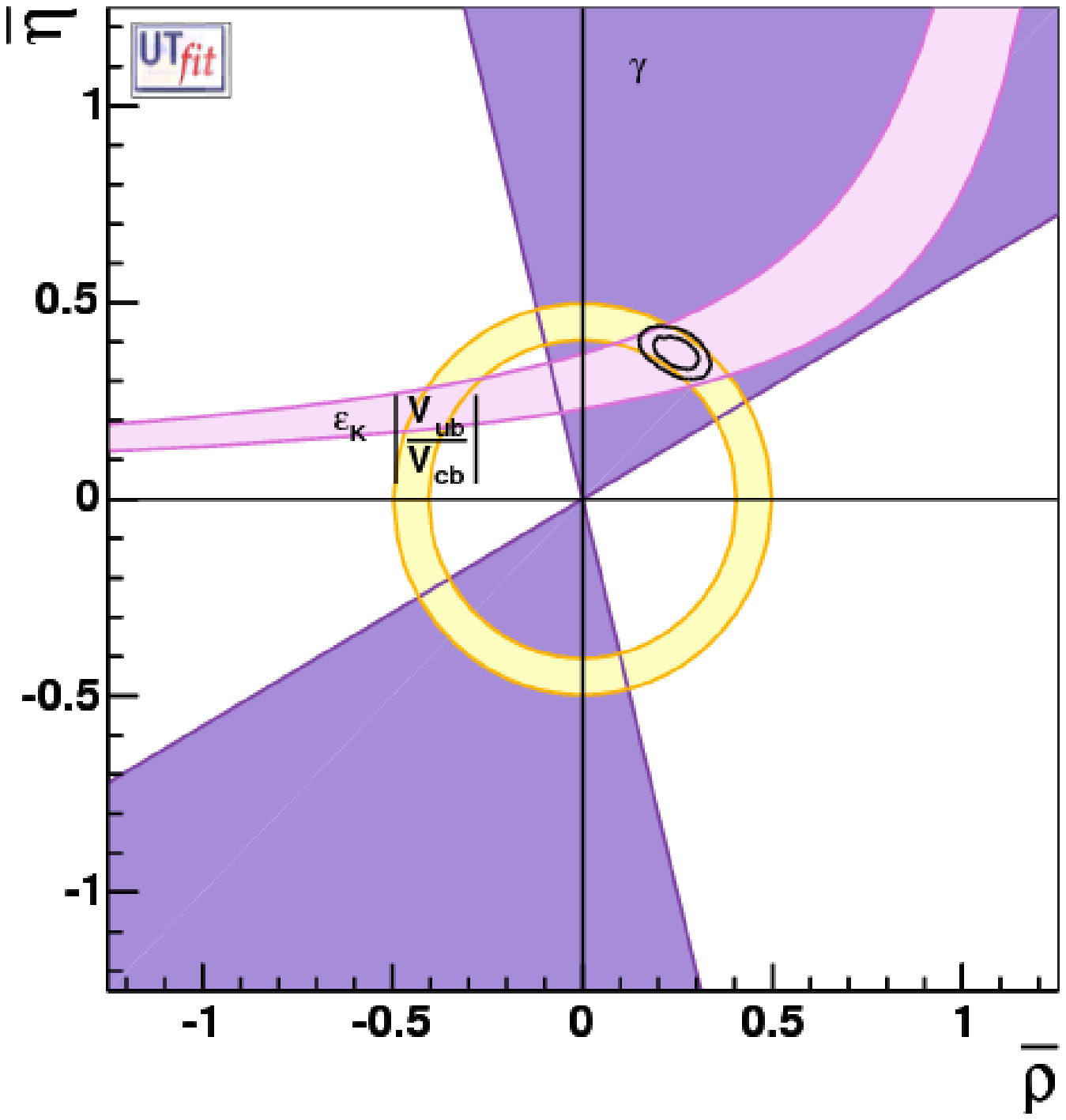}} 
\caption{\textit{From top to bottom and from left to right, the
    p.d.f.'s for $C_{B_d}$, $\phi_{B_d}$,  $\phi_{B_d}$ vs. $C_{B_d}$,
    $\phi_{B_d}$ vs. $\gamma$ and the $\rhobar-\etabar$ plane. The
    darker and the lighter zones correspond respectively to 68\% and
    95\% of the area. These results are obtained in the presence of NP
    in $B^0_d-\bar{B}^0_d$ mixing only.}}
\label{fig:Bdonlynp}
\end{center}
\end{figure}

For the case in which NP only enters $B_d - \bar B_d$ mixing, the
results are given in Fig.~\ref{fig:Bdonlynp} and in
Tab.~\ref{tab:npseparate}. The main difference with the results in the
previous section is that one can use $\varepsilon_K$ to eliminate the
solutions with negative $\gamma$.

\begin{table}[hbt]
\begin{center}
\begin{tabular}{|c|c|c|}
\hline
\multicolumn{3}{|c|}{Generalized {\utfit} analysis in the presence of NP}
\\
\hline
& $|\Delta S|=2$ \textbf{only} &   $|\Delta B_d|=2$ \textbf{only}\\
\hline
\multicolumn{3}{|c|}{UT parameters} \\
\hline
$\rhobar$ & 0.267 $\pm$ 0.056 [0.145,0.368] & 0.246 $\pm$ 0.038 [0.166,0.317]\\
$\etabar$ & 0.319 $\pm$ 0.034 [0.257,0.387]& 0.372 $\pm$ 0.028 [0.318,0.424]\\
\hline
$\sin 2 \beta$ & 0.730 $\pm$ 0.028 [0.674,0.781] & 0.794 $\pm$ 0.033
[0.727,0.854] \\ 
$\gamma$ $[^\circ]$ & $50 \pm 9$ [35,69] & $56 \pm 5$
[46,67]\\ 
$\alpha$ $[^\circ]$& $107 \pm 9$ [87,123]&  $97 \pm 6$
[86,108]\\ 
$2\beta+\gamma$ $[^\circ]$& $100 \pm 10$ [80,118]&  $109
\pm 6$[96,120] \\ 
Im $\lambda_t$ [$\times 10^{-5}$]& $12.6\pm1.4$ [10.1,15.4] & $14.9 \pm 1.0$
[13.0,16.8]  \\ 
$\Delta m_s$ [ps$^{-1}$]& $22.6\pm4.2$ [15.2,30.8] & $17.4 \pm 2.0$ 
[14.8,22.4] \\
\hline 
\multicolumn{3}{|c|}{NP related parameters} \\
\hline
$C_{B_d}$ & 1 & 1.25 $\pm$ 0.21 [0.84,1.69]\\
$\phi_{B_d}$ $[^\circ]$ & 0 & $-4.6 \pm 2.0$ [-8.5,-0.7]\\
\hline
$C_{\epsilon_K}$ & $1.10\pm0.21$ [0.73,1.59]& 1\\
\hline
\end{tabular}
\caption{\textit{Results of the NP generalized analysis on UT
    parameters, when only NP contributions to the $|\Delta S|=2$ (left)
    and $|\Delta B_d|=2$ (right) processes are considered.  The values for
    $C_{B_d}$, $\phi_{B_d}$ and $C_{\epsilon_K}$ are reported.  The
    quoted errors represent 68$\%$ [95\%] probability ranges. }}
\label{tab:npseparate} 
\end{center}
\end{table}

\section{Minimal Flavour Violation models}
\label{sec:uut}

We now specialize to the case of MFV. Making the basic assumption that
the only source of flavour and CP violation is in the Yukawa
couplings \cite{gino}, it can be shown that:
\begin{enumerate}
\item The phase of $|\Delta B|=2$ amplitudes is unaffected by NP, and so
  is the ratio $\Delta m_s/\Delta m_d$. This allows the determination
  of the Universal Unitarity Triangle independent on NP effects, based
  on $\vert V_{ub}/V_{cb}\vert$, $\gamma$, $A_{CP}(B \to J/\Psi K^{(*)})$, $\beta$ from $B \to D^0h^0$,
  $\alpha$, and $\Delta m_s/\Delta m_d$ \cite{uut}.
\item For one-Higgs-doublet models, and for two-Higgs-doublet models
  at low $\tan \beta$, all NP effects in the UT analysis amount to a
  redefinition of the top box contribution to $|\Delta F|=2$ processes
  $S_0(x_t) \to S_0(x_t) + \delta S_0$. 
\item For two-Higgs-doublet models with large $\tan \beta$, NP enters
  in a similar way with respect to the low $\tan \beta$ case, but this
  time one cannot relate the parameter redefining $S_0(x_t)$ in the
  $B$ sector to the similar term in the $K$ sector. Therefore, two
  different redefinitions must be made for the $B$ and $K$ sectors:
  $S_0(x_t) \to S_0(x_t) + \delta S_0^{B,K}$.
\end{enumerate}

We perform three different analyses, corresponding to the points 1.-3.
above. First, we present the determination of the UUT, which is
independent of NP contributions (Sec.~\ref{subsec:UUT}) in the context
of MFV models. Then we add to the analysis the NP parameter $\delta
S_0$ and constrain it, together with $\rhobar$ and $\etabar$, using
also the neutral meson mixing amplitudes. Finally, we consider the
case $\delta S_0^B \neq \delta S_0^K$ and determine the constraints on
$\rhobar$, $\etabar$ and these NP parameters. We take $\delta
S_0$, $\delta S_0^B$ and $\delta S_0^K$ to be flatly distributed in a
range much larger than the experimentally allowed region.

\subsection{Universal Unitarity Triangle}
\label{subsec:UUT}

\begin{figure}[t]
\begin{center}
\includegraphics[width=0.8\textwidth]{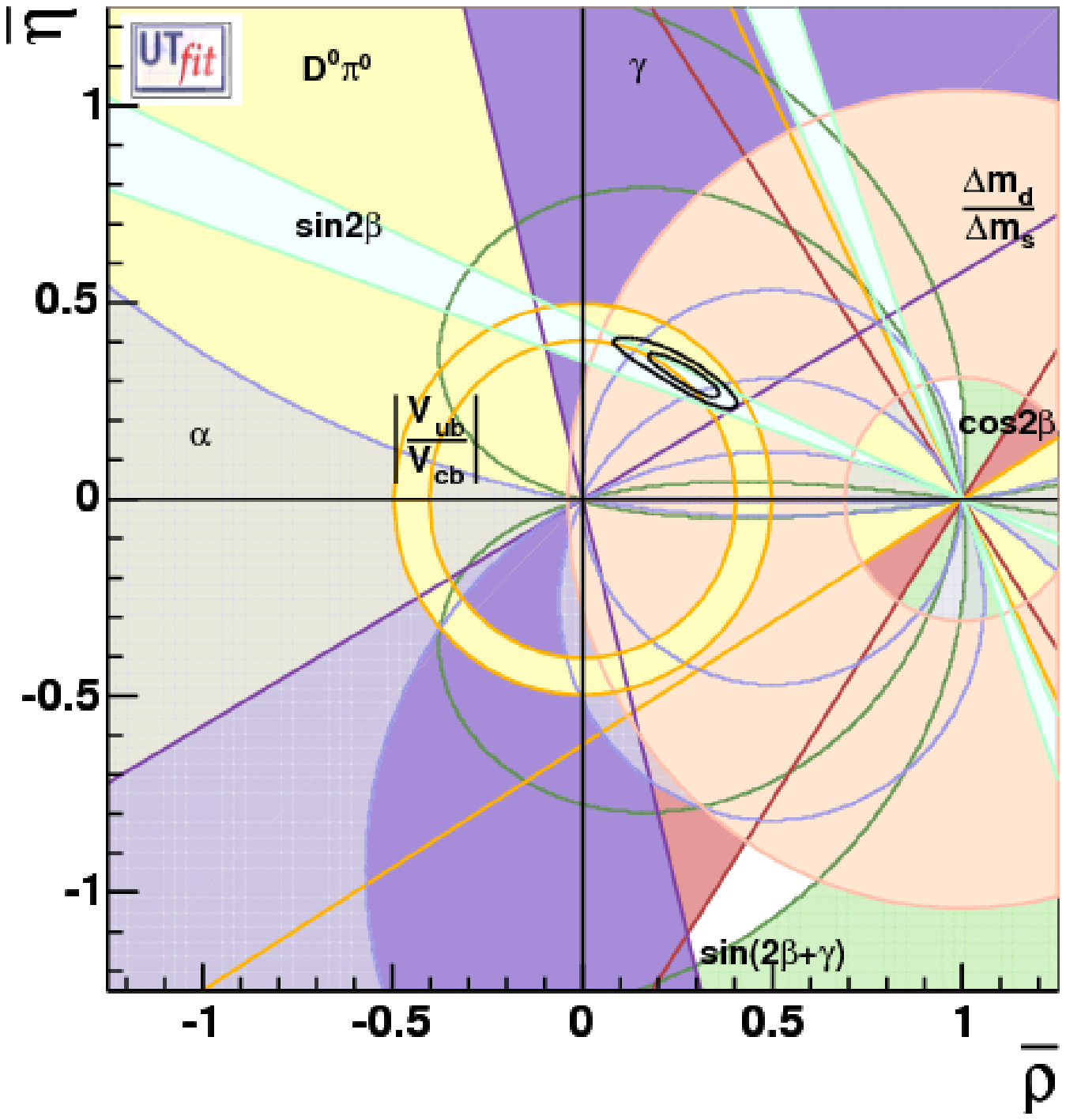} 
\caption{%
\textit{The selected region on $\rhobar$-$\etabar$ plane obtained from 
the determination of the UUT.}}
\label{fig:uut}
\end{center}
\end{figure}

\begin{figure}[hp]
\begin{center}
{\includegraphics[width=0.45\textwidth]{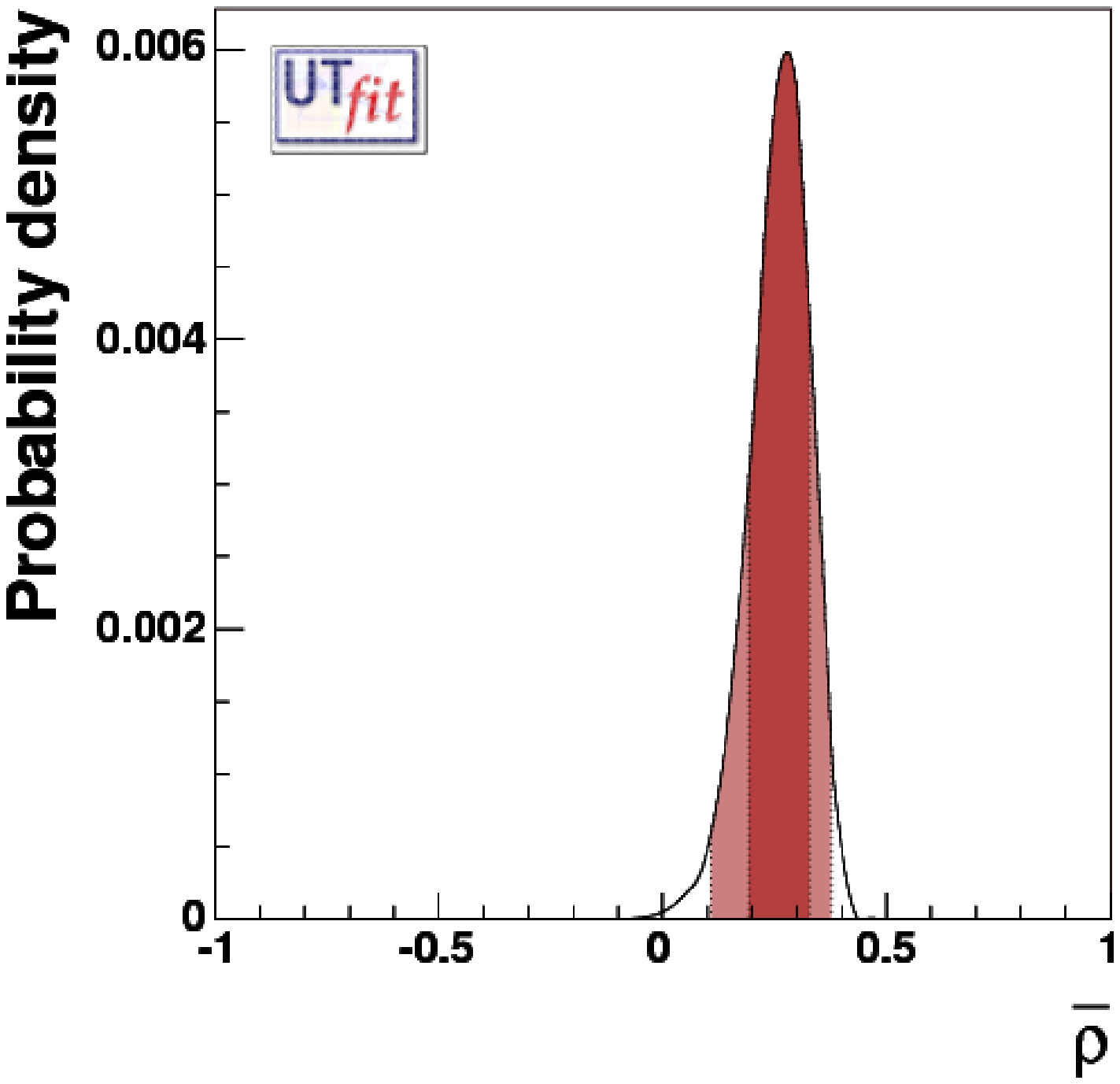}}
{\includegraphics[width=0.45\textwidth]{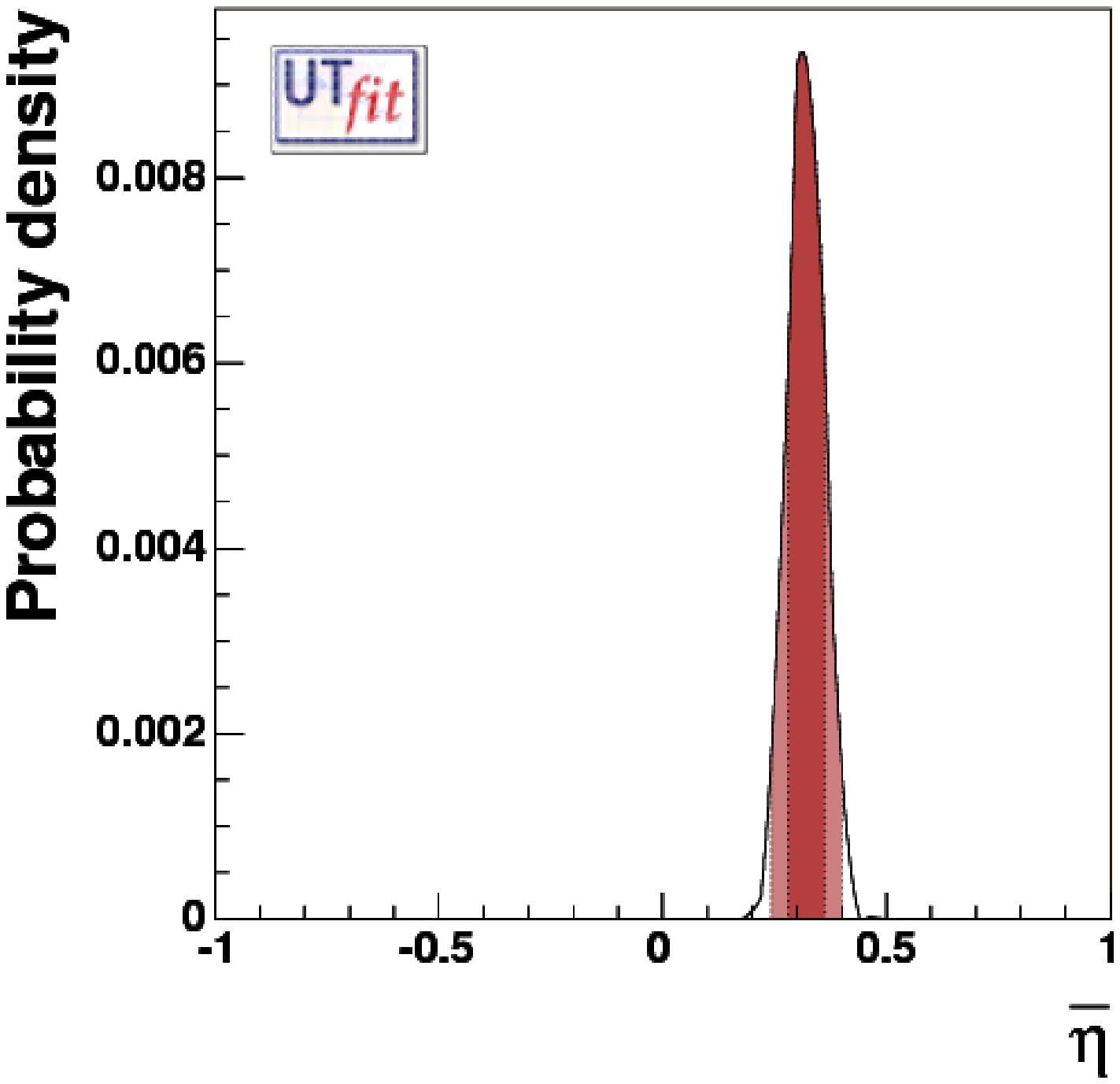}}\\
{\includegraphics[width=0.45\textwidth]{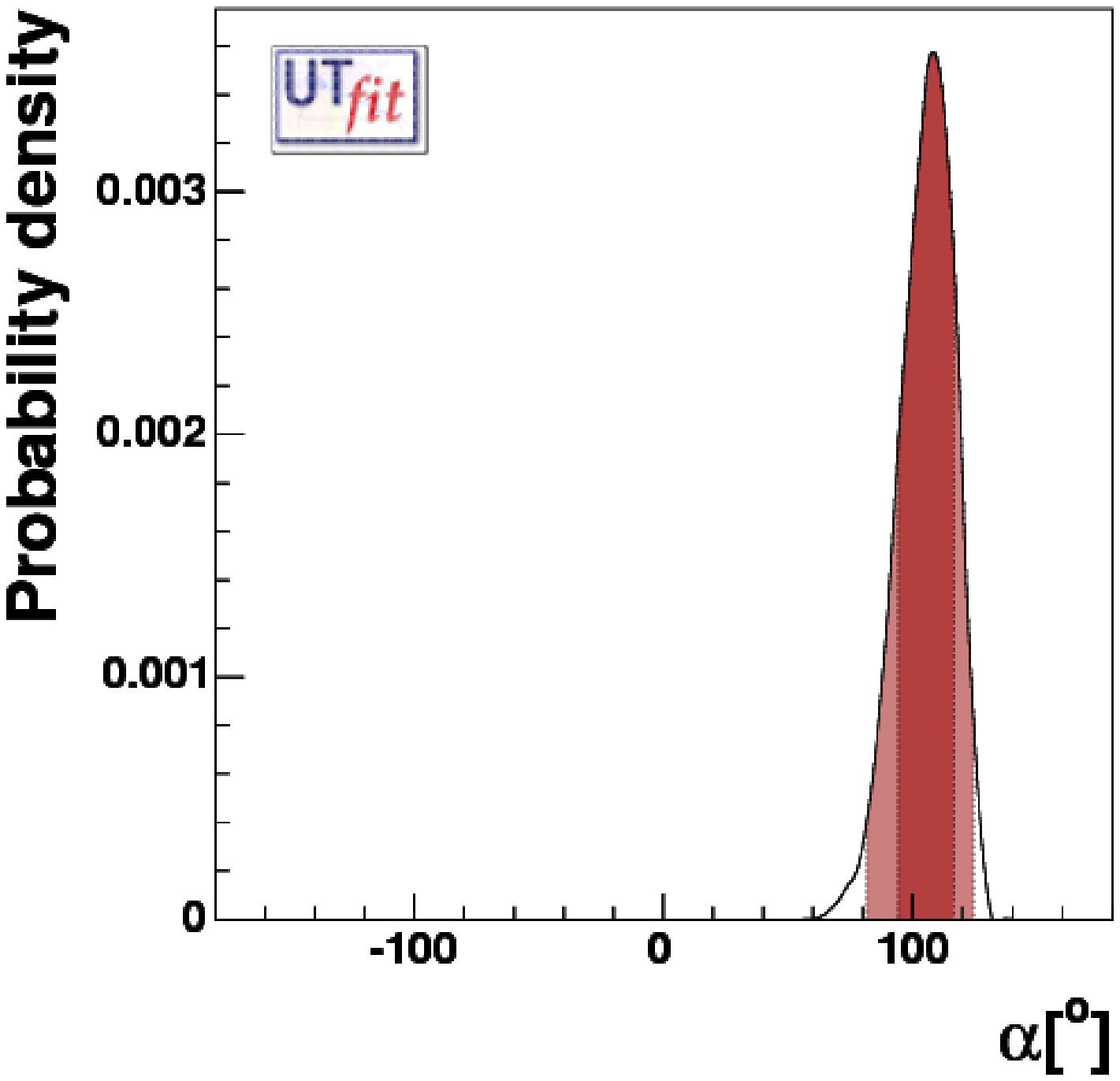}}
{\includegraphics[width=0.45\textwidth]{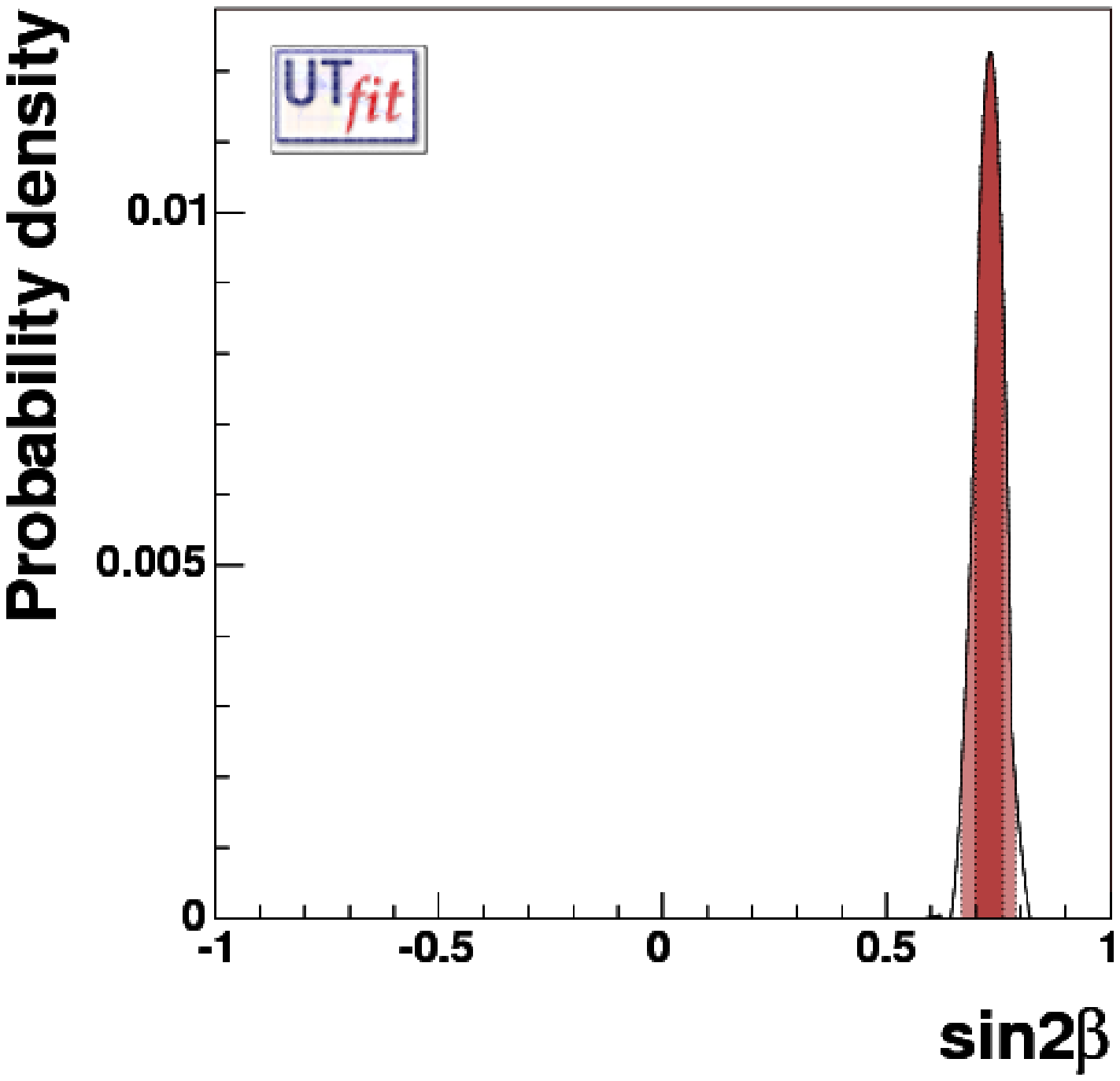}} \\
{\includegraphics[width=0.45\textwidth]{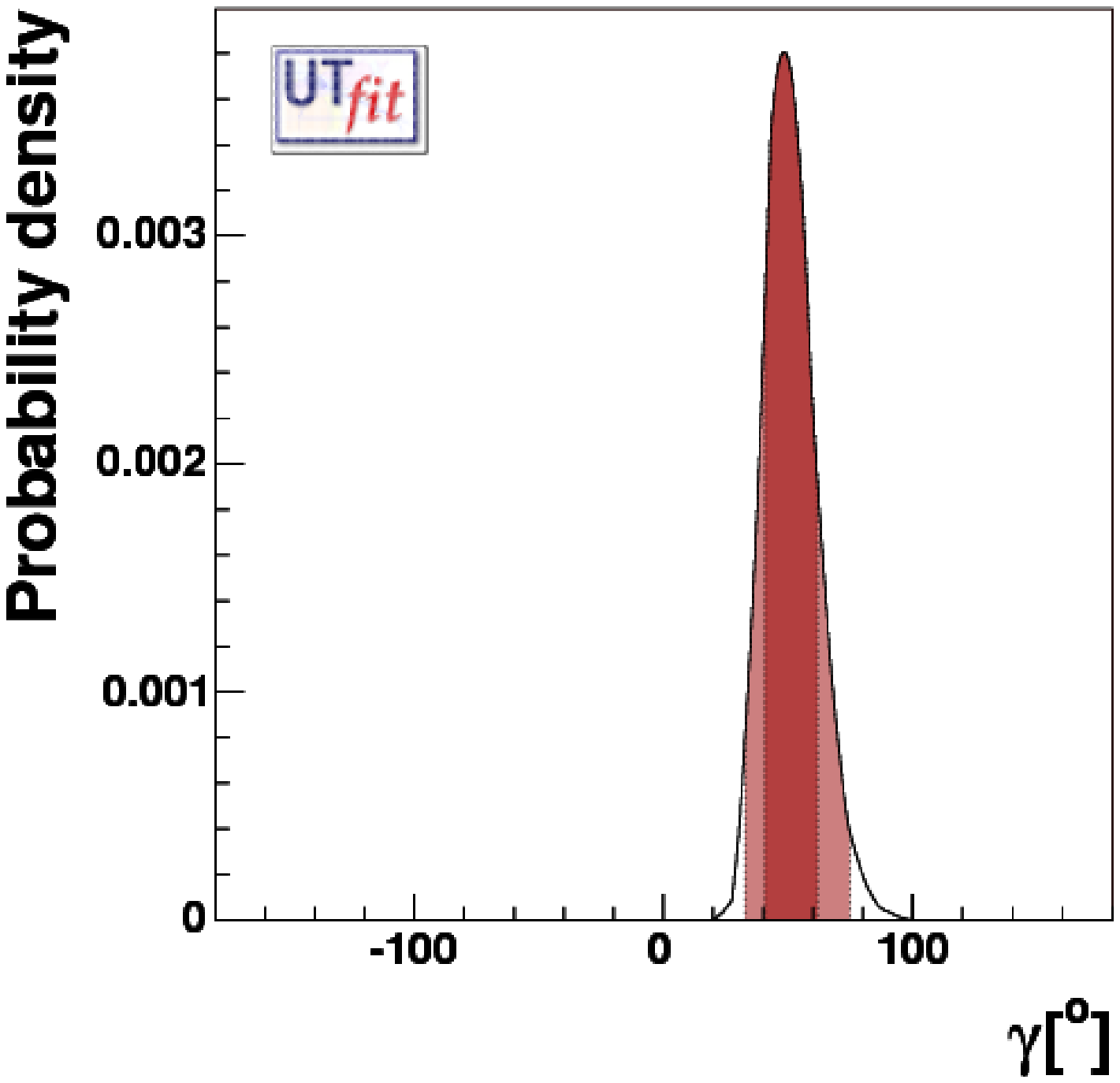}}
{\includegraphics[width=0.45\textwidth]{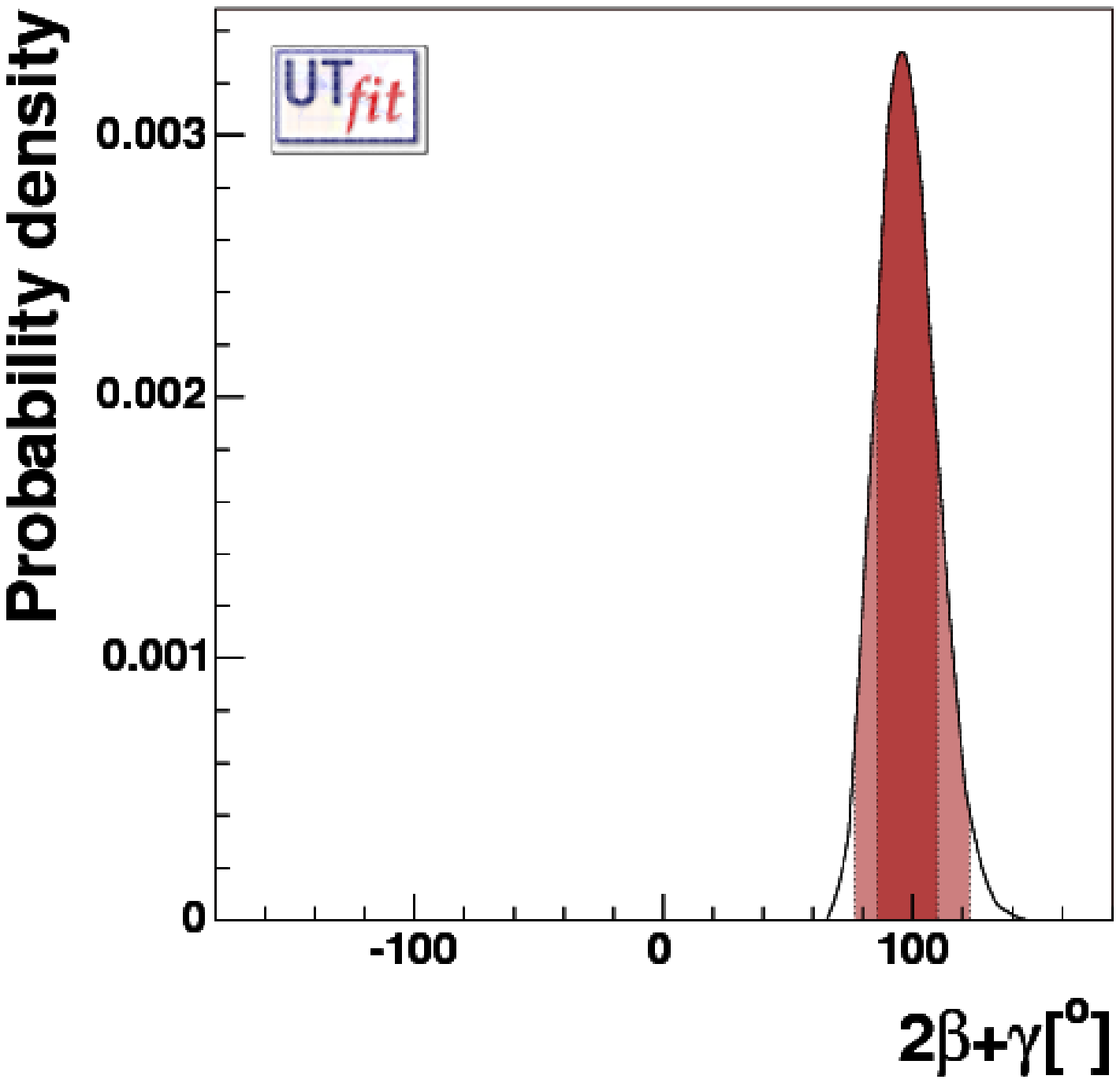}}
\caption{\textit{From top to bottom and from left to right, the p.d.f.'s for
$\rhobar$, $\etabar$,
    $\alpha$, $\snb$, $\gamma$ and $2\beta+\gamma$. The red (darker) and the
    yellow (lighter) zones correspond respectively to 68\% and 95\% of
    the area. These results are obtained from the UUT analysis.}}
\label{fig:uut1d}
\end{center}
\end{figure}

In Fig.~\ref{fig:uut} we show the allowed region in the
$\rhobar-\etabar$ plane for the UUT, and in Fig.~\ref{fig:uut1d} we
plot the p.d.f.'s for several UT quantities. The corresponding values
and ranges are reported in Tab.~\ref{tab:uut}. The most important
differences with respect to the general case are that i) the lower
bound on $\dms$ forbids the solution in the third quadrant, and ii)
the constraint from $\sin 2 \beta$ is now effective, so that we are
left with a region very similar to the SM one (for the reader's
convenience, we also report results of the SM UT analysis in
Tab.~\ref{tab:uut}). The values in Tab.~\ref{tab:uut} are the starting
point for any study of rare decays and CP violation in MFV models. See
Ref.~\cite{BurasMFV} for a recent analysis based on the results of
this work.

\begin{table}[hbt]
\begin{center}
\begin{tabular}{ccccc}
\hline
\multicolumn{5}{c}{Universal Unitarity Triangle analysis}
\\
\hline
& UUT ($68\%$)  & UUT ($95\%$) & SM ($68\%$) & SM ($95\%$)\\
\hline
\multicolumn{5}{c}{UT parameters} \\
\hline
$\rhobar$                   & 0.259 $\pm$ 0.068 & $ [0.107,\,0.376]$ & 0.216$ \pm$ 0.036  & $[0.143,\,0.288]$\\
$\etabar$                   & 0.320 $\pm$ 0.042 & $ [0.241,\,0.399]$ & 0.342$ \pm $0.022  & $[0.300,\,0.385]$\\
$\snb$                      & 0.728 $\pm$ 0.031 & $ [0.668,\,0.778]$ & 0.735$ \pm$ 0.024  & $[0.688,\,0.781]$\\
$\alpha[^{\circ}]$          & 105  $\pm$ 11   & $ [81,\,124]$  &  98.5$ \pm$ 5.7    & $[87.1,\,109.8]$\\
$\gamma[^{\circ}]$          & 51  $\pm$ 10   & $ [33,\,75]$   &  57.6$ \pm$ 5.5    & $[46.8,\,68.7]$\\
$(2\beta+\gamma)[^{\circ}]$ & 98 $\pm$ 12   & $ [77,\,123]$  & 105.3$ \pm$ 8.1    & $[89.6,\,121.4]$\\
Im $\lambda_t$ [$\times 10^{-5}$]& 12.7 $\pm$ 1.7 & $ [9.7,\,15.9]$ & 13.5 $\pm$ 0.8 & $[12.0,\,15.0]$  \\ 
$\Delta m_s$ [ps$^{-1}$] & 20.6 $\pm$ 5.6 & $ [10.6,\,32.6]$  & 20.0 $\pm$ 1.8 & $ [15.5,\,24.2]$  \\
\hline 
\end{tabular}
\caption{\textit{Results of the UUT analysis. For convenience, the SM results from
Ref.~\cite{utfit2005} are also reported.}}
\label{tab:uut} 
\end{center}
\end{table}

\subsection{Constraints on NP contributions in MFV models}
\label{sec:gino}

We now determine the allowed ranges of NP contributions to $|\Delta F|=2$
processes, both in the small and large $\tan \beta$ regime. Furthermore, using
the conventions of Ref.~\cite{gino}, we quantify the scale of NP that can
be probed with the UT analysis.

\begin{figure}[hbt]
\begin{center}
\includegraphics[width=0.45\textwidth]{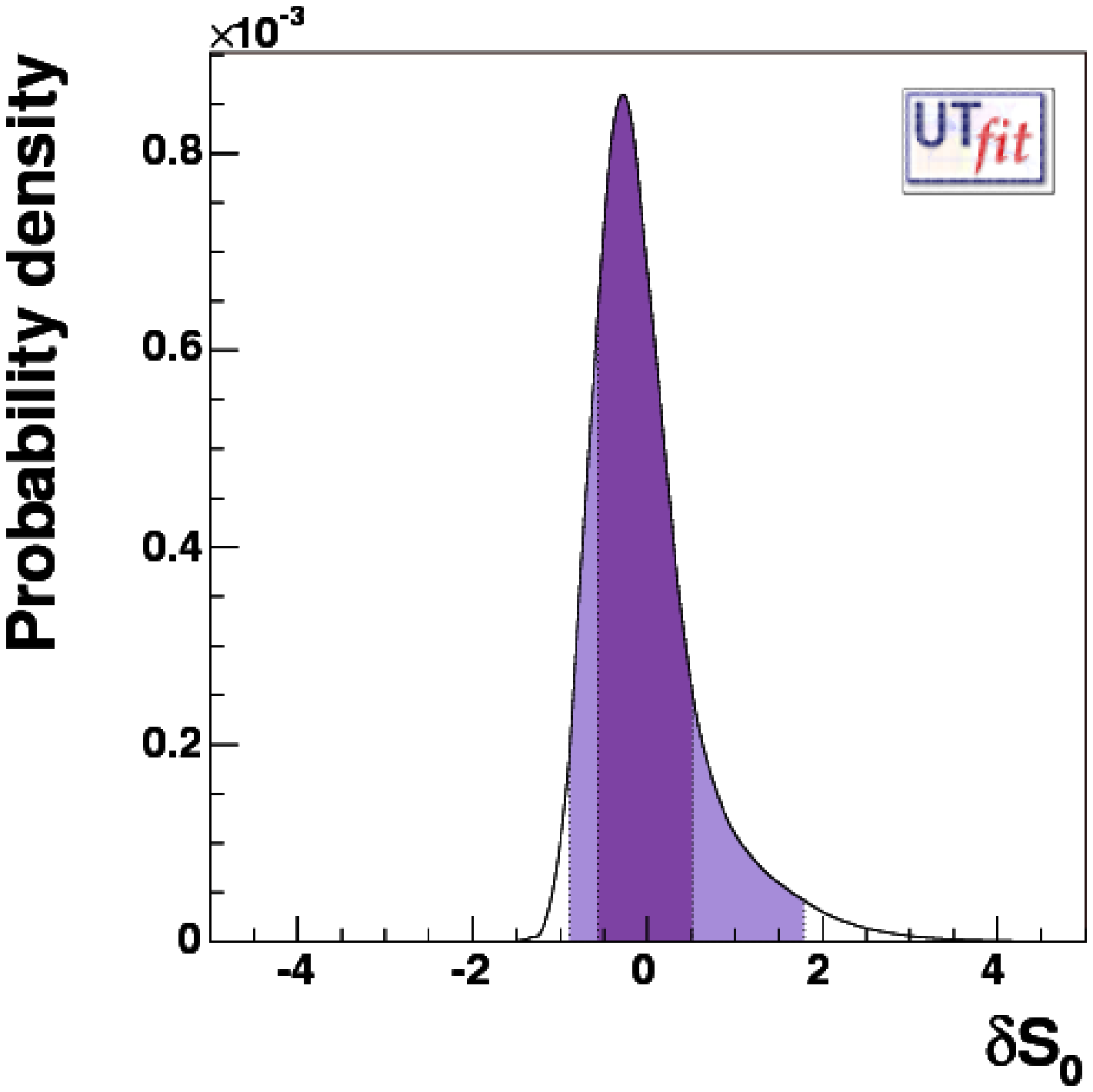}
\includegraphics[width=0.45\textwidth]{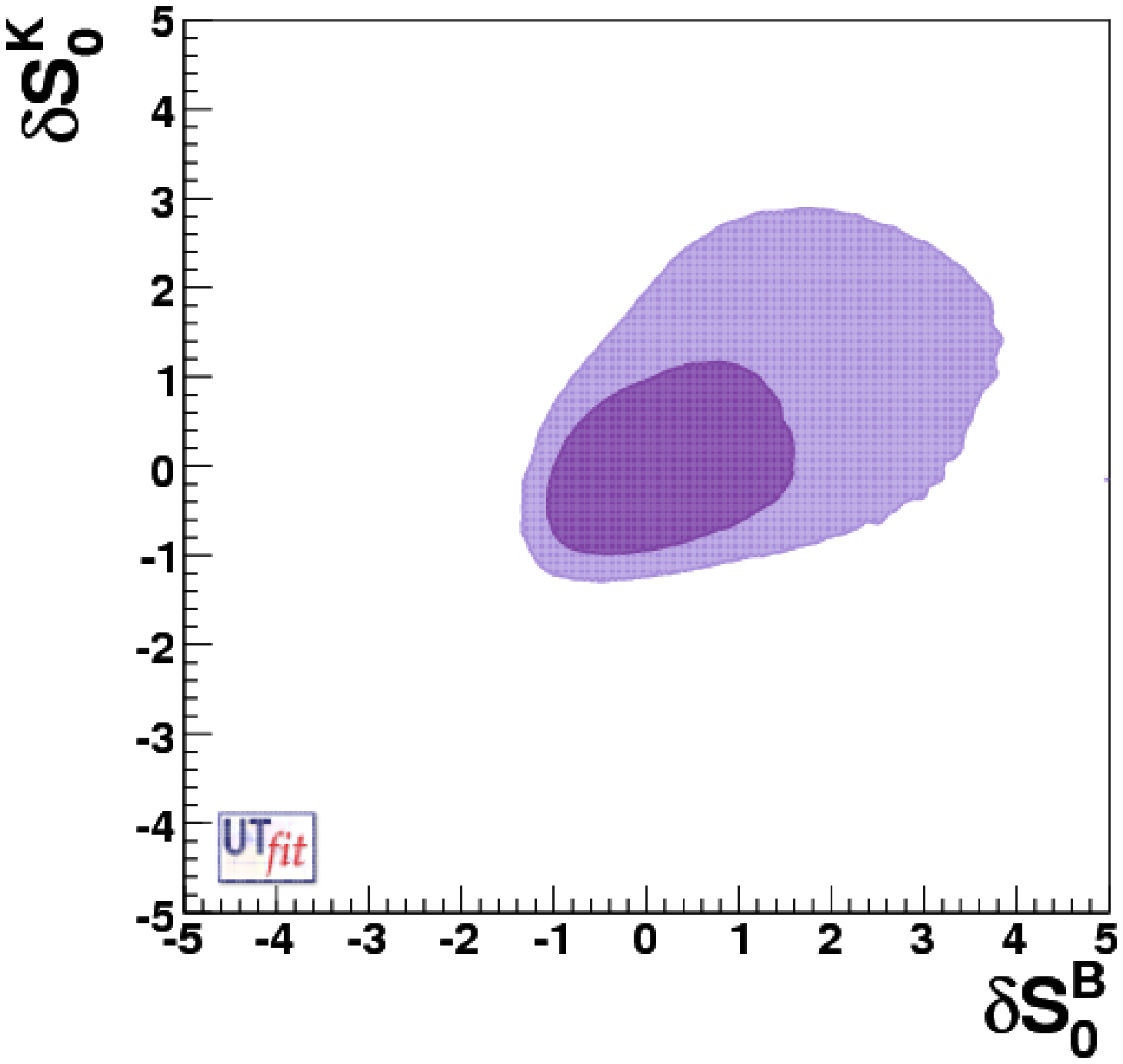}
\includegraphics[width=0.45\textwidth]{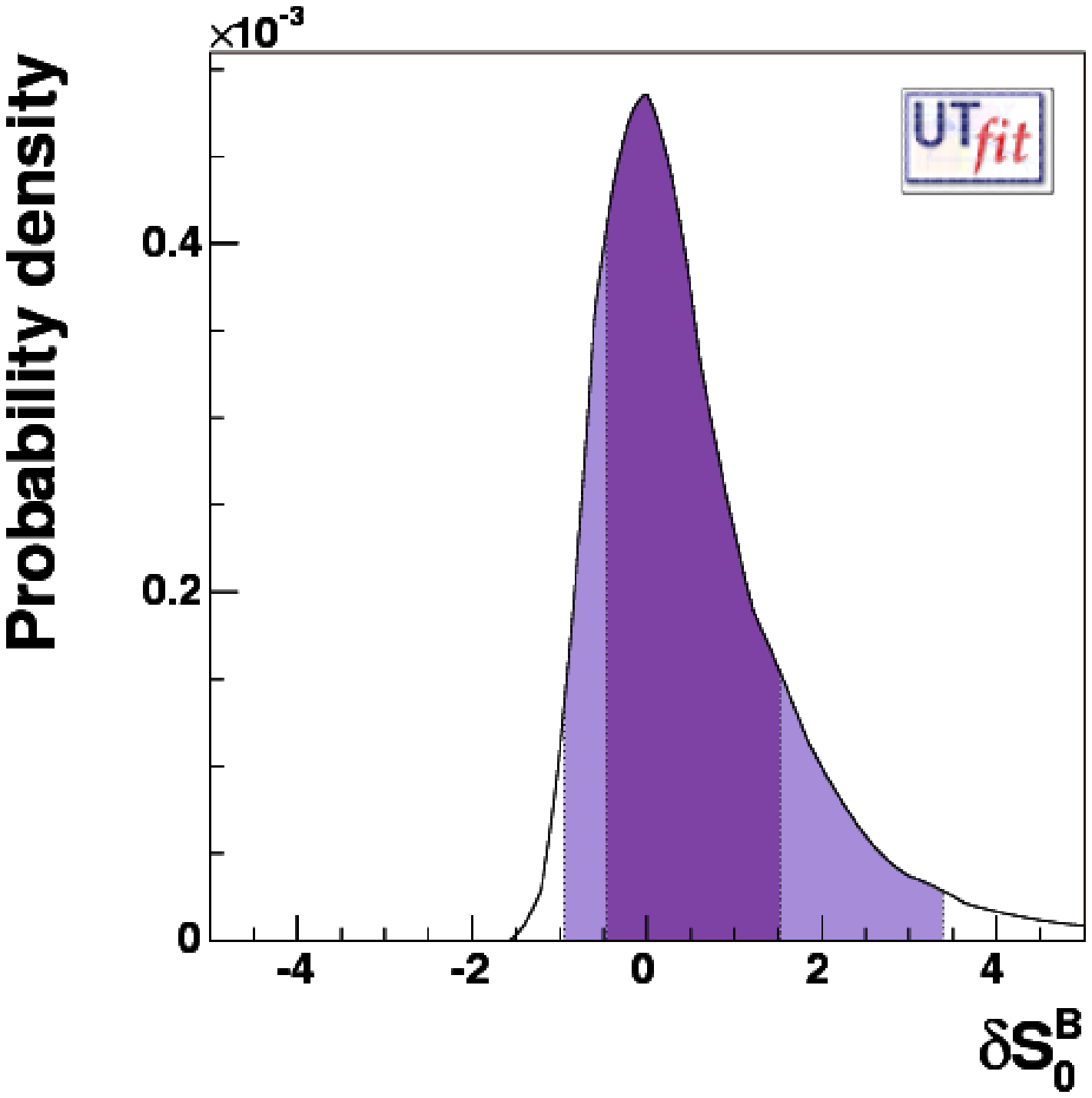}
\includegraphics[width=0.45\textwidth]{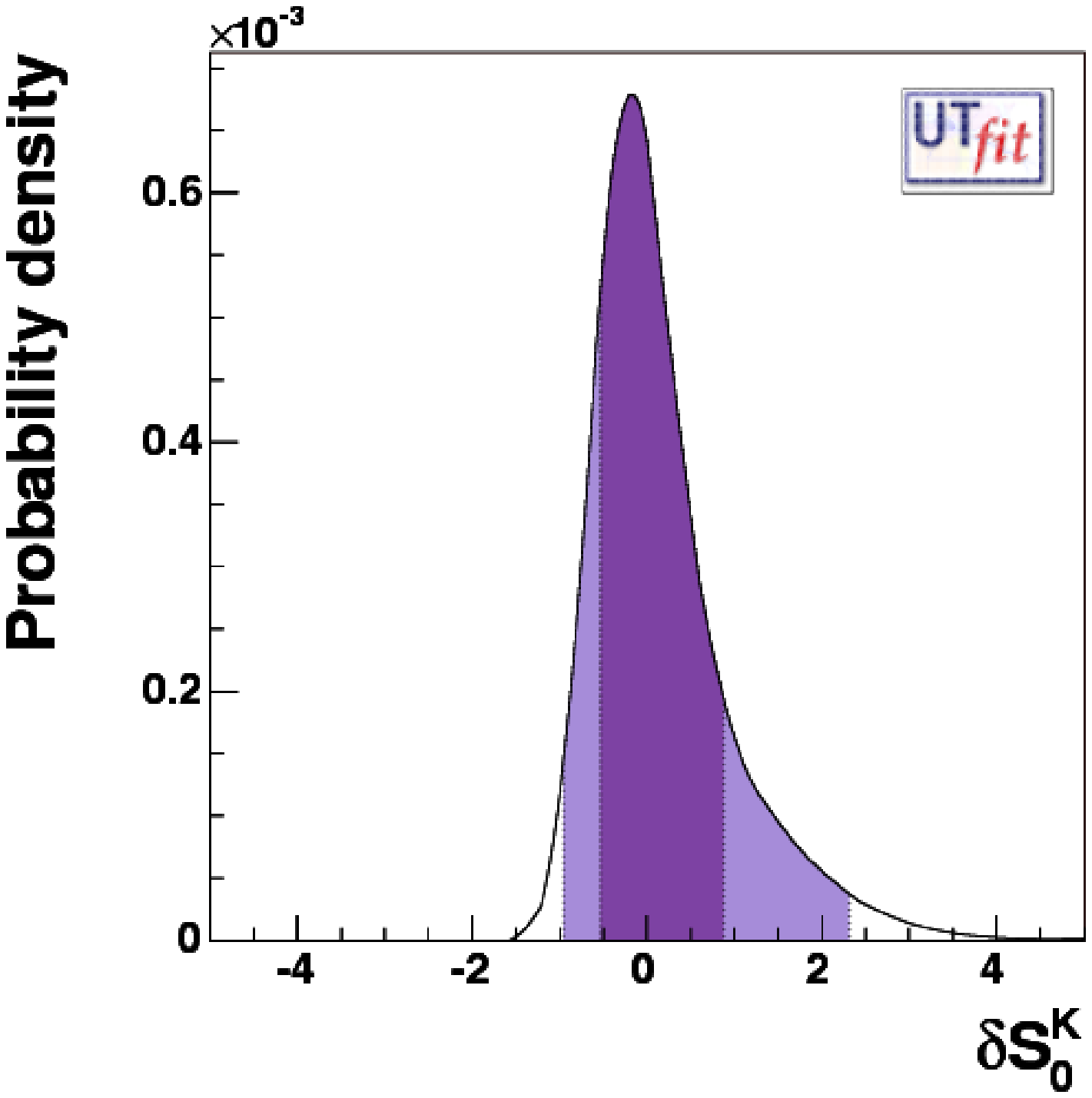}
\caption{\textit{P.d.f. of $\delta S_0$ (top-left), $\delta S_0^K$ vs
    $\delta S_0^B$ (top-right), $\delta S_0^B$ (bottom-left) and
    $\delta S_0^K$ (bottom-right). See the text for details.}}
\label{fig:deltaS0}
\end{center}
\end{figure}

Let us start by considering MFV models with one Higgs doublet or
low/moderate $\tan \beta$. In this case, all NP effects in $|\Delta
F|=2$ transitions are due to the effective Hamiltonian\footnote{Here
  and in the rest of this section we follow the notation of
  Ref.~\cite{gino}.}
\begin{equation}
\label{eq:npscaledef}
\frac{a}{\Lambda^2} \frac{1}{2}\left( \bar Q_L \lambda_{FC}\gamma_\mu
Q_L\right)^2\,,
\end{equation} 
with $(\lambda_{FC})_{ij}=Y_t^2 V^*_{ti}V_{tj}$ for $i\neq j$ and zero
otherwise, $Y_t$ the top quark Yukawa coupling, $\Lambda$ the scale of
NP and $a$ an unknown (but real) Wilson coefficient. The value of $a$
can range from order one for strongly interacting extensions of the SM
to much smaller values for weakly interacting theories and/or symmetry
suppressions analogous to the GIM mechanism in the SM. It is now trivial
to project this onto the SM $|\Delta F|=2$ effective Hamiltonian: it
amounts only to a modification of the top quark contribution to box
diagrams.  Normalizing the NP Wilson coefficient to the SM
\textit{effective electroweak scale}\footnote{i.e. the scale obtained
  by writing the SM contribution to $|\Delta F|=2$ transitions in the
  form of Eq.~(\ref{eq:npscaledef}) with coefficients $a$ of order
  one.} $\Lambda_0=Y_t \sin^2 \theta_W M_W/\alpha \approx 2.4$ TeV, we
obtain
\begin{equation}
S_0(x_t)\to S_0(x_t) + \delta S_0\,,\qquad \delta S_0 =  4 a \left(
\frac{\Lambda_0}{\Lambda}\right)^2.
\end{equation}  
We can therefore determine simultaneously the shape of the UT and
$\delta S_0$ from the standard UT analysis.  Then, choosing as
reference values $a=\pm 1$, we can translate the constraints on
$\delta S_0$ into a lower bound on $\Lambda$. We obtain (see
Fig.~\ref{fig:deltaS0}):
\begin{equation}
\Lambda  >  \left\{
\begin{array}{l} 
3.6 \mathrm{~TeV~@95\%~Prob.~for~positive~} \delta S_0 \\
5.1 \mathrm{~TeV~@95\%~Prob.~for~negative~} \delta S_0 \\
\end{array}\right.
\end{equation}

Also in this case, we can obtain predictions for UT parameters,
together with a constraint on NP contributions (see
Tab.~\ref{tab:mfv}).

\begin{table}[hbt!]
\begin{center}
\begin{tabular}{ccccc}
\hline
\multicolumn{5}{c}{Minimal Flavour Violation analysis} \\
\hline
& \multicolumn{2}{c}{low/moderate $\tan \beta$} & \multicolumn{2}{c}{large $\tan \beta$}  \\
\hline
& $68\%$  & $95\%$ & $68\%$ & $95\%$\\
\hline
$\rhobar$                   & 0.216 $\pm$ 0.058 & $ [0.109,\,0.361]$ 
& 0.231$ \pm$ 0.067  & $[0.112,\,0.375]$\\
$\etabar$                   & 0.351 $\pm$ 0.032 & $ [0.265,\,0.406]$ 
& 0.347$ \pm $0.036  & $[0.254,\,0.404]$\\
$\snb$                      & 0.733 $\pm$ 0.027 & $ [0.679,\,0.781]$ 
& 0.731$ \pm$ 0.027  & $[0.673,\,0.781]$\\
$\alpha[^{\circ}]$          & 98.6  $\pm$ 9.5   & $ [81.6,\,121.7]$  
&  101$ \pm$ 11    & $[82,\,124]$\\
$\gamma[^{\circ}]$          & 57.6  $\pm$ 9.1   & $ [35.7,\,79.1]$   
&  55$ \pm$ 11    & $[34,\,74]$\\
$(2\beta+\gamma)[^{\circ}]$ & 104 $\pm$ 10   & $ [80,\,122]$  
& 102$ \pm$ 12    & $[77,\,121]$\\
Im $\lambda_t$ [$\times 10^{-5}$] & 13.6 $\pm$ 1.4 & $ [10.1,\,16.0]$ 
& 13.4 $\pm$ 1.9 & $[9.7,\,16.3]$  \\ 
$\Delta m_s$ [ps$^{-1}$] & 19.5 $\pm$ 2.6 & $ [15.0,\,31.7]$  
& 22.6 $\pm$ 5.4 & $ [15.5,\,35.1]$  \\
\hline 
\end{tabular}
\caption{\textit{Results for UT parameters from the MFV generalized
    analysis.}}
\label{tab:mfv} 
\end{center}
\end{table}

In the case of large $\tan \beta$, the situation changes since the
bottom Yukawa coupling is not negligible anymore, and it can
distinguish transitions involving $b$ quarks from those
involving only light quarks.  This spoils the correlation of $|\Delta
B|=2$ with $|\Delta S|=2$ amplitudes, so that two uncorrelated
parameters $\delta S_0^B$ and $\delta S_0^K$ are required in this case,
to take into account NP contributions to $B_{d,s}$--$\bar B_{d,s}$ and
$K$--$\bar K$ mixing. In a global fit, made by using all the available
inputs, $\Delta m_d$ and $\Delta m_d/\Delta m_s$ determine the value
of $\delta S_0^B$, $\epsilonk$ fixes $\delta S_0^K$, while $\rhobar$
and $\etabar$ are given by the combination of all the other
constraints.

Performing this analysis, we bound the UT parameters as given in
Tab.~\ref{tab:mfv}, limiting the NP scale to be:
\begin{eqnarray}
\Lambda & > & \left\{
\begin{array}{l} 
2.6 \mathrm{~TeV~@95\%~Prob.~for~positive~} \delta S_0^B \\
4.9 \mathrm{~TeV~@95\%~Prob.~for~negative~} \delta S_0^B \\
\end{array}\right.
\mathrm{from~} B_{d,s}-\bar B_{d,s} \mathrm{~mixing}  \nonumber \\
\Lambda & > & \left\{
\begin{array}{l} 
3.2 \mathrm{~TeV~@95\%~Prob.~for~positive~} \delta S_0^K \\
4.9 \mathrm{~TeV~@95\%~Prob.~for~negative~} \delta S_0^K \\
\end{array}\right.
\mathrm{from~} K-\bar K \mathrm{~mixing}  \nonumber \\
\end{eqnarray}
The output distributions for $\delta S_0^B$ and $\delta S_0^K$ are
given in Fig.~\ref{fig:deltaS0}.

It is instructive to observe the two-dimensional plot of $\delta
S_0^B$ vs. $\delta S_0^K$ in Fig.~\ref{fig:deltaS0}: within models
with only one Higgs doublet or with small $\tan \beta$, the two
$\delta$'s are bound to lie on the line $\delta
S_0^B=\delta S_0^K$. The correlation coefficient $R$ provides a
measure of this relation. We find $R=0.52$ giving no compelling
indication on the value of $\tan \beta$.

\begin{figure}[!hbt]
\begin{center}
\includegraphics[width=0.8\textwidth]{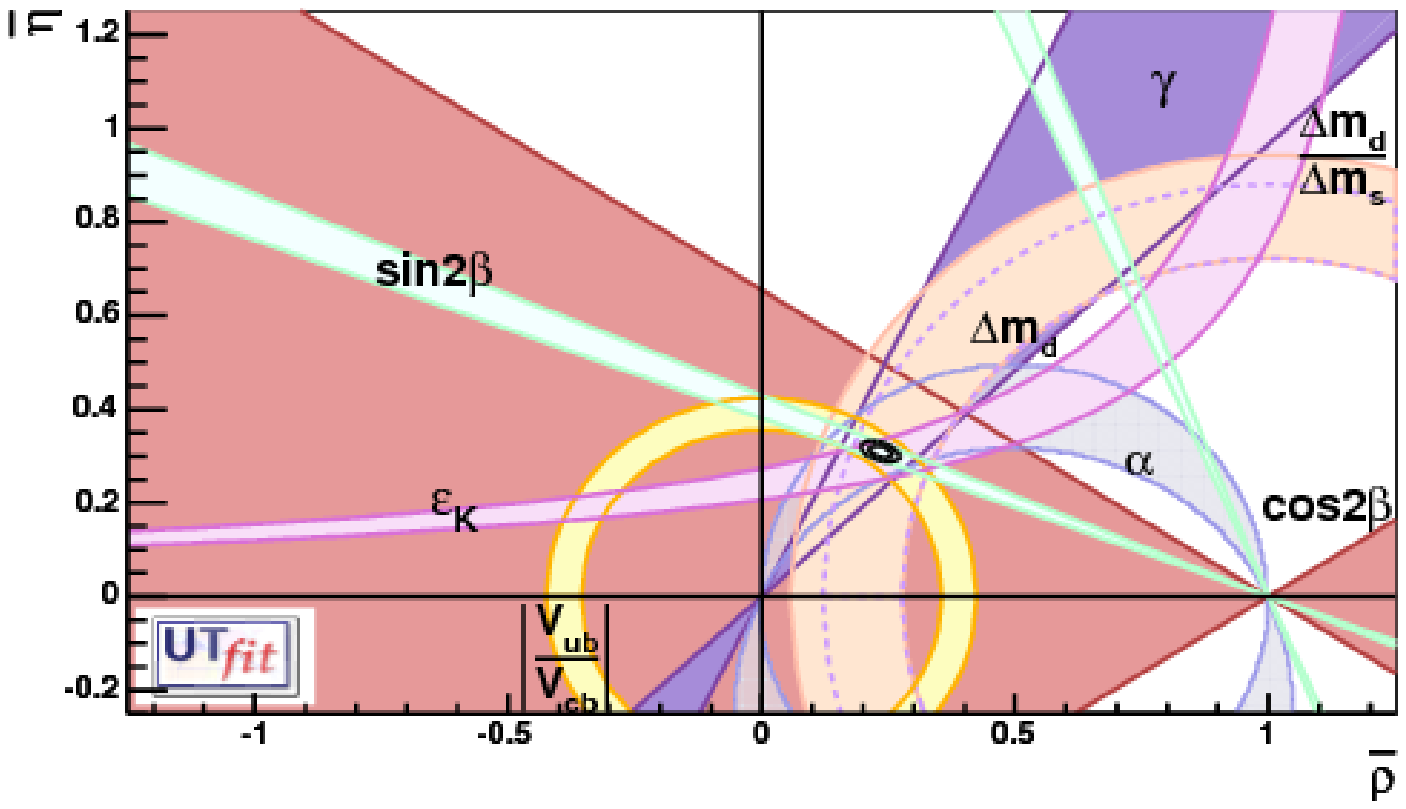} 
\caption{%
\it The selected region on $\rhobar$-$\etabar$ plane obtained from 
the Standard Model UT analysis in the ``year 2010'' scenario.}
\label{fig:SM2010}
\end{center}
\end{figure}
\section{Model Independent constraints on New Physics in the 
  $\mathbf{\vert \Delta F \vert}$=2 sector in year 2010}
\label{sec:ckm2010}

\begin{table}[hbt]
\begin{center}
\begin{tabular}{@{}llll}
\hline
       Observable                       &   projected value $\&$ error              \\ \hline
      $\snb$                            &   0.695 $\pm$ 0.010 (1.4$\%$)              \\ 
      $\alpha[^{\circ}]$                           &   104 $\pm$ 5                                  \\ 
      $\gamma[^{\circ}]$ (DK)                       &   54 $\pm$ 5                                    \\ \hline  
      $\BK$                             &   0.930 $\pm$ 0.047 (5$\%$)               \\ 
      $\fbssqbs$ [MeV]                   &   0.276 $\pm$ 0.014 (5$\%$)               \\ 
      $\xi$                             &   1.200 $\pm$ 0.037 (3$\%$)                  \\ \hline  
      $\vcb$-(incl+excl) ($10^{-3}$)    &   41.7 $\pm$ 0.4 (0.9$\%$)                    \\ 
      $\vub$-(incl+excl) ($10^{-4}$)    &   36.4 $\pm$ 1.6 (4.2$\%$)                    \\ \hline  
      $\Delta m_d$ [ps$^{-1}$]          &   0.503 $\pm$ 0.003 (0.6$\%$)                    \\ 
      $m_t$ [GeV]                        &   171 $\pm$ 3.0                             \\ 
      $\lambda$                  &   0.2240 $\pm$ 0.0008                                 \\  \hline
      $\Delta m_s$ [ps$^{-1}$]          &   20.5 $\pm$ 0.3                     \\ 
      $\sin 2\chi_s\, (J/\psi \phi)$       &    0.031 $\pm$ 0.045                          \\
      $(\gamma-2\chi_s)[^{\circ}]\,(D_s K)$ &   51 $\pm$ 10                                    \\ \hline  
\end{tabular}    
\caption{\textit{Projected values and errors in year 2010 for the most
  relevant quantities entering in the UT analysis. 
  The central values are chosen such that the constraints are
  perfectly compatible within the SM. 
}} 
\label{tab:ckm2010} 
\end{center}
\end{table}
\begin{figure}[ht!]
\begin{center}
\includegraphics[width=0.32\textwidth]{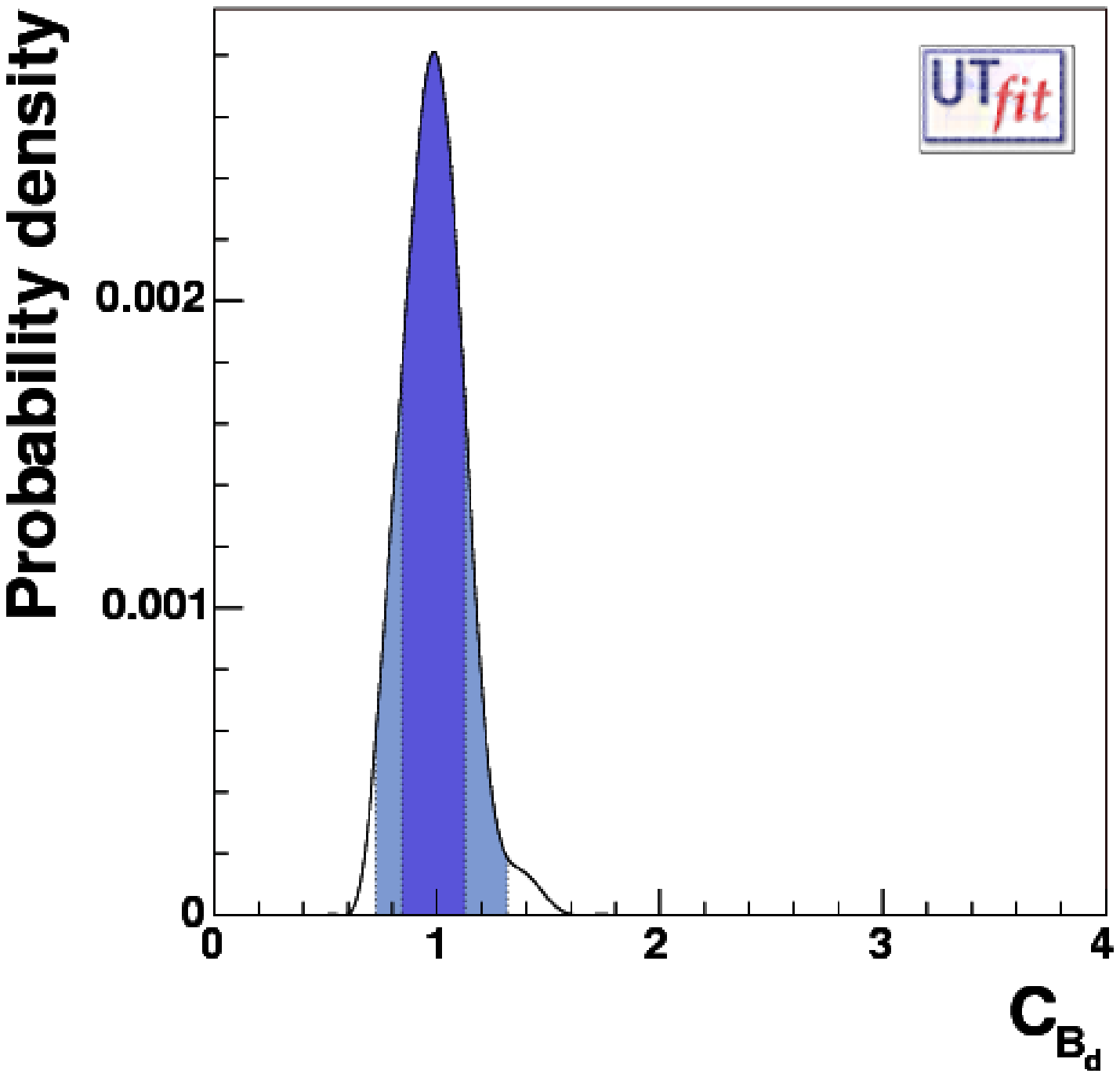} 
\includegraphics[width=0.32\textwidth]{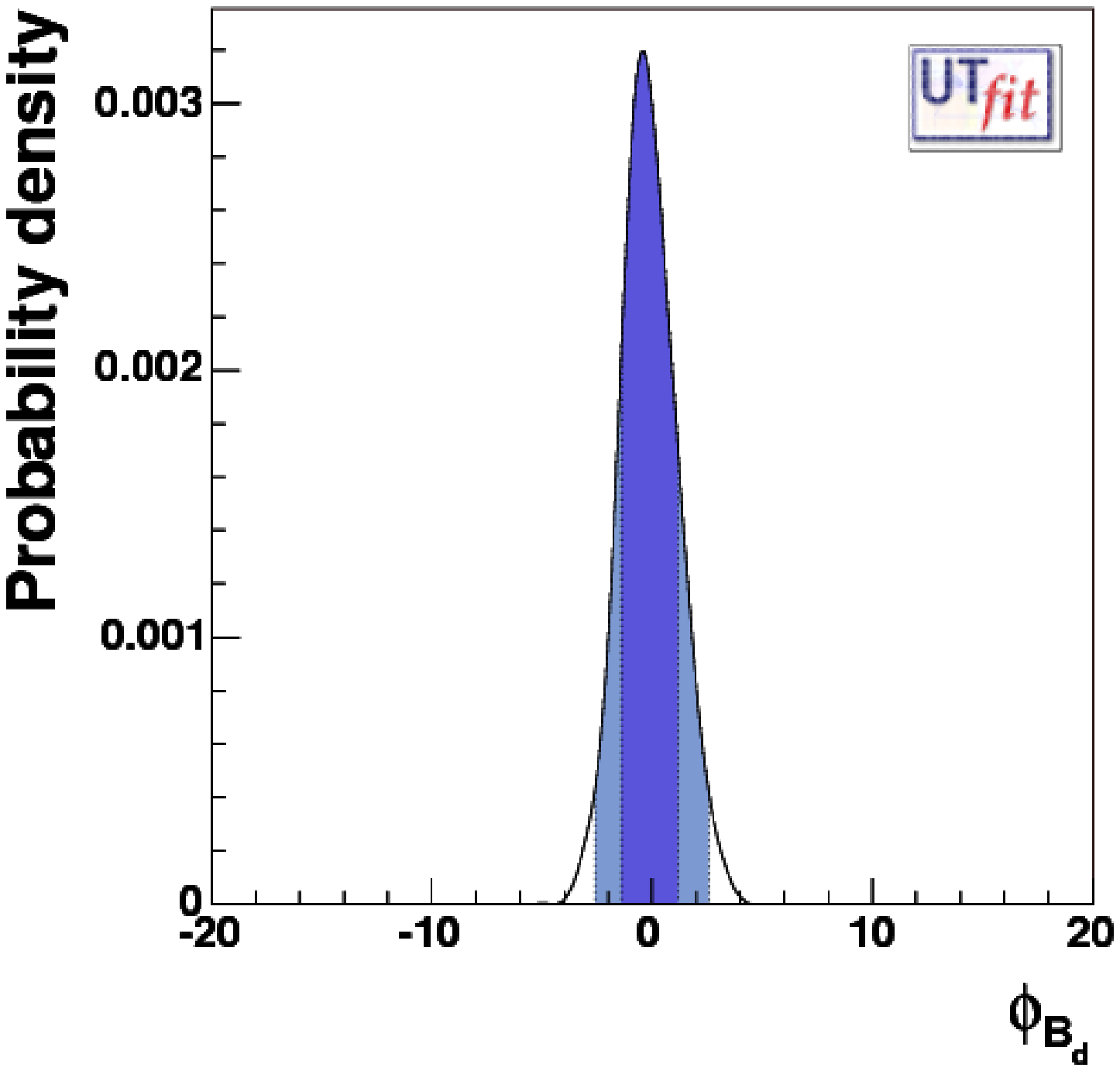}
\includegraphics[width=0.32\textwidth]{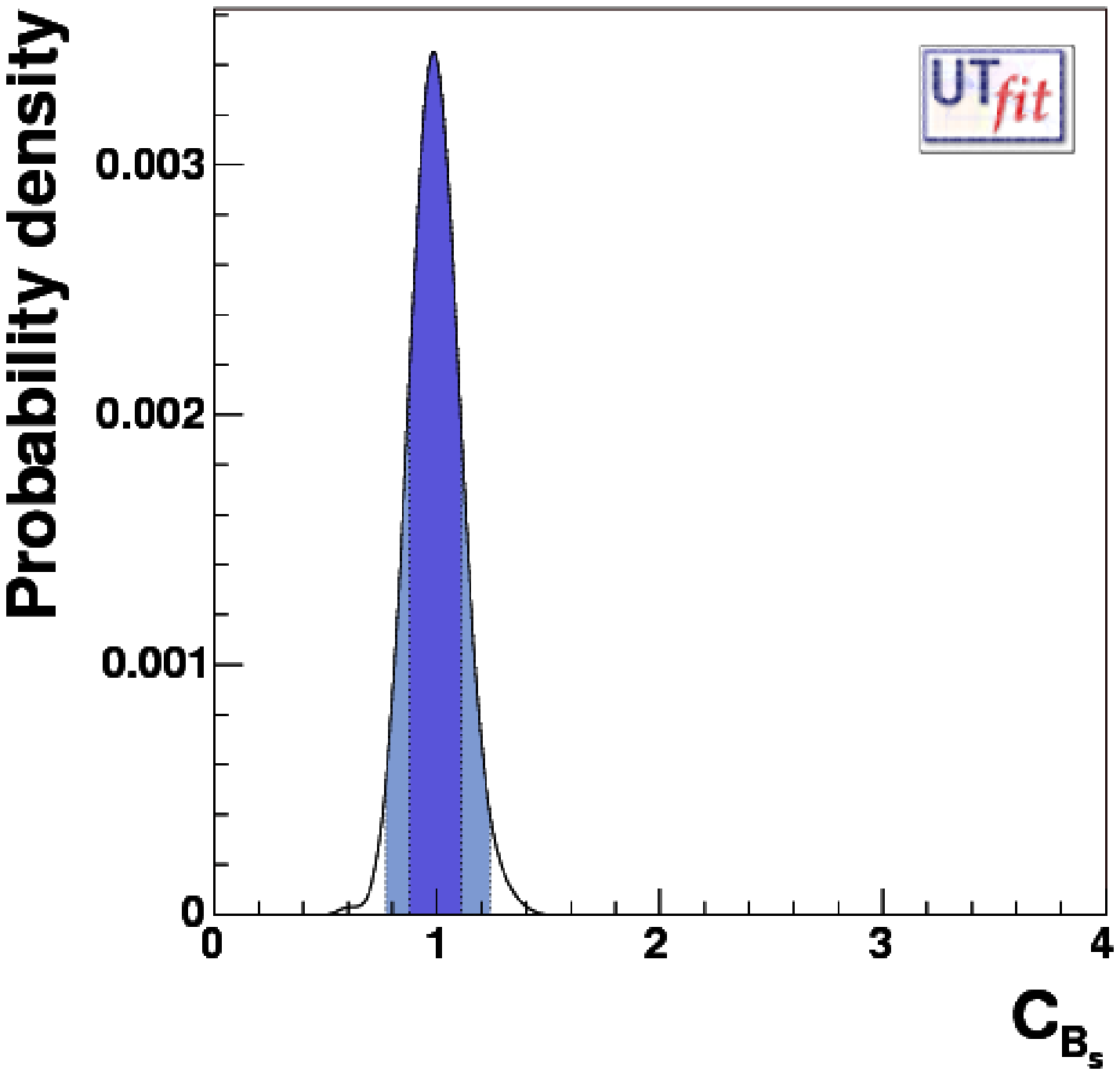}
\includegraphics[width=0.32\textwidth]{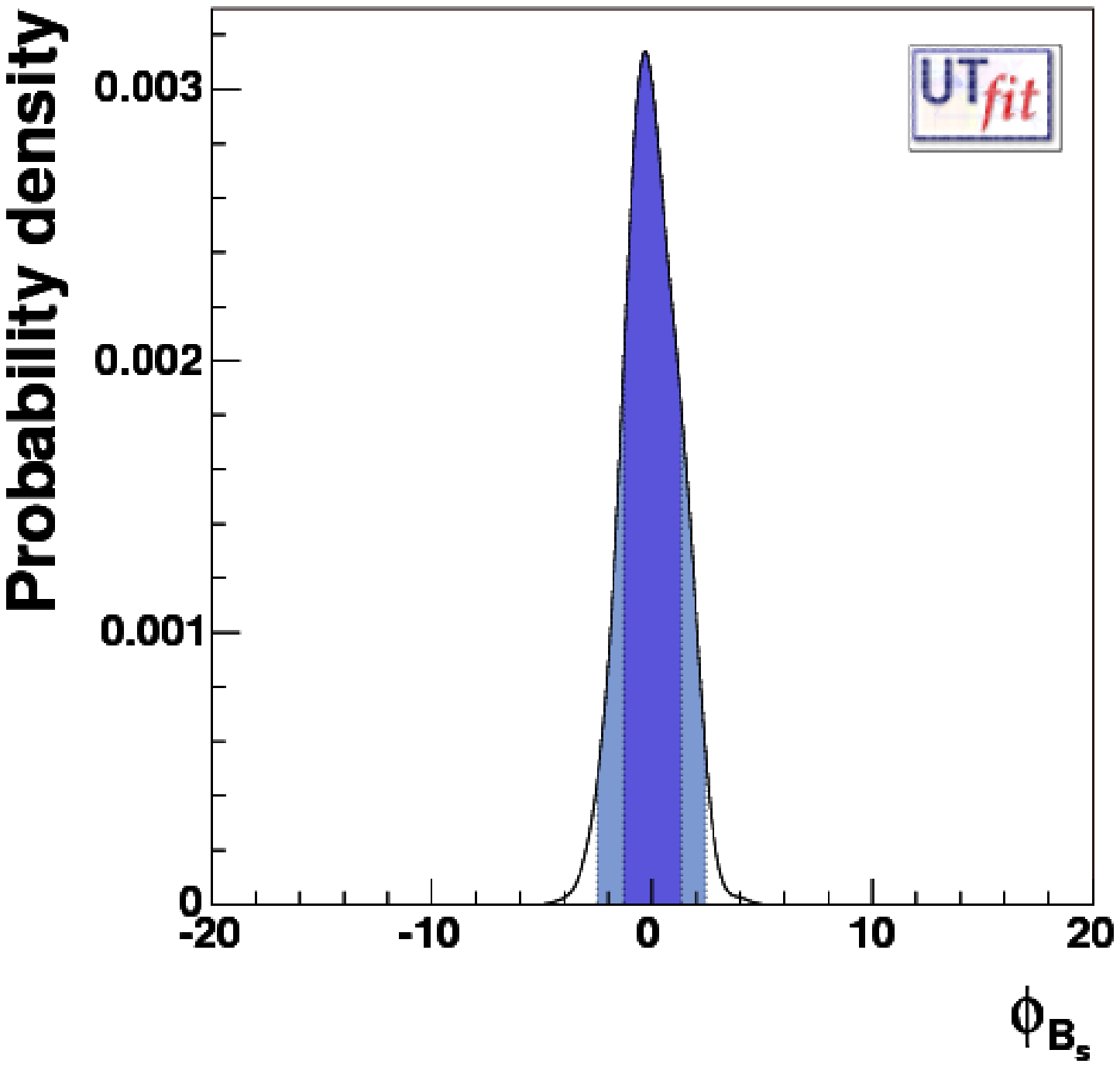}
\includegraphics[width=0.32\textwidth]{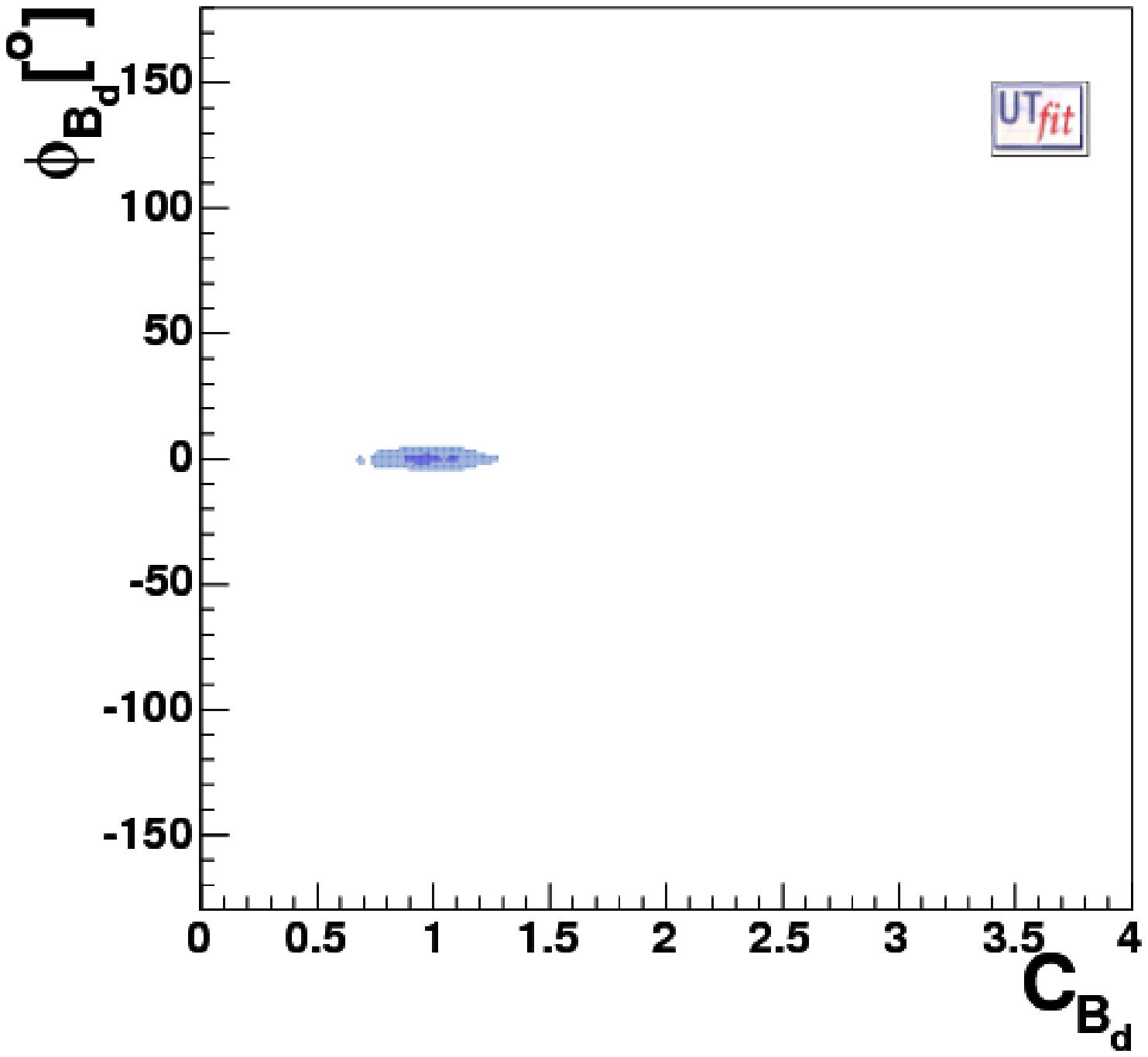}
\includegraphics[width=0.32\textwidth]{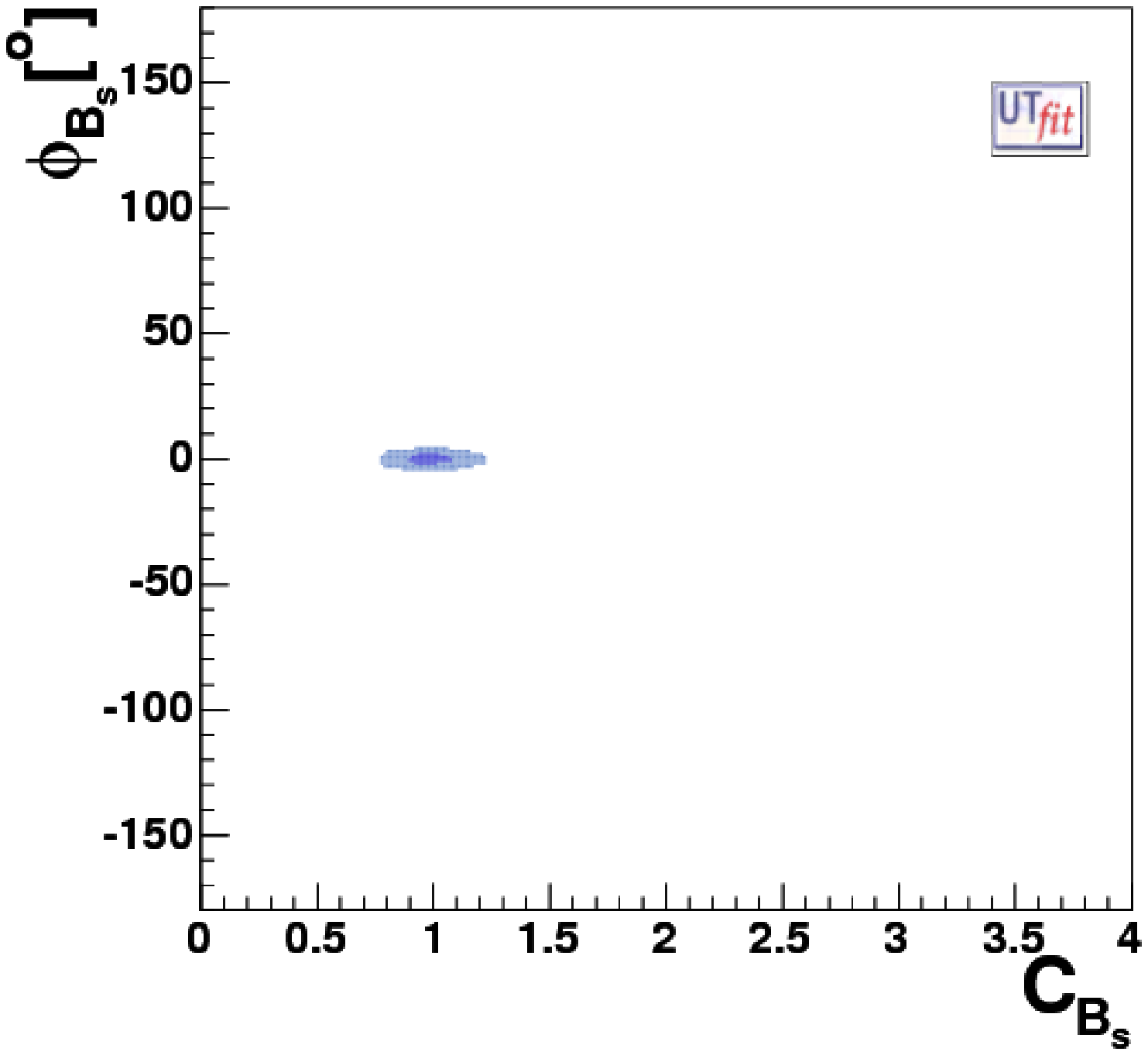}
\includegraphics[width=0.32\textwidth]{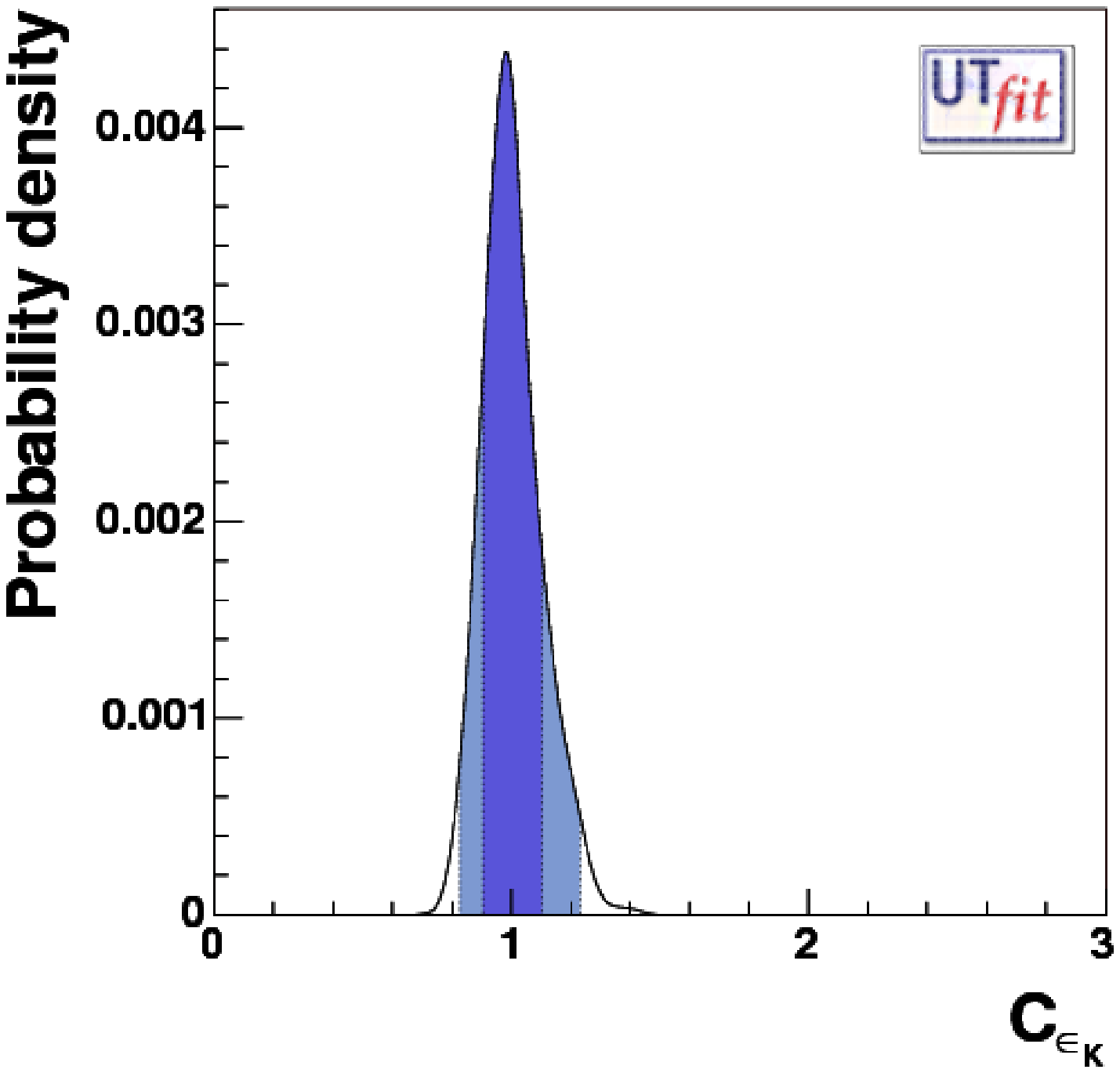}
\caption{%
  \textit{Projected situation in year 2010. p.d.f.  distributions for
    NP parameters: from top to bottom and from left to right,
    distributions of $C_{B_d}$ and $\phi_{B_d}$, distributions of
    $C_{B_s}$ and $\phi_{B_s}$, 2D distributions of
    $\phi_{B_d}~vs.~C_{B_d}$ and $\phi_{B_s}~vs.~C_{B_s}$, and
    distribution of $C_{\epsilon_K}$.  Dark (light) area corresponds
    to the $68\%$ ($95\%$) probability region.}\label{fig:ckm2010-NP}}
\end{center}
\end{figure}

We present an exercise on the knowledge of the UT parameters within the
generalized NP analysis in a possible scenario in year 2010. At this
date, the $B$ factories will have completed their data analysis and the LHCb
experiment will have started running.

For this exercise we have assumed a total integrated luminosity for
the $B$ factories of 2 ab$^{-1}$ and two years of data taking at LHCb,
with an integrated luminosity of 4 fb$^{-1}$.
At that time the lattice community will have produced the final
numbers from the Tera-Flops machines.
The 2010 projected values and errors for the
quantities which are most relevant in UT analysis are given in
Tab.~\ref{tab:ckm2010}.  Lattice parameters are taken from
\cite{workshop}, while the extrapolation of the errors on the
experimental measurements is taken from
\cite{tdrl,superb,colle,lhcbtdr}.  In addition to the improvements of
existing measurements, we have added new measurements in the $B_s$
sector from LHCb. In particular the determination of
\begin{center}
\begin{tabular}{rcll}
  $\Delta m_s^\mathrm{exp}$ & = & $C_{B_s} \Delta m_s^\mathrm{SM}$ & $\mathrm{from}~B_s \to D_s \pi\,,$ \\
  $\sin 2 \chi_s^\mathrm{exp}$  &=& $\sin (2 \chi_s^\mathrm{SM} + 2\phi_{B_s})$ & $\mathrm{from}~B_s \to J/\psi \phi\,,$ \\
  $(\gamma -2 \chi_s)^\mathrm{exp}$ &=& $\gamma^\mathrm{SM} - 2 \chi_s^\mathrm{SM} - 2\phi_{B_s}$ & $\mathrm{from}~B_s \to D_s K\,.$
  \nonumber
\end{tabular}
\end{center}
\noindent
The central values for the different observables have been generated
in the SM starting from an arbitrarily chosen value of $\rhobar$ and
$\etabar$, so that they are all compatible with each other and the
result of the fit is fully ``SM like''.  The reason for this procedure
is to investigate whether, in the ``worst case'' scenario of perfect
confirmation of the SM, one can asymptotically reduce the errors on
the NP related quantities introduced in the previous sections, and
translate the derived constraint into a lower limit on the
energy scale for NP particles. We start from a picture of what the UT
analysis should look like in 2010. In Fig.~\ref{fig:SM2010} we show
the selected region in the $\rhobar$--$\etabar$ plane, while in the
third column of Tab.~\ref{tab:SMMFV2010} we quote the uncertainties on
the various quantities from the UT analysis in the Standard Model.
This should be taken as a reference for the approaches beyond the
Standard Model that follow.
\begin{table}[ht]
\begin{center}
\begin{tabular}{|c|c|c|c|}
\hline
\multicolumn{4}{|c|}{UT analysis in 2010}\\
\hline
Observable & Input error & \multicolumn{2}{c|}{Output error}\\
\hline
                           &        & SM UT & UUT \\
\cline{3-4}
      $\rhobar$                       &         -         &   0.015   &   0.021         \\
      $\etabar$                       &         -         &   0.007   &   0.010        \\
\hline
      $\snb$                          &       0.010       &   0.009   &   0.009       \\ 
      $\alpha[^{\circ}]$              &         5         &   2.1     &   3.5        \\ 
      $\gamma[^{\circ}]$              &         5         &   2.0     &   3.2        \\ 
      $2\beta+\gamma[^{\circ}]$       &         -         &   2.3     &   3.4        \\ 
\hline
\end{tabular}    
\caption {\textit{Projected uncertainties in year 2010 for the SM UT and the UUT analysis, using all the 
inputs of Tab.~\ref{tab:ckm2010}.}}
\label{tab:SMMFV2010} 
\end{center}
\end{table}

Moving from the SM analysis to the model independent approach of
Sec.~\ref{sec:generalC}, we expect in the future a sizable improvement
of the knowledge of the NP parameters:
\begin{eqnarray}
C_{B_d} = 0.98 \pm 0.14 \qquad && \qquad \phi_{B_d} = (-0.1 \pm 1.3)^\circ  \nonumber \\
C_{B_s} = 0.99 \pm 0.12 \qquad && \qquad \phi_{B_s} = ( 0.0 \pm 1.3)^\circ  \nonumber \\
C_{\epsilon_K} = 1.00 \pm 0.10 \qquad &&
\end{eqnarray}
as shown in Fig.~\ref{fig:ckm2010-NP}.

\begin{figure}[tb!]
\begin{center}
\includegraphics[width=0.8\textwidth]{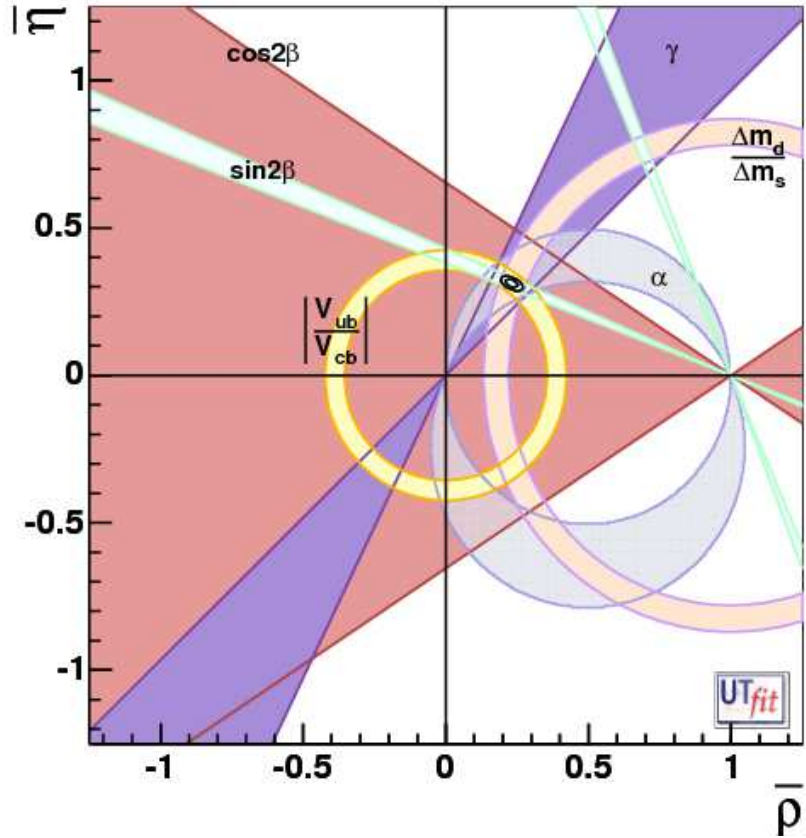} 
\caption{%
\textit{The selected region on $\rhobar$-$\etabar$ plane obtained from 
the UUT analysis in the ``year 2010'' scenario.}}
\label{fig:MFV2010}
\end{center}
\end{figure}

\begin{figure}[tb!]
\begin{center}
\includegraphics[width=0.32\textwidth]{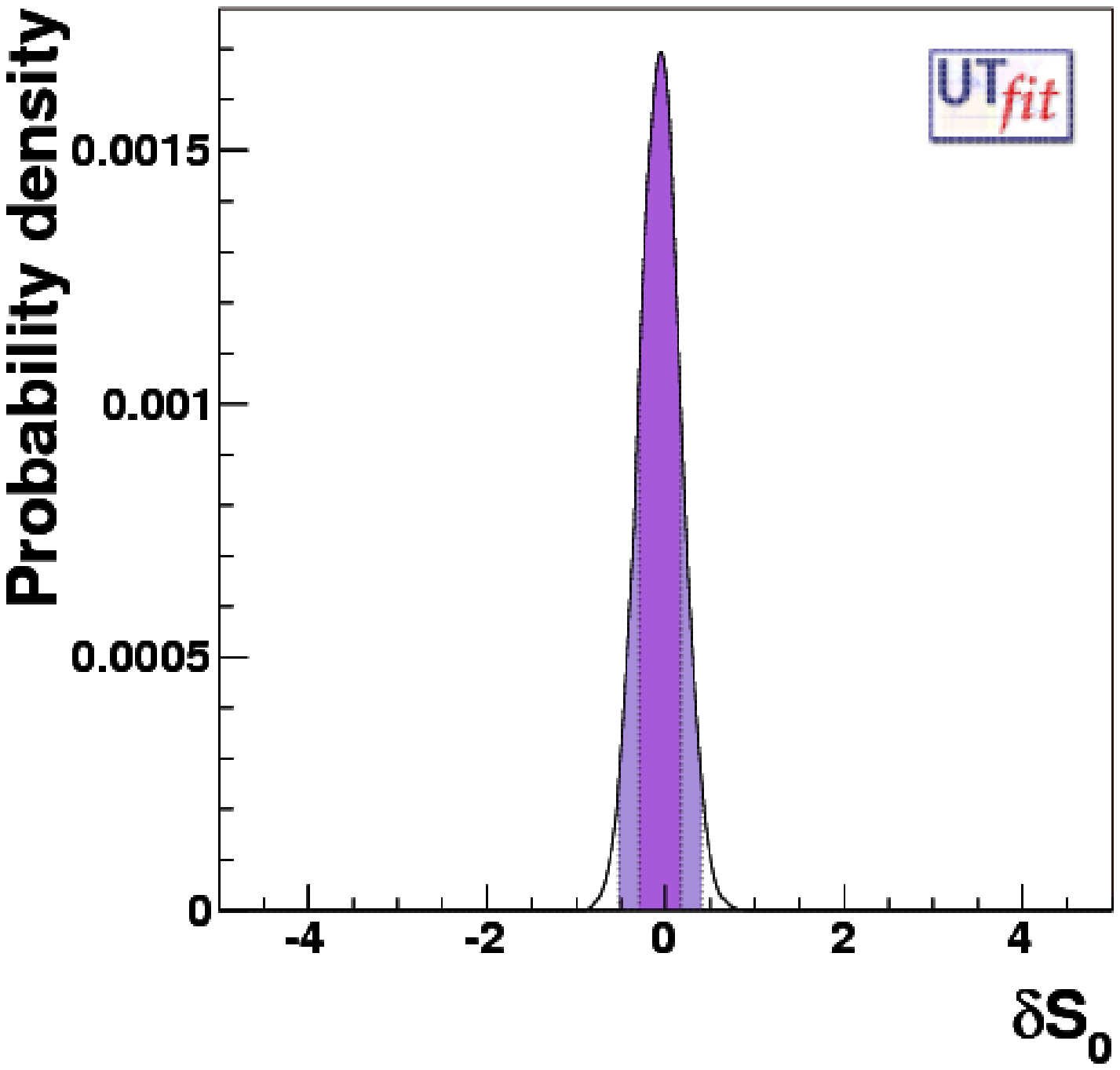}
\includegraphics[width=0.32\textwidth]{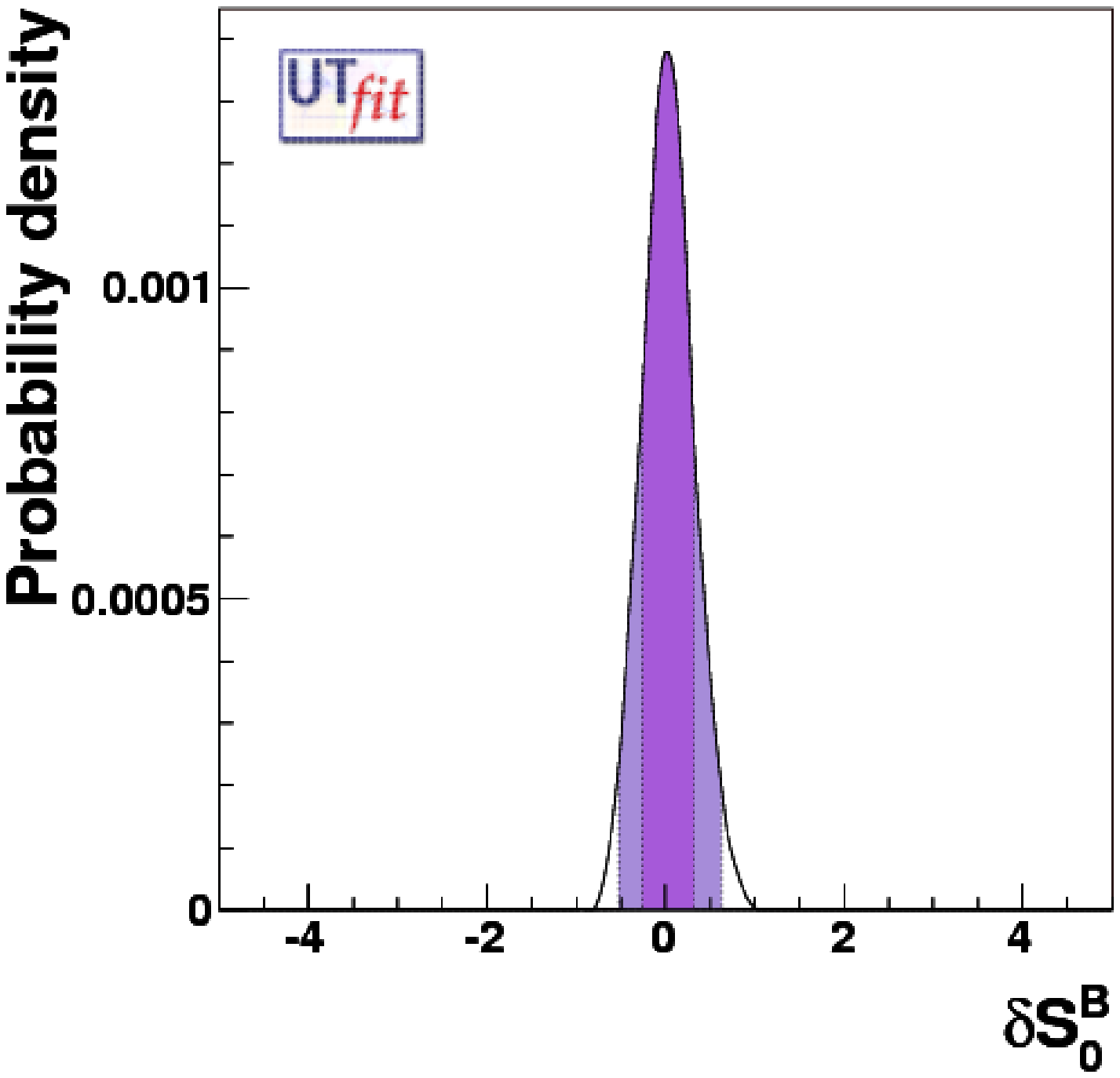}
\includegraphics[width=0.32\textwidth]{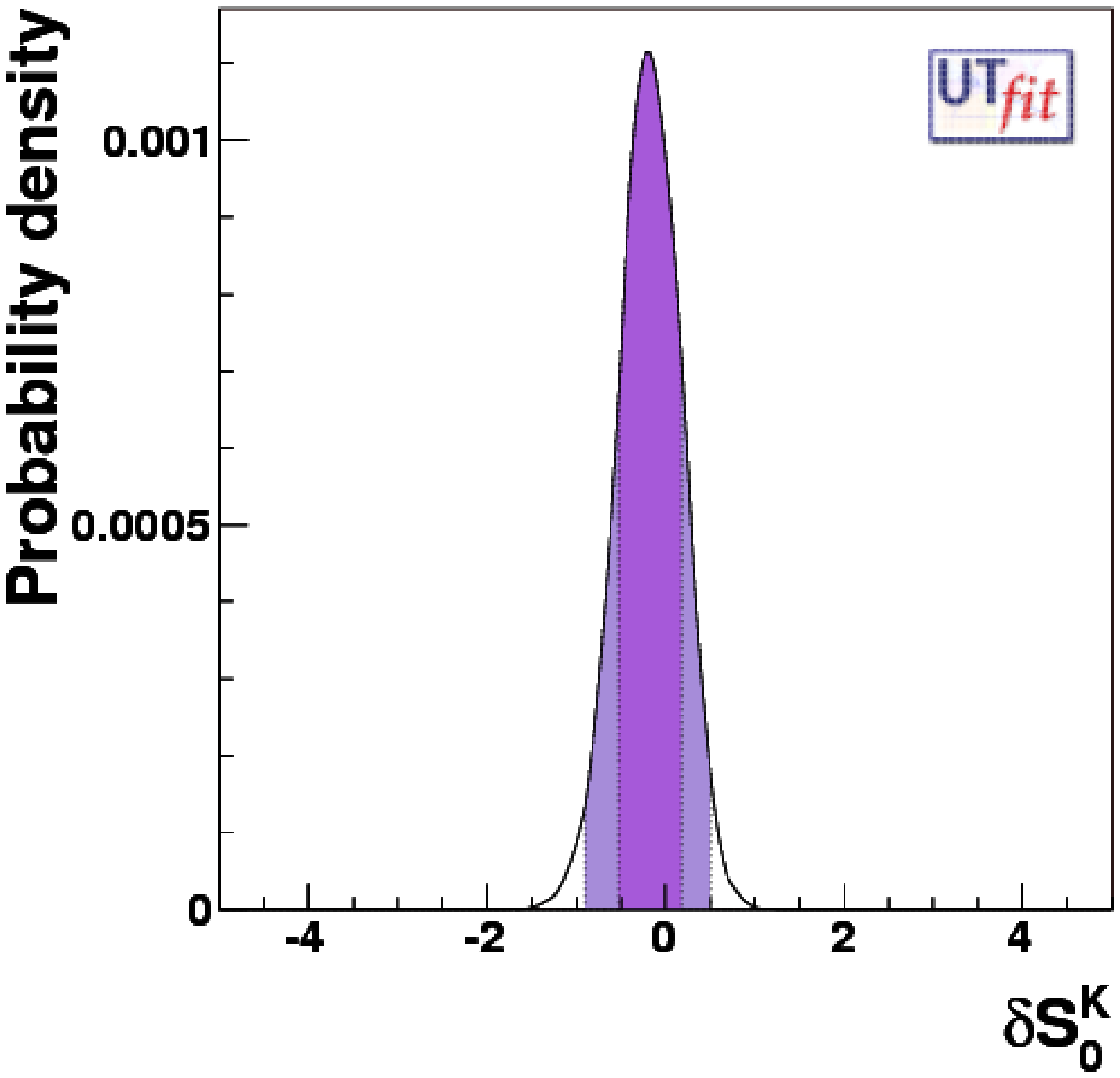}
\caption{\textit{P.d.f. of $\delta S_0$ from MFV fit in the 2010 scenario,
  in the case of small (left) and large values of $\tan \beta$, from
  $B_{d,s}$--$\bar B_{d,s}$ (center) and $K$--$\bar K$ (right)
  mixing.}}
\label{fig:deltaS02010}
\end{center}
\end{figure}

In the same future scenario, one can repeat the MFV analysis, both determining
UT parameters using the UUT approach and adding the information from (NP
sensitive) mixing quantities to bound the NP scale. The expected errors on the
relevant UT quantities are summarized in the forth and fifth columns of Tab.~\ref{tab:SMMFV2010}.

If no evidence of violation of the Standard Model will emerge from
$B$ physics in the era of direct NP search at LHC, this generalized
approach will replace the present UT analysis as the default
procedure. So, it is important to remark the fact that the
generalization of the analysis costs an increasing of about $10\%$ of
the errors, which is not a huge price to pay if compared to the gain
in terms of the larger physics scenario.\footnote{Of course, the
  output error on $\Delta m_s$ is also affected by the absence of
  $\Delta m_d$ in the fit.}  In Fig.~\ref{fig:MFV2010}, we give a hint
of what the UUT analysis would look like in 2010.

In this framework, one should expect to increase the lower bound on
$\Lambda$ when the NP sensitive quantities are added to the UUT fit.
To have a quantitative example of the expected improvement, we used the
input listed above for the 2010 scenario, obtaining the $\delta S_0$
distributions shown in Fig.~\ref{fig:deltaS02010}.  From these
distributions, we get $\Lambda > 7.5$($6.6$) TeV at $95\%$
probability, in the case of positive (negative) value of $\delta S_0$,
in the case of MFV models with one Higgs doublet or low/moderate $\tan
\beta$. For the case of large $\tan \beta$, we get
\begin{eqnarray}
\Lambda & > & \left\{
\begin{array}{l} 
6.0 \mathrm{~TeV~@95\%~Prob.~for~positive~} \delta S_0^B \\
6.6 \mathrm{~TeV~@95\%~Prob.~for~negative~} \delta S_0^B \\
\end{array}\right.
\mathrm{from~} B_{d,s}-\bar B_{d,s} \mathrm{~mixing}  \nonumber \\
\Lambda & > & \left\{
\begin{array}{l} 
6.8 \mathrm{~TeV~@95\%~Prob.~for~positive~} \delta S_0^K \\
5.1 \mathrm{~TeV~@95\%~Prob.~for~negative~} \delta S_0^K \\
\end{array}\right.
\mathrm{from~} K-\bar K \mathrm{~mixing}  
\end{eqnarray}

\section{Conclusions}

We have performed a model-independent analysis of the UT in general
extensions of the SM with loop-mediated contributions to FCNC
processes. Going beyond the pure tree-level determination of the UT
already presented in Ref.~\cite{utfit2005}, we have shown how the
recent measurements performed at $B$ factories allow for a simultaneous
determination of the CKM parameters together with the NP contributions
to $|\Delta F|=2$ processes. We have found strong constraints on NP contributions
that can be as large as the SM ones only if the SM and NP amplitudes have
the same weak phase.

Motivated by this result, which points towards models with MFV, we
have analyzed in detail the UUT. By putting together all
the available information, it is possible to determine the UT
parameters almost as accurately as in the SM case and to constrain the
additional NP parameters. In this way, we probe dimension-six
operators up to a scale of 5 TeV, to be compared with the SM reference
scale of 2.4 TeV and to the sensitivity of other rare processes, which
reaches scales of $9$--$12$ TeV in the case of $b \to s
\gamma$~\cite{gino}.

Finally, we have presented a possible scenario for the UT analysis in
five years from now, taking into account foreseeable progress in
theory and experiment, under the pessimistic assumption that the SM
perfectly agrees with the data. This exercise allows us to assess the
sensitivity to NP that we can expect in the near future.  The
impressive accuracy we can reach in this kind of analyses shows the
great potential of flavour studies in investigating the structure of
NP.

\section*{Acknowledgments}
We thank the authors of ref.~\cite{pirgiolo} for pointing out to us an
error in Fig.~\ref{fig:achilleplot} and the referee for suggesting a
revision of the discussion in Sec.~\ref{subsec:dFeq1}.  This work was
supported in part by the EU network "The quest for unification" under
the contract MRTN-CT-2004-503369.

\section*{Note Added}

While completing this work, we became aware of ref.~\cite{pirgiolo}, which
contains an analysis similar to the one performed in Sec.~\ref{sec:generalC}.

\end{document}